\documentclass[referee,doublespacing,usenatbib,usegraphicx]{mn2e}
\def\apj{ApJ}
\def\apjs{ApJS}
\def\apjl{ApJL}
\def\aap{A\&A}

\def\mnras{MNRAS}

\def\sci{Sci}
\def\jcp{{J.\ Chem.\ Phys.\ }}
\def\pra{{Phys.\ Rev.\ A }}
\def\nat{Nat}
\def\jjaapp{{J.\ Appl.\ Phys.\ }}

\begin{document}

\title[Radiative Feedback by the First Stars]{Does Radiative Feedback
by the First Stars  Promote or Prevent
Second Generation Star Formation?}

\author[K. Ahn \& P. R. Shapiro]{Kyungjin Ahn\thanks{Email:
    kjahn@astro.as.utexas.edu} and Paul R. Shapiro\thanks{Email:
    shapiro@astro.as.utexas.edu} \\ 
Department of Astronomy, The University of Texas at Austin, 1
University Station C1400, Austin, TX 78712, USA}

\maketitle
\begin{abstract}
We study the effect of starlight from the first stars on the ability
of other minihaloes in their neighbourhood to form additional stars.  The 
first stars in the $\Lambda$CDM universe are believed to have formed
in minihaloes 
of total mass $\sim 10^{5-6}\,M_\odot$  at redshifts $z\ga 20$, when molecular
hydrogen ($\rm H_2$) formed and cooled the dense gas at their centres,
leading to 
gravitational collapse.  Simulations suggest that the Population III 
(Pop III) stars thus formed were massive ($\sim 100\,M_\odot$) and luminous
enough in ionizing radiation to cause an ionization front (I-front)
to sweep outward, through their host minihalo and beyond, into the
intergalactic medium.  Our previous work suggested that this I-front
was trapped when it encountered other, nearby minihaloes, and that it
failed to penetrate the dense gas at their centres within the lifetime
of the Pop III stars ($\la 3\,\rm Myrs$).  The question of what the dynamical
consequences were for these target minihaloes, of their exposure to the
ionizing and dissociating starlight from the Pop III star requires 
further study, however.  Towards this end, we have performed a series
of detailed, 1D, radiation-hydrodynamical simulations to answer the 
question of whether star formation in these surrounding minihaloes was
triggered or suppressed by radiation from the first stars.  We have
varied the distance to the source (and, hence, the flux) and the
mass and evolutionary stage of the target haloes to quantify this
effect.  We find: (1) trapping of the I-front and its transformation
from R-type to D-type, preceded by a shock front; (2) photoevaporation
of the ionized gas (i.e. all gas originally located outside the
trapping radius); (3) formation of an $\rm H_2$ precursor shell which leads
the I-front, stimulated by partial photoionization; and (4) the shock-
induced formation of $\rm H_2$ in the minihalo neutral core when the shock
speeds up and partially ionizes the gas.  The fate of the neutral core
is mostly determined by the response of the core to this shock front,
which leads to molecular cooling and collapse that, when compared to
the same halo without external radiation,
is either: 
(a) expedited, (b) delayed, (c) unaltered, or (d) reversed or prevented,
depending upon the flux (i.e. distance to the source) and the halo
mass and evolutionary stage. 
When collapse is expedited, star formation in neighbouring minihaloes
or in merging subhaloes within the host minihalo
sometimes occurs {\em within} the lifetime of the first star.
Roughly speaking, most haloes that were
destined to cool, collapse, and form stars in the absence of external
radiation are found to do so even when exposed to the first Pop III
star in their neighbourhood, while those that would not have done so
are still not able to.  
A widely held view
that the first Pop III stars must exert either positive or negative feedback
on the formation of the stars in neighbouring minihaloes should, therefore,
be revisited.   
\end{abstract}

\section{Introduction}
\label{sec:Secondstar-Intro}

Cosmological minihaloes at high redshift -- i.e. dark-matter dominated
haloes with virial temperatures $T_{\rm vir} < 10^4 \,\rm K$, with
masses above the 
Jeans mass in the intergalactic medium (IGM) before reionization
($10^4 \la M/M_\odot \la 10^8$) -- are believed to have been the sites
of the first 
star formation in the universe.  To form a star, the gas inside these haloes
must first have cooled radiatively and compressed, 
so that the baryonic component could
become self-gravitating and gravitational collapse could ensue.  For
the neutral 
gas of H and He at $T < 10^4\,\rm K$ inside minihaloes, this requires that a 
sufficient trace abundance of $\rm H_2$ molecules formed to cool the gas by atomic
collisional excitation of the rotational-vibrational lines of $\rm H_2$ .  
The formation of this trace abundance of $\rm H_2$  proceeds via the creation
of intermediaries, $\rm H^-$ or $\rm H_{2}^{+}$, which act as catalysts, which in turn
requires the presence of a trace ionized fraction, in the following
two-step gas-phase reactions (see, e.g., 
\citealt{1968ApJ...154..891P,1967Natur.216..976S,1984ApJ...280..465L,1987ApJ...318...32S};
\citealt*{1994ApJ...427...25S}, henceforth, ``SGB94''; \citealt{1998A&A...335..403G}):
\begin{eqnarray}
&&{\rm H + e^- \rightarrow H^- + \gamma},\nonumber \\
&&{\rm H^- + H \rightarrow H_2 + e^-},
\label{eq:solomon}
\end{eqnarray}
and
\begin{eqnarray}
&&{\rm H + H^+ \rightarrow H_{2}^{+} + \gamma},\nonumber \\
&&{\rm H_{2}^{+} + H \rightarrow H_2 + H^+}.
\label{eq:solomon2}
\end{eqnarray}
Unless there is a strong
destruction mechanism for $\rm H^-$ (e.g. cosmic microwave background
at $z\ga 100$), the former (equation \ref{eq:solomon}) is generally the dominant
process for $\rm H_2$ formation.

Gas-dynamical simulations of the Cold Dark Matter (CDM) universe suggest
that the first stars formed in this way when the dense gas at the centres
of minihaloes of mass $M \sim 10^{5 - 6}\, M_\odot$ cooled and
collapsed gravitationally 
at redshifts $z \ga 20$
(e.g. \citealt*{2000ApJ...540...39A,2002Sci...295...93A};
\citealt*{1999ApJ...527L...5B,2002ApJ...564...23B};
\citealt{2003ApJ...592..645Y};
\citealt*{2001ApJ...548..509M,2003MNRAS.338..273M};
\citealt{2006astro.ph..6106Y}). 
 This work and others further suggest that these
stars were massive ($M_* \ga 100 \,M_\odot$), hot
($T_{\rm eff} \simeq 10^5 \,\rm K$), and short-lived ($t_* \la 3 \,\rm
Myrs$), thus copious emitters of ionizing and dissociating radiation.

These stars constitute the Population
III (Pop III) stars, or zero metallicity stars, which are believed to
have exerted a strong, radiative feedback on 
their environment. The details of this
feedback and even the overall sign (i.e. negative or positive) are
poorly understood. 
Once the ionizing radiation escaped from its halo of origin, it created H II
regions in the IGM, beginning the process of cosmic reionization.  The
photoheating which accompanies this photoionization raises the gas pressure
in the IGM, thereby preventing baryons from collapsing gravitationally
out of the IGM into new minihaloes when they form inside the H II regions,
an effect known as ``Jeans-mass filtering''
(SGB94; \citealt{1998MNRAS.296...44G}; \citealt{2003MNRAS.346..456O}).
Inside the H II regions, whenever the I-fronts encounter pre-existing
minihaloes, those minihaloes are subject to photoevaporation
(\citealt*[henceforth, SIR]{2004MNRAS.348..753S}; 
\citealt*[henceforth, ISR]{2005MNRAS.361..405I}).
A strong background of UV photons in the Lyman-Werner (LW) bands of
$\rm H_2$ also builds up which can
dissociate molecular hydrogen inside minihaloes 
even in the neutral regions of the IGM, thereby disabling further
collapse and, thence, star formation
(e.g. \citealt{1999ApJ...518...64O}; \citealt*{2000ApJ...534...11H};
\citealt{2001ApJ...546..635O}).
This conclusion changes, however, if some additional sources of partial
ionization existed to stimulate $\rm H_2$ formation without heating the gas to
the usually high temperature of fully photoionized gas ($\sim 10^4
\,\rm K$) at which
collisional dissociation occurs, such as X-rays from miniquasars
\citep*{1996ApJ...467..522H} or if stellar sources  
create a partially-ionized boundary layer outside of intergalactic H
II regions \citep*{2001ApJ...560..580R}. Such positive feedback
effects, however, may have been only temporary, because photoheating
would soon become effective as 
background flux builds up over time \citep*{2006MNRAS.368.1301M}.

The study of feedback effects has been limited mainly
by technical difficulties.
\citet{2000ApJ...534...11H} studied the feedback of
LW, ultraviolet (UV), and X-ray backgrounds on minihaloes without
allowing hydrodynamic evolution. 
\citet{2001ApJ...560..580R} studied 
the radiative feedback effect of stellar sources only on a static,
uniform 
IGM. \citet*{2002ApJ...575...33R,2002ApJ...575...49R} studied stellar
feedback more self-consistently by performing cosmological hydrodynamic
simulations with 
radiative transfer, but the resolution of these simulations is not
adequate for resolving minihaloes. 
\citet*{2001ApJ...548..509M, 2003MNRAS.338..273M} also performed
cosmological hydrodynamic simulations, 
with higher resolution, but radiative feedback was
treated assuming the optically thin limit, which overestimates the
ionization efficiency, 
especially in the
high density regions which would initially
be easily protected from ionizing radiation due to
their high optical depth.
The first self-consistent, radiation-hydrodynamical simulations of the
feedback effect of external starlight on cosmological minihaloes were
those of SIR and ISR, who studied the
encounter between the intergalactic I-fronts that reionized the
universe and individual minihaloes along their path. These simulations
used Eulerian, grid-based hydrodynamics with radiative transfer and
adaptive mesh refinement (AMR) to ``zoom-in'' with very high
resolution, to demonstrate that the I-fronts from external ionizing
sources are trapped when they encounter minihaloes, slowing down and
transforming from weak, R-type to D-type, preceded by a shock. The gas
on the ionized side of these I-fronts was found to be evaporated in a
supersonic wind, and, if the radiative source continued to shine for a
long enough time, the I-front eventually penetrated the minihaloes
entirely and expelled all of the gas. These simulations elucidated
the impact of the I-front and the physical effects of ionizing
radiation on minihalo gas, quantifying the timescales and photon
consumption required to complete the photoevaporation. They did not, however,
address the aftermath of ``interrupted'' evaporation, when
the source turns off before evaporation is finished.

Recent studies by \citet{2005ApJ...628L...5O},
\citet*{2006ApJ...639..621A}, and \citet*{2006astro.ph..4148M}
addressed this question for minihaloes 
exposed to the radiation from the first
Pop III star in their neighbourhood, instead of the effect 
of either a steadily-driven I-front during global reionization or a
uniform global
background. 
The results of \citet{2005ApJ...628L...5O}
and \citet{2006astro.ph..4148M} are seriously misleading, however,
since they did not account properly for the optical depth to hydrogen
ionizing photons.

\citet{2005ApJ...628L...5O} assumed that the UV radiation from the
first Pop III star that formed inside a minihalo in some region would
fully ionize the gas in the neighbouring minihaloes. Using 3D
hydrodynamics simulations, they found that, when the star turned off,
$\rm H_2$ molecules formed in the dense gas that remained at the
centre of the neighbouring minihalo, fast enough to cool the gas
radiatively and cause gravitational collapse leading to more star
formation. The $\rm H_2$ formation mechanism was the same as that 
described by \citet{1987ApJ...318...32S}, in which ionized gas of
primordial composition at a temperature $T\ga 10^4 \,\rm K$ cools
radiatively and recombines out of ionization equilibrium, enabling an
enhanced residual ionized fraction to drive reaction (1) (and [2], as
well) as the temperature falls below the level at which collisional
dissociation suppresses molecule formation. As a result,
\citet{2005ApJ...628L...5O} concluded that the radiative feedback of
the first Pop III stars was positive, triggering a second generation
of star formation in the minihaloes surrounding the one that hosted the
first star.

\citet*{2006astro.ph..4148M} also used 3D hydrodynamics simulations to
consider the fate of the gas in the relic H II regions created by the
first Pop III stars. they concluded that the radiative feedback of the
first stars could be either negative or positive and  estimated a
critical UV intensity which 
would mark the transition from negative to positive feedback.
\citet{2006astro.ph..4148M}, however, studied this effect
only in the optically thin limit, as had also been done by
\citet{2001ApJ...548..509M, 2003MNRAS.338..273M}. The main mechanisms
of the positive feedback effect in \citet{2005ApJ...628L...5O} and
 \citet{2006astro.ph..4148M} are, therefore, identical.

\citet{2006ApJ...639..621A}, on
the other hand, performed a high-resolution ray-tracing calculation to
track the position of the I-front created by the first Pop III star as
it swept outward in the density field of a 3D cosmological SPH
simulation of primordial star formation in the $\Lambda$CDM universe
over the lifetime of the star. When this I-front encountered the
minihaloes in the neighbourhood of the one which hosted the first Pop
III star, it was trapped by the minihalo gas before it could reach the
high-density 
region (core), due to the minihalo's high column density of neutral hydrogen.
This is consistent with the results of SIR and ISR mentioned above.
According to \citet{2006ApJ...639..621A}, 
in fact, the lifetime of the Pop III star is less than the evaporation
times determined by SIR and ISR for the relevant minihalo masses and
flux levels in this case, so the neutral gas in the core is never
ionized by the I-front.
It seems that the initial
assumption of full ionization of nearby haloes by
\citet{2005ApJ...628L...5O} and the optically thin limit assumed by 
\citet{2006astro.ph..4148M} are invalid.

The final fate of this protected neutral core, however, is still
unclear, because the I-front tracking calculations by
\citet{2006ApJ...639..621A} 
did not include the
hydrodynamical response of the minihalo gas to its ionization,
a full treatment of radiative transfer or the
primordial chemistry 
involving $\rm H_2$. One might naively expect that the nett effect would be
negative, because heating from photoionization would ultimately expel
most of gas from minihaloes, although the results of SIR and ISR,
again, show that this minihalo evaporation 
would not be complete within the lifetime of the Pop III star.
On the other hand, partial
ionization beyond the I-front
by hard photons from a Pop III star might 
be able to {\em promote}
${\rm H_2}$ formation, 
once the dissociating UV radiation from the star is turned off,
which would then lead to a cooling and
collapsing core. This issue can be addressed only by a fully coupled
calculation of radiative transfer, chemistry, and hydrodynamics, which
will be the focus of this paper.

We shall attempt to answer the following questions: Does the light
from the first Pop III star in some neighbourhood promote or prevent
the formation of more Pop III stars in the surrounding minihaloes? More
specifically, do the neutral cores of these nearby minihaloes, which
are shielded from the ionizing radiation from the external Pop III
star, subsequently cool and collapse gravitationally, as they must in
order to form stars, or are they prevented from doing so?
Towards this end, we simulate 
the evolution of these target haloes under the influence
of an external Pop III star using the
1-D spherical, Lagrangian, radiation-hydrodynamics
code we have developed.
We adopt a $120\,M_\odot$ Pop III
star as a source, and place different mass haloes at different
distances to explore a wide range of the
parameter space for this problem. Masses of
target haloes are chosen to 
span the range from those too low for haloes to cool and
collapse by ${\rm H_2}$ cooling without external radiation to those
massive enough to do so on their own.

Our calculation is the first self-consistent gas-dynamical
calculation of the
feedback effects of a single Pop III star on nearby haloes. A
similar approach by 1-D radiation-hydrodynamics calculation has been
performed by \citet{2001MNRAS.326.1353K}. Their work, however, focuses
on the effect of a steady
global background from quasars and from stars with surface
temperatures $T_{*} \sim 10^{4} \,{\rm K}$, rather than a single,
short-lived Pop III star with $T_{*} \sim 10^{5} \,{\rm K}$. 
In addition, while we were preparing this manuscript, a study which is
similar to our work was reported by \citet{2006ApJ...645L..93S}, where
a 3D radiation-hydrodynamics calculation with SPH particles was
performed\footnote{A new preprint by \citet*{2006astro.ph..6019A} has
  also appeared which addresses this issue. We will discuss this
  further in Section \ref{sub:abel}}. A major difference of their work from
ours is that they 
focus on the subclumps of the halo which hosts the first Pop III star,
while 
we focus on external 
minihaloes in the neighbourhood of such a host halo. We also
apply a more accurate treatment of ${\rm H_2}$ self-shielding, as well
as a more complete chemistry network of neutral and ionic species of
H, He, and ${\rm H_2}$.
A more
fundamental difference from these previous studies is our finding of a
novel ${\rm 
  H_2}$ formation mechanism: {\em collisional ionization of
  pre-I-front gas by a 
  shock detached from a D-type I-front}. This mechanism occurs at the 
centre of target haloes, which would otherwise remain very
neutral. This mechanism creates new electrons abundant enough to
promote further ${\rm H_2}$ formation, which can even expedite the
core collapse.

In Section \ref{sec:code} we describe the details of the 1-D spherical
radiation-hydrodynamics code we have developed. Some details left out
in Section \ref{sec:code}
will be described in Appendices. 
In Section \ref{sec:Initial-Setup},
we describe the initial setup of our problem. 
We briefly describe a test case in Section
\ref{sec:opt-thin}, where we let a minihalo evolve from an initially
ionized state, to show that
our code reproduces the result of
\citet{2005ApJ...628L...5O} in that case.
In Section \ref{sec:mcm} and Section \ref{sec:2star-Result}, we present the main
results of our full radiation-hydrodynamics calculation. We 
summarize our results in Section \ref{sec:2star-Discussion}.  Throughout
this paper, we use the $\Lambda$CDM cosmological parameters,
($\Omega_\Lambda$, $\Omega_0$, $\Omega_b$, $h$) =
($0.73$, $0.27$, $0.043$, $0.7$), 
consistent with the {\em WMAP} first-year data
\citep{2003ApJS..148..175S}\footnote{As we do not perform a
  statistical study, our result is independent of the cosmic density power
  spectrum. The three-year {\em WMAP} data does not show a big
  discrepancy in the set of cosmological parameters of the interest in
  this paper \citep{2006astro.ph..3449S}. The change in
  $\sigma_8$ and the index of the primordial power
  spectrum $n$ would translate to $\sim 1.4$ redshift delay of
  structure formation and reionization \citep{2006ApJ...644L.101A}}.

\section{Numerical Method: 1-D spherical, radiation-hydrodynamics
  with primordial chemistry network}
\label{sec:code}

In this section, we describe in detail the 1-D spherical, Lagrangian,
radiation-hydrodynamics code we have developed
for both dark and baryonic matter.
We describe how hydrodynamics, dark matter dynamics,
radiative transfer, radiative heating and cooling, and finally the
nonequilibrium chemistry are handled.
The finite differencing scheme, reaction rates, and certain other
details not treated in this section will be described in Appendices.
We include the neutral
and ionic species of  
H, He and ${\rm H_{2}}$, namely H, $\rm H^+$, He, ${\rm He^{+}}$,
${\rm He^{++}}$, $\rm H^-$, $\rm H_2$,
$\rm H_{2}^+$ and $e^-$, in order to treat the primordial chemistry fully.
As deuterium and lithium exist in a negligible amount, we neglect ${\rm
  D}$ and ${\rm Li}$ species%
\footnote{D and Li components have usually been neglected due to their
  relatively 
low abundance, hence the negligible contribution to cooling (e.g.
\citealt{1984ApJ...280..465L,1987ApJ...318...32S}). Recent
studies by \citet{2005MNRAS.364.1378N} and \citet{2006MNRAS.366..247J},
however, show that enough HD is generated in
strongly-shocked, ionized primordial gas which then can cool
below the temperature of $\sim 100 \,\rm K$ already achieved by $\rm
H_2$ cooling alone, down
to the temperature of the CMB.
As the HD cooling process is negligible if gas remains neutral
(e.g. \citet{2006MNRAS.366..247J}), however, 
we may neglect the HD cooling process in our calculation as long as
we are interested in the centre of target haloes which remains mostly
neutral at any time. We will discuss this issue further in Section
\ref{sec:2star-Discussion}.%
}.

\subsection{Hydrodynamic Conservation Equations}
\label{sub:2star-Hydrodynamics}

The baryonic gas obeys inviscid fluid conservation equations,
\begin{equation}
\frac{\partial\rho}{\partial t}+\frac{\partial}{r^{2}\partial
  r}(r^{2}(\rho u))=0,
\label{eq:realhydro_mass}
\end{equation}
\begin{equation}
\frac{\partial}{\partial t}(\rho u)+\frac{\partial}{\partial r}(p+\rho
u^{2})+\frac{2}{r}\rho u^{2}=-\rho\frac{Gm}{r^{2}},
\label{eq:realhydro_momentum}
\end{equation}
\begin{equation}
\frac{De}{Dt}=-\frac{p}{\rho}\frac{\partial}{r^{2}\partial
  r}(r^{2}u)+\frac{\Gamma-\Lambda}{\rho},
\label{eq:realhydro_energy}
\end{equation}
 where $e\equiv(3p)/(2\rho)$ is the internal energy per unit baryon
mass, $\Gamma$ is the external heating rate, and $\Lambda$ is the
radiative cooling rate. Note that all the variables in equations
(\ref{eq:realhydro_mass}) 
- (\ref{eq:realhydro_energy}) denote baryonic properties, except
for $m$, the mass enclosed by a radius $r$, which is composed of
both dark and baryonic matter.

We do not change the adiabatic index $\gamma$ throughout the
simulation. As long as monatomic species, H and He, dominate the abundance,
$\gamma=5/3$ is the right value to use.
This ratio of specific heats,
$\gamma$, can change significantly, however, if
a large fraction of H is converted into molecules. For example,
the three-body ${\rm H_2}$ formation process,
\begin{equation}
{\rm H+H+H \rightarrow H_2 + H},
\label{threebody}
\end{equation}
will occur vigorously when $n_{\rm H}\ga 10^8 \,{\rm cm\, s^{-1}}$ and $T\la
10^{3} \,{\rm K}$, which will invalidate the use of a constant $\gamma$.
To circumvent such a problem, when such high density occurs, we simply
stop the simulation. This process is, nevertheless, important in forming the
protostellar molecular cloud
(e.g. \citealt{2002Sci...295...93A}). This issue will be further discussed in
Section \ref{sec:2star-Result}, when we define the criterion for the
collapse of cooling regions.

The shock is treated using the usual artificial viscosity technique
(e.g. \citealt{artvis}). The pressure $p$ in equations
(\ref{eq:realhydro_momentum}) and (\ref{eq:realhydro_energy}) contains
the artificial viscosity term. The details of this implementation are
described in Appendix A.

\subsection{Dark Matter Dynamics}
\label{sub:fluid-approx}

Gravity is contributed both by the dark matter and the baryonic components.
Let us first focus on the dark matter component. In order to treat
the dark matter gravity under spherical symmetry, almost all previous
studies have used either a frozen dark matter potential or a set of
self-gravitating dark matter shells in radial motion only (e.g.
\citealt{1995ApJ...442..480T}). Both methods have their own limitations.
The frozen potential approximation cannot address the effect of a
possible evolution of the gravitational potential. The radial-only dark
matter approximation suffers from the lack of any tangential motion,
producing a virialized structure whose
central density profile is much steeper ($\rho\propto r^{-\beta}$ with
$\beta \ge 2$; see e.g. ) than that of haloes in cosmological,
3-D N-body simulations ($\beta \approx 1$, as found in
\citealt*{1997ApJ...490..493N}).

In order to treat the dynamics of dark matter more accurately than
these previous treatments, we use the 
the fluid approximation we have developed and reported elsewhere
\citep{2005MNRAS.363.1092A}. We briefly summarize its
derivation here; for a detailed description, see
\citet{2005MNRAS.363.1092A}. Collisionless CDM particles are described by the
collisionless Boltzmann equation. When integrated, it yields an
infinite set of conservation equations, which is called the BBGKY
hierarchy (e.g. \citealt{1987gady.book.....B}). However, CDM N-body
simulations show that virialized haloes are well approximated by
spherical symmetry. These simulations also show that the velocity
dispersions are highly isotropic: radial dispersion is almost the same
as the tangential dispersion. These two conditions make it possible to
truncate the hierarchy of equations to a good approximation, which
then yields only three sets of conservation equations. Amazingly
enough, these equations are identical to the normal fluid conservation
equations for the adiabatic index $\gamma=5/3$ gas:
\begin{equation}
\frac{\partial\rho_d}{\partial t}+\frac{\partial}{r^{2}\partial
  r}(r^{2}(\rho_d u_d))=0,
\label{eq:DM_mass}
\end{equation}
\begin{equation}
\frac{\partial}{\partial t}(\rho_d u_d)+\frac{\partial}{\partial r}(p_d+\rho_d
u_{d}^{2})+\frac{2}{r}\rho_d u_{d}^{2}=-\rho_d \frac{Gm}{r^{2}},
\label{eq:DM_momentum}
\end{equation}
\begin{equation}
\frac{De_d}{Dt}=-\frac{p_d}{\rho_d}\frac{\partial}{r^{2}\partial
  r}(r^{2}u_d),
\label{eq:DM_energy}
\end{equation}
where the subscript $d$ represents dark matter, the effective pressure
$p_d \equiv \rho_d \left\langle u_d - \left\langle u_d\right\rangle
\right\rangle^2$ is the product of the dark matter density and the
velocity dispersion at a given radius, and the effective internal
energy per dark matter mass $e_d \equiv 3 p_d/ 2 \rho_d$.
We use these effective fluid
conservation equations (equation \ref{eq:DM_mass},
\ref{eq:DM_momentum}, 
\ref{eq:DM_energy}) to handle the motion of dark matter particles.

Note that dark
matter shells in this code represent a collection of dark matter particles
in spherical bins, in order to describe {}``coarse-grained'' properties
such as density ($\rho_d$) and the effective pressure ($p_d$).
As these coarse-grained variables follow the usual fluid conservation
equations, the hyperbolicity of these equations leads to the formation
of an effective ``shock.'' The location of this shock will determine the
effective ``post-shock'' region. This post-shock region 
corresponds to the dark matter shell-crossing
region. Because of the presence of this effective shock, we also use
the artificial viscosity technique. This collisional behaviour of our
coarse-grained dark matter shells originates from our choice of
physical variable.
For further details, the reader is referred to
\citet{2005MNRAS.363.1092A} and \citet{2003RMxAC..18....4A} for
description and
application of our fluid approximation.

The mass enclosed by a dark matter shell of radius $r$,
\begin{equation}
m(<r)=m_{{\rm DM}}(<r)+m_{{\rm bary}}(<r),
\label{eq:totalmass}
\end{equation}
enters equations (\ref{eq:realhydro_momentum}) and
 (\ref{eq:DM_momentum}). When computing $m(<r)$,
we properly take account of the
mismatch of the location of dark matter shells and baryon shells.

\subsection{Radiative transfer}
A full, multi-frequency, radiative transfer calculation is performed
in the code. Since ${\rm H_2}$ cooling
is of prime importance here, we first pay
special attention to calculating the optical depth to UV
dissociating photons in the LW bands and 
the corresponding ${\rm H_2}$ self-shielding
function. We then describe how we calculate the optical depth
associated with any other species depending upon the location of the
radiation source. The finite difference scheme for the calculation of
radiative rates is described in the Appendix A.

\subsubsection{Photodissociation of $\rm H_2$ and Self-Shielding}
\label{sub:ss}
Hydrogen molecules are photodissociated
when a UV photon in the LW bands between $11\,\rm eV$ and $13.6\,\rm
eV$ excites $\rm H_2$ to an excited electronic state from which
dissociation sometimes occurs.
When the column density of ${\rm H_2}$ becomes high
enough ($N_{\rm H_2}\ga 10^{14} \,\rm cm^{-2}$), the optical depth
to photons in these Lyman-Werner bands can be high, so ${\rm H_2}$ can
``self-shield'' from dissociating photons.
Exact calculation of this self-shielding
requires a full treatment of all 76 Lyman-Werner lines, even when only
the lowest energy level transitions are included. 
Such a calculation is feasible under simplified conditions such as
a radiative transfer problem through a static medium
(e.g. \citealt{2000ApJ...534...11H};
\citealt{2001ApJ...560..580R}). Unfortunately, for combined calculations of
radiative transfer and hydrodynamics, such a full treatment
is computationally very expensive. 

Under certain circumstances, however,
one can use a pre-computed self-shielding function expressed in terms of
the molecular column density $N_{{\rm
    H_{2}}}$ and the temperature $T$ of gas, which saves a great
amount of computation time. 
In a \emph{cold, static} medium, for instance, one can use a self-shielding
function provided by \citet{1996ApJ...468..269D}:
\begin{equation}
F_{{\rm shield}}={\rm min}\left[1,\left(\frac{N_{{\rm
        H_{2}}}}{10^{14}{\rm
      cm^{-2}}}\right)^{-3/4}\right].
\label{eq:DB_shield_factor_cold}
\end{equation}
The photodissociation rate
is then given by
\begin{equation}
k_{{\rm H_{2}}}=1.38\times10^{9}\left(J_{\nu}\right)_{h\nu=12.87{\rm
    eV}}F_{{\rm shield}},
\label{eq:DB_rate}
\end{equation}
where $\left(J_{\nu}\right)_{h\nu=12.87{\rm eV}}$ ($\rm
erg\,s^{-1}\,cm^{-2}\,Hz^{-1}\,sr^{-1}$) is the mean intensity in the
spectral region of the LW bands.
This approximation has been widely used in the study of high redshift
structure formation
(e.g.
\citealt{2001MNRAS.326.1353K, 2001MNRAS.321..385G,2003ApJ...592..645Y,
  2004ApJ...613..631K}).

The problem with equation (\ref{eq:DB_shield_factor_cold}) is that when the
gas temperature is high or gas has motion along the line of sight to
the source, the thermal and velocity
broadening of the LW bands caused by the Doppler effect
can significantly reduce 
the optical depth. A better treatment for thermal broadening is also given
by
\citet{1996ApJ...468..269D}, now in terms of the molecular column
density $N_{\rm H_{2}}$ and the velocity-spread parameter $b$ of the gas:
\begin{eqnarray}
F_{{\rm shield}}= \frac{0.965}{(1+x/b_5)^2} + 
\frac{0.035}{(1+x)^{0.5}} \nonumber \\
\times \exp[-8.5\times 10^{-4}(1+x)^{0.5}],
\label{eq:DB_shield_factor}
\end{eqnarray}
where $x\equiv N_{\rm H_2}/5\times 10^{14}\, {\rm cm}^{-2}$, $b_5\equiv
b/10^5 {\rm cm \, s^{-1}} $, and $b=1.29 \times 10^4
\left(T_K/A\right)^{1/2} {\rm cm \, s^{-1}},$ where $A$ is the atomic
  weight \citep{1978ppim.book.....S}. For ${\rm H_2}$, $b=9.12 \,{\rm
    km \, s^{-1}} \left(T/10^4 \, {\rm K}\right)^{1/2}$.

In the problem treated in this paper, we frequently find  $T\approx
10^3 - 5\times 10^3 \, {\rm K}$ in the gas parcel (shell)
which contributes most of the ${\rm H_2}$ column density. We also find
that this gas parcel usually moves at $v \approx 2-5 \, {\rm km \,
  s^{-1}}$ (see Section \ref{sub:H2shell}). The combined effect of the
thermal broadening and the 
Doppler shift on the shielding function, then, may be well approximated
by a thermally broadened shielding function with $T\approx 10^4\,{\rm
  K}$. Throughout this paper, therefore, we use equation
(\ref{eq:DB_shield_factor}) with $T= 10^4\,{\rm K}$ to calculate
the self-shielding. For the photo-dissociation rate, we use equation
(\ref{eq:DB_rate}).

\begin{figure}
\includegraphics[%
  width=84mm]{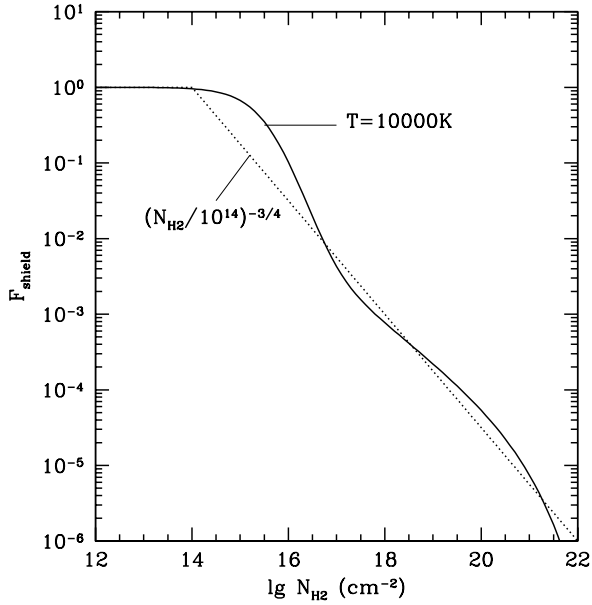}

\caption{Power-law self-shielding function for cold, static gas
  vs. self-shielding function for hot gas at $T=10^4\, {\rm K}$. The
  problem of interest to us resides in the sensitive region, $N_{\rm
  H_2} \approx 10^{14} - 10^{16} \, {\rm cm}^{-2}$, where the biggest
  discrepancy exists.
\label{fig-DB}}
\end{figure}

We show in Fig. \ref{fig-DB} how much the static, cold shielding function
(equation \ref{eq:DB_shield_factor_cold}) 
may overestimate the self-shielding in our problem, by comparing this to the
thermally-broadened shielding function (equation
\ref{eq:DB_shield_factor}) at $T= 10^4\,{\rm K}$. The biggest
discrepancy between these two shielding functions exists for $N_{\rm
  H_2} \approx 10^{14} - 10^{16} \, {\rm cm}^{-2}$. Interestingly
enough, the ${\rm H_2}$ column density in our problem usually resides
in this regime. It is crucial, therefore, to take into
account the effects of 
thermal broadening and Doppler shift carefully, as we do in this
paper.

\subsubsection{External Source}
\label{sub:External-source}

Since our calculations are 1-D, spherically-symmetric, we have assumed
the external radiation source contributes a radial flux
$F_{\nu}^{{\rm ext}}(r)$ at frequency $\nu$ and radius $r$, measured
from the minihalo centre, given by
\begin{equation}
F_{\nu}^{{\rm ext}}(r)=
    \frac{L_{\nu}^{{\rm ext}}}{4\pi D^{2}}e^{-\tau_{\nu}(>r)},
\label{eq:fnu_ext}
\end{equation}
 where $L_{\nu}^{{\rm ext}}$ is the source luminosity, and
 $\tau_{\nu}(>r)$ is the optical depth along the radial direction from
 radius $r$ to the source located at a distance $r=D$. 

The radiative rate of species $i$ at radius $r$ is then given by
\begin{equation}
k_{i}(r)=\int_{0}^{\infty}d\nu\frac{\sigma_{i,\nu}4\pi J_{\nu}(r)}{h\nu}
        =\int_{0}^{\infty}d\nu\frac{\sigma_{i,\nu}F_{\nu}^{{\rm
      ext}}(r)}{h\nu},
\label{eq:rad_rate_ext_body}
\end{equation}
where we have used the fact that $4\pi J_{\nu} = F_{\nu}^{\rm ext}$,
as long as the external radiation can be approximated as a 1D planar flux.
In practice, one calculates this rate in a given grid-cell -- i.e. spherical
shell -- with finite thickness. If such a grid-cell has a
small optical depth, $F_{\nu}^{{\rm ext}}$ is almost constant across
the grid, so one could take the grid-centred value of $F_{\nu}^{{\rm
ext}}$ to calculate $k_{i}(r)$. This naive scheme, however, does
not yield an accurate result when a grid-cell is optically thick, where
$F_{\nu}^{{\rm ext}}$ may vary significantly over the cell width.
This problem occurs frequently for solving radiative transfer through
optically thick media, where individual cells have large optical depth.
In order to resolve this problem, we
use a ``photon-conserving'' scheme like that described by
\citet{1999MNRAS.309..287R} and \citet{1999ApJ...523...66A}. The
details of our implementation of this 
scheme are described
in the Appendix A.

\subsection{Heating and Cooling}
\label{sub:heating-cooling}
\subsubsection{Photoheating}
\label{sub:photo-heating}

Photoheating results from thermalization of the residual kinetic energy
of electrons after they are photoionized. In general, the photoheating
function is described by 
\begin{eqnarray}
\Gamma=\sum_{i}\Gamma_{i}&=&\sum_{i}n_{i}\int_{0}^{\infty}d\nu\frac{4\pi
  J_{\nu}\sigma_{\nu}}{h\nu}(h\nu-h\nu_{i,{\rm th}}) \nonumber \\
&=&\sum_{i}n_{i}\int_{0}^{\infty}d\nu\frac{
  F_{\nu}^{\rm ext}\sigma_{\nu}}{h\nu}(h\nu-h\nu_{i,{\rm th}}),
\label{eq:generic_photo_heat}
\end{eqnarray}
where $h \nu_{i,{\rm th}}$ is the threshold energy over which the
residual photon energy is converted into the kinetic energy of
electrons, and the nett heating 
function $\Gamma$ is the sum of individual heating functions
(\{$\Gamma_{i}$\}). 
In finite-differencing equation
(\ref{eq:generic_photo_heat}), we also use the photon-conserving
scheme as we do for equation (\ref{eq:rad_rate_ext_body}). This
prevents cells with large optical depth from obtaining unphysically high
heating rates. See Appendix A for details.

\subsubsection{Radiative cooling}

Cooling occurs through various processes. For atomic species, it comes
from collisional excitation, collisional ionization, recombination,
free-free emission, and CMB photons scattering off free electrons
(Compton cooling/heating). For atomic H and He, cooling is dominated
by collisional excitation (for $T\la 2\times 10^{5}{\rm K}$) and free-free
emission (for $T\ga 2\times 10^{5}{\rm K}$). The atomic cooling rate
decreases rapidly at $T\la 10^{4}{\rm K}$, as there are no collisions
energetic enough to cause excitation. It is difficult, therefore,
to cool gas below $T\approx10^{4}{\rm K}$ solely by atomic cooling
of primordial gas. 

Molecular hydrogen (${\rm H_{2}}$), however, is able to cool gas below
$T\approx10^{4}{\rm K}$,
down to $T\approx100{\rm K}$,
by collisional excitation of rotational-vibrational lines by H atoms.
An important question to address is how much ${\rm H_{2}}$ is created,
maintained, or destroyed under the
influence of an ionizing and dissociating radiation field. 
Even a small
fraction, $n_{\rm H_{2}}/n_{\rm H}\ga 10^{-4}$, is sometimes
enough to cool 
gas below $10^{4}{\rm K}$ (e.g. see \citealt{1987ApJ...318...32S}).

We use cooling rates in the parametrized forms given by
\citet{1997NewA....2..209A}, except for the hydrogen molecular
cooling. For ${\rm H_2}$ cooling, we use the fit given by
\citet{1998A&A...335..403G}, where the low density cooling rate has
been updated significantly from the previously used rate by
\citet{1984ApJ...280..465L}, which suffers from the uncertainties
associated with the only collisional coefficients available at that
time. At low densities, $n_{\rm H} \la 10^2 {\rm cm^{-3}}$, the cooling rate
of \citet{1984ApJ...280..465L} is bigger by an order of magnitude than
that of \citet{1998A&A...335..403G} at $T\approx 1000 K$.

\subsection{Nonequilibrium chemistry}
\label{sub:noneq_chem}

The general rate equation for the abundance of
species $i$ is given by 
\begin{equation}
\frac{\partial n_{i}}{\partial t}=C_{i}(T,\{ n_{j}\})-D_{i}(T,\{
n_{j}\})n_{i},
\label{eq:verygeneric_rate_eq}
\end{equation}
where $C_{i}$ is the collective source term for the creation of species
$i$, and the second term is the collective {}``sink'' term for
the destruction of species $i$. 
The processes included and adopted are shown in Table
\ref{table:rates} in Appendix B. Most of the rate coefficients are those
from the fits by \citet{1987ApJ...318...32S}, with a few updates.

We also adopt the rate solving scheme proposed by \citet{1997NewA....2..181A}.
It is well known that coupled rate equations in the form of equation
(\ref{eq:verygeneric_rate_eq}) are {}``stiff'' differential equations,
whose numerical solution suffers from instability if explicit ODE
solvers are used. \citet{1997NewA....2..181A} show that their implicit,
backward difference scheme provides enough stability. Accuracy of
the solution is achieved by updating each species in some specific
order, rather than updating all species simultaneously from their values
at the last time step. In addition, the abundance of the relatively
fast reactions of ${\rm H^{-}}$ and ${\rm H_{2}^{+}}$ are approximated
by their equilibrium values, which are expressed by simple algebraic
equations. See the Appendix A for the corresponding
finite-differencing scheme.

We will frequently quote our results in terms of the fractional
number density of species $i$, 
$y_{i}\equiv\frac{n_{i}}{n_{{\rm H}}}$, 
where $n_{{\rm H}}$ is the number density of the total atomic hydrogen
atoms.
We use $x$, however, to denote the fractional electron number
density, $y_{e}$, which is a measure of the ionized fraction.

\subsection{Code tests}
\label{sec:codetest}

We tested our code against the following
problems which have analytic solutions:

(A) the self-similar, spherical, cosmological infall and accretion
shock resulting from a point-mass perturbation in an Einstein-de
Sitter universe of gas and collisionless dark matter
\citep{1985ApJS...58...39B}; 

(B) the self-similar blast wave which results from a strong, adiabatic
point explosion in a uniform gas -- the Sedov solution
(\citealt{1959sdmm.book.....S}) 

(C) the propagation of an I-front from a steady point-source in a
uniform, static medium

(D) the gas-dynamical expansion of an H II region from a point source
in a uniform gas \citep{1966ApJ...143..700L}

(E) the gas-dynamical expansion-phase of the H II region from a
point-source in a nonuniform gas whose density varies with distance
$r$ from the source as $r^{-w}$, $w=3/2$ \citep*{1990ApJ...349..126F}.

Our code passed all the tests described above with an acceptable
accuracy. Test results are described in Appendix C.

\section{The Simulations}

\subsection{Initial Setup}
\label{sec:Initial-Setup}

We now describe the initial setup for the problem of radiative feedback
effects of Pop III stars on nearby haloes at $z\approx20$.
The first stars form inside rare, high density peaks at high
redshift. We 
place target haloes of different mass $M=[2.5\times 10^{4}, \, 
5\times 10^{4},\,
10^{5},\,
2\times 10^{5},\,
4\times 10^{5},\,
8\times 10^{5}]\,
M_{\odot}$ at different
locations from the source, with proper distance
$D=\{180,\,\,360,\,\,540,\,\,1000\}\,{\rm pc}$, 
which are all assumed to be affected directly by the radiation field
from the source Pop III star of mass $M_{*}=120\,
M_{\odot}$\footnote{The additional case of $D=50\,\rm pc$, $F_0=600$,
  $M=5.5\times 10^5 \,M_\odot$, will be discussed separately in Section
  \ref{sub:abel} with regard to 
the case in which the target minihalo is merging with the minihalo
which hosts the star, separated by less
  than its virial radius from the star}. 
We expose the target halo to this radiation field for the lifetime of
the star, $t_{*}(120\, M_{\odot})\simeq2.5\,{\rm Myrs}$
(\citealt{2002A&A...382...28S}). 
The source Pop III star is assumed to be located in a halo of mass
$M\simeq10^{6}M_{\odot}.$ Time is measured from the arrival of the
stellar radiation at the location of the target minihalo.

This setup is well justified by the cosmological simulations
by \citet{2006ApJ...639..621A}. 
A cosmological gas and N-body simulation of structure formation in the
$\Lambda$CDM universe on small scales by a GADGET/SPH code was used to
identify the site at which the first Pop III star would form.
This occurred at $z=20$, at the location of 
 the highest density SPH
particle in the
simulation box, located within a halo of mass
$M\simeq10^{6}M_{\odot}.$ 
This provided the initial density field for the I-front tracking
calculations in \citet{2006ApJ...639..621A}. The
I-front from this first star
escaped from the host halo 
quickly with high escape fraction, traveling as a supersonic, 
weak R-type front.
By the end of the lifetime of the star ($\sim[3-2]\,{\rm Myrs}$)
for stellar masses in the range $M_* \sim[80-200]M_{\odot}$, the star's
H II region had reached a maximum 
radius of about $3\,{\rm kpc}$.

We approximate the spectral energy distribution (SED) of the source
star by a blackbody spectrum.
A Pop III star of mass $M_{*}\approx120\, M_{\odot}$, according to
\citet{2002A&A...382...28S}, has the time-average effective temperature
$T_{{\rm eff}}\approx10^{5}{\rm K}$ and luminosity
$L=\int_{0}^{\infty}d\nu L_{\nu}\approx10^{6.243} L_{\odot}$.
The corresponding
ionizing photon luminosity with this blackbody spectrum is 
$Q_* \equiv \int_{\nu_{\rm H}}^{\infty}d\nu L_{\nu}/h \nu
=1.5\times
10^{50} \rm s^{-1}$, where $h\nu_{\rm H}\equiv 13.6\,{\rm eV}$ is the
hydrogen ionization threshold energy. We assume
that the source radiates with these time-averaged values throughout
its lifetime, then stops. As the photons escape in a time scale short
compared to the lifetime of the star and the escape fraction is
high, we simply ignore the effect of the intervening gas (e.g. optical
depth from the host halo and the IGM) and assume that the bare
radiation field hits the edge of 
target haloes directly. 

As we fix the luminosity of the source, different
distances correspond to different fluxes. 
We express the
frequency-integrated ionizing photon flux, 
$F$ in units of $\rm 10^{50}\, s^{-1}\, kpc^{-2}$, to give the
dimensionless flux,
$F_0 \equiv N_{\rm
  ph,50}/D_{\rm kpc}^2 = N_{\rm ph,56}/D_{\rm Mpc}^2$, where 
$N_{\rm ph,50}$ is the ionizing photon luminosity (in units of $\rm
10^{50}\,s^{-1}$) and $D_{\rm kpc}$ ($D_{\rm Mpc}$) is the distance in
units of kpc (Mpc), respectively.
The value $F_0\approx 1$ is typical for minihaloes encountered by
intergalactic I-fronts during global
reionization (e.g. see
\citealt{2004MNRAS.348..753S}). Interestingly enough, $F_0$ for our
``small-scale'' problem has a similar value. The Pop III star in
our problem has $N_{\rm ph,50}\equiv
Q_*/10^{50}\,s^{-1}=1.5$. 
For distances $180\,{\rm pc}$,
$360\,{\rm pc}$, $540\,{\rm pc}$ and $1000\,{\rm pc}$, $F_0$
corresponds to 46.3, 11.6, 5.14 and 1.5,
respectively.

\subsection{Initial Halo Structure}
\label{sub:IHS}
For the initial halo
structure, we adopt the minimum-energy
truncated isothermal sphere (TIS) model
(\citealt*{1999MNRAS.307..203S}; \citealt{2001MNRAS.325..468I}), which
will be described further in Section \ref{sub:phase1}.
The thermodynamic
properties and chemical abundances of the gas in these
target haloes, however, is
somewhat ambiguous. 
The density and virial temperature of these haloes
are higher than those of the IGM in general, which drives their
chemical abundances to change from the IGM equilibrium state to a new
equilibrium state. The most notable feature is the change of $y_{\rm
  H_2}$ and $x$. The IGM equilibrium value of the electron abundance,
$x\approx 10^{-4}$, is high enough to promote ${\rm H_2}$ formation
inside minihaloes to yield a high molecule fraction, $y_{\rm H_2}\approx
10^{-4} - 10^{-3}$. 
At the density of gas in the halo core,
this newly created ${\rm H_2}$ is capable of cooling the
minihalo gas to $T\approx 100\, {\rm K}$, and depending on the virial
temperature, the minihalo may, therefore, undergo a runaway collapse.

The time for this evolution of the target halo gas is short compared
to the age of the universe when the first star forms in their
neighbourhood. As a result, it is likely that the target haloes are
exposed to the ionizing and dissociating radiation from that first
star as they are in the midst of evolving, with fine-tuning required
to catch all of them in a particular stage of this evolution.
As the evolutionary ``phase'' of our target haloes is uncertain, we adopt two
different phases as our representative initial conditions. 
  In Phase I,
  chemical abundances have not yet evolved away from their IGM
  equilibrium values. This stage is characterized by low $\rm H_2$
  fraction, $y_{\rm 
  H_2}\sim 2\times 10^{-6}$ and high electron fraction, $x\sim
  10^{-4}$. 
 Phase II is the state which is reached, after
  allowing the Phase I minihalo to evolve chemically, thermodynamically
  and hydrodynamically for a few million years (a small fraction of a Hubble
  time, $t_{\rm H}=186\,\rm Myrs$ at $z=20$), until 
  the electron fraction has
  decreased to $x\sim 10^{-5}$. Phase II is characterized by high $\rm H_2$
  fraction, $y_{\rm 
  H_2}\sim 10^{-4} - 10^{-3}$, and cooling-induced compression of the
core relative to Phase I, by a 
  factor between 1 and 20, higher for higher minihalo mass.

\subsubsection{Phase I: Unevolved Halo with IGM chemical abundance in
  hydrostatic equilibrium}
\label{sub:phase1}
The first phase we choose is the initial state we assumed above,
namely the nonsingular TIS structure with IGM chemical
abundances. This phase is 
characterized by gas in hydrostatic equilibrium, with
the truncation radius (outer boundary of the halo)
\begin{eqnarray}
r_{t}&=&102.3
\left(\frac{\Omega_{0}}{0.27}\right)^{-1/3}
\left(\frac{h}{0.7}\right)^{-2/3}      \nonumber \\
&&\times \left(\frac{M}{2\cdot10^{5}M_{\odot}}\right)^{1/3} 
\left(\frac{1+z}{1+20}\right)^{-1}\,{\rm pc},
\end{eqnarray}
the virial temperature
\begin{eqnarray}
T&=&593.5
\left(\frac{\mu}{1.22}\right)
\left(\frac{\Omega_{0}}{0.27}\right)^{1/3}
\left(\frac{h}{0.7}\right)^{2/3} \nonumber \\
&&\times \left(\frac{M}{2\cdot10^{5}M_{\odot}}\right)^{2/3}
\left(\frac{1+z}{1+20}\right)\,{\rm K},
\label{eq:tvir}
\end{eqnarray}
where $\mu$ is the mean molecular weight (1.22 for neutral gas and
0.59 for ionized gas) and the central density
\begin{equation}
\rho_{0}=4.144\times10^{-22}
\left(\frac{\Omega_{0}}{0.27}\right)
\left(\frac{h}{0.7}\right)^{2}
\left(\frac{1+z}{1+20}\right)^{3}\,
{\rm  g\,\, cm^{-3}},
\end{equation}
which can also be expressed in terms of the hydrogen number density by
\begin{eqnarray}
n_{{\rm H,0}}&=&\frac{X(\Omega_{b}/\Omega_{0})\rho_{0}}{m_{{\rm
      H}}} \nonumber \\
&=&30\left(\frac{X}{0.76}\right)\left(\frac{\Omega_{b}}{0.043}\right)\left(\frac{h}{0.7}\right)^{2}\left(\frac{1+z}{1+20}\right)^{3}{\rm \, cm^{-3}},
\end{eqnarray}
where $X$ is the hydrogen mass fraction in the baryon component. This
central density is about $1.8\times 10^4 \,\overline{\rho}(z),$ where
$\overline{\rho}(z)$ is the mean matter density at redshift $z$, while 
at $r=r_{\rm tr}$, $\rho=35\,\overline{\rho}(z)$. For more details,
see \citet{1999MNRAS.307..203S} and \citet{2001MNRAS.325..468I}.

We assign chemical abundances that reflect the IGM
equilibrium state, which is characterized by high electron fraction --
high enough to promote ${\rm H_2}$ formation under the right conditions --
and low ${\rm H_2}$ fraction -- low enough to contribute negligible molecular
cooling. We adopt $y_{\rm H}=1$, $y_{{\rm He}}=0.0789$,
$x\simeq y_{{\rm H^{+}}}=10^{-4}$, 
$y_{{\rm H_{2}}}=2\times10^{-6}$, and $\{y_{i}\}=0$ for other species
(see, e.g. SGB94; \citealt{2001ApJ...560..580R}).

\subsubsection{Phase II: Evolved Halo with Recombining and Cooling Core}
\label{sub:phase2}
The second initial condition we choose is the evolved state (Phase II)
reached by allowing the system to evolve from Phase I initial
conditions before the arrival of radiation from the Pop III star.
In particular, we follow this evolution until
the central electron 
fraction has dropped to $10^{-5}$ by recombination from Phase I. 
We choose this condition because it is now characterized by high
molecule and low electron fraction, contrary to Phase I. The fate of
this halo will then mainly be determined by how easily this abundant
${\rm H_2}$ is 
protected against dissociating radiation after the star turns on.
The answer will also depend upon how much change has occurred
hydrodynamically, because in some cases the halo core may have cooled
and collapsed significantly enough to be unaffected by the feedback from
{\em late} irradiation.

\begin{figure*}
\includegraphics[%
  width=68mm]{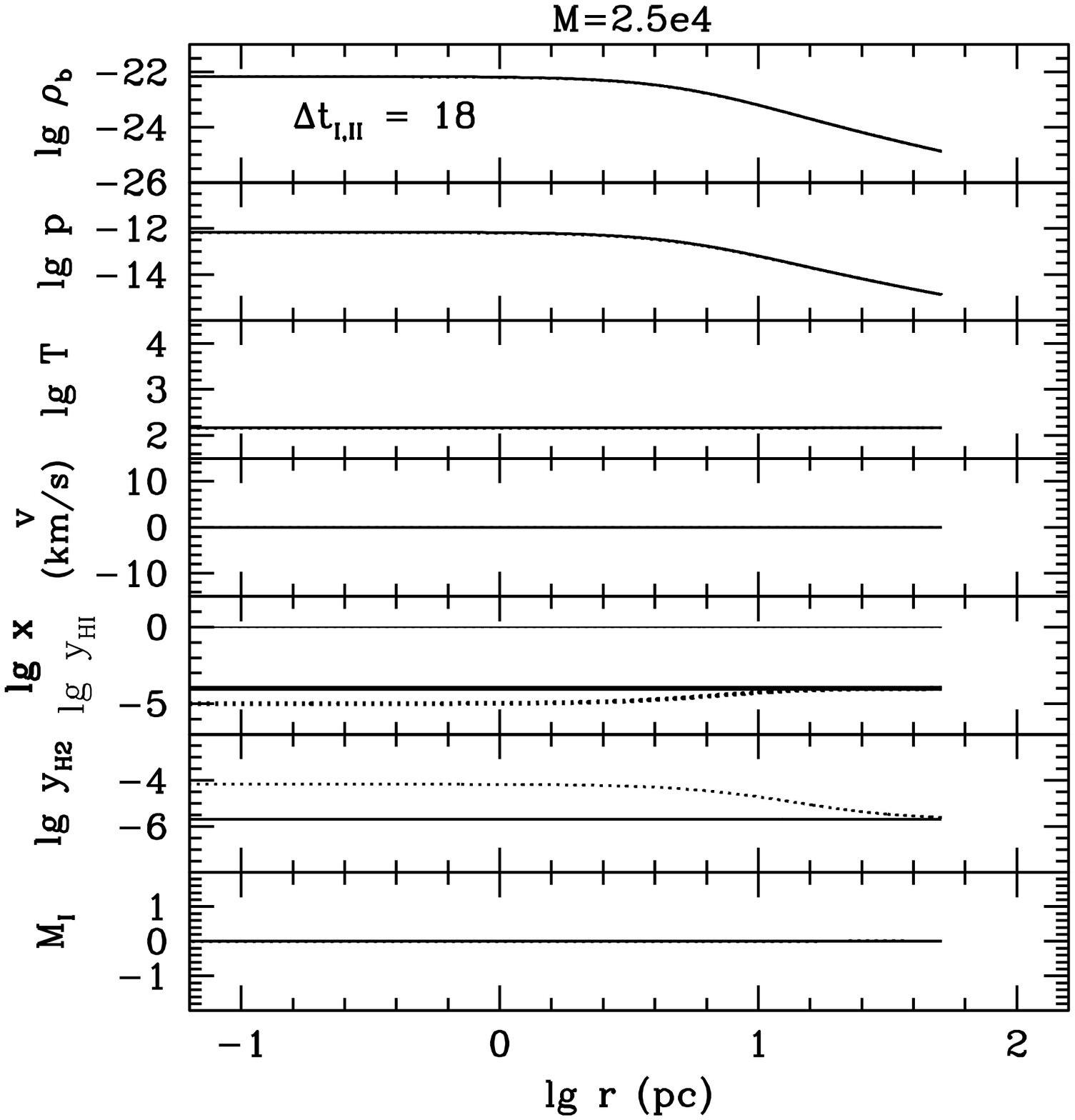}
\includegraphics[%
  width=68mm]{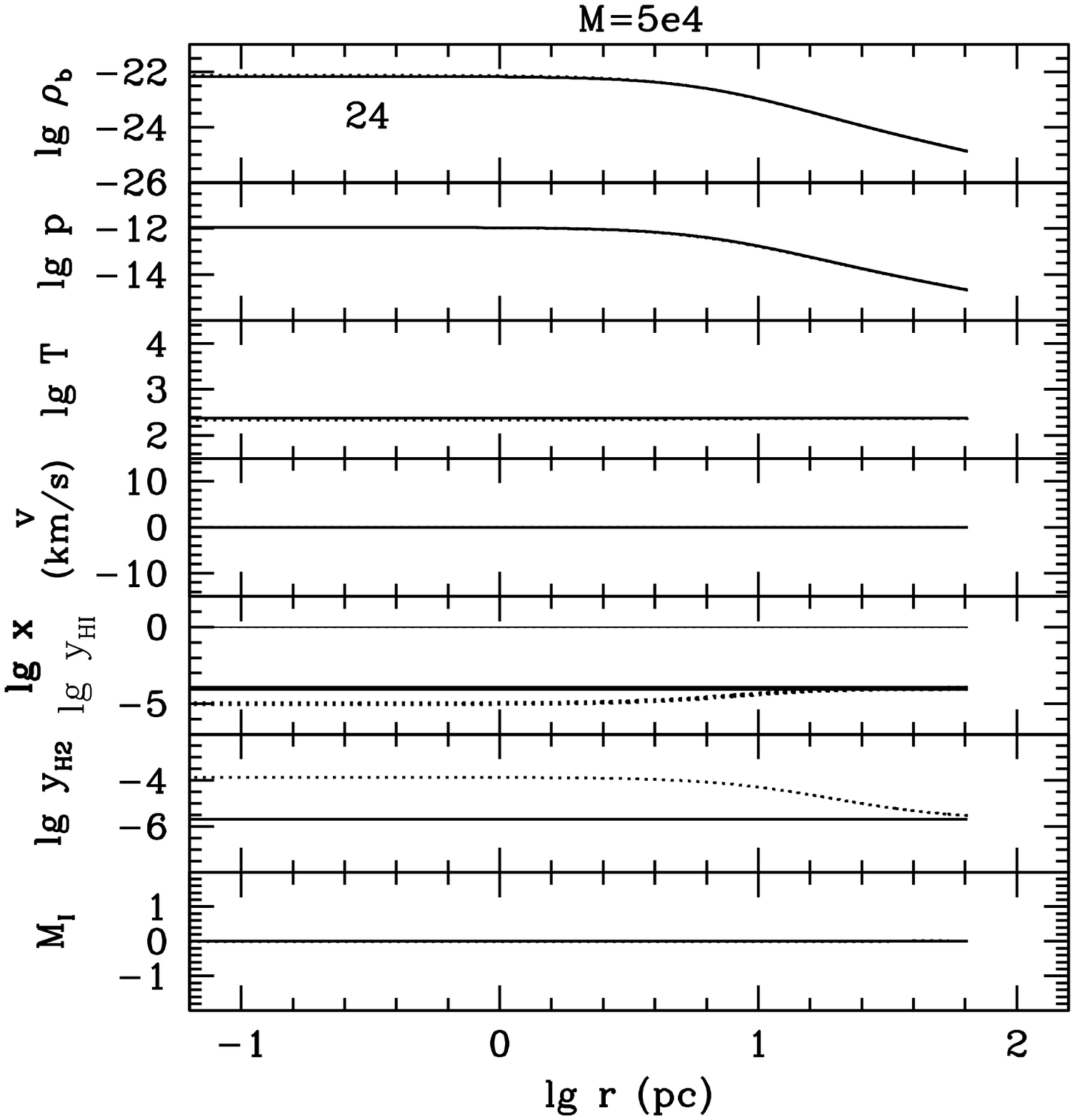}
\includegraphics[%
  width=68mm]{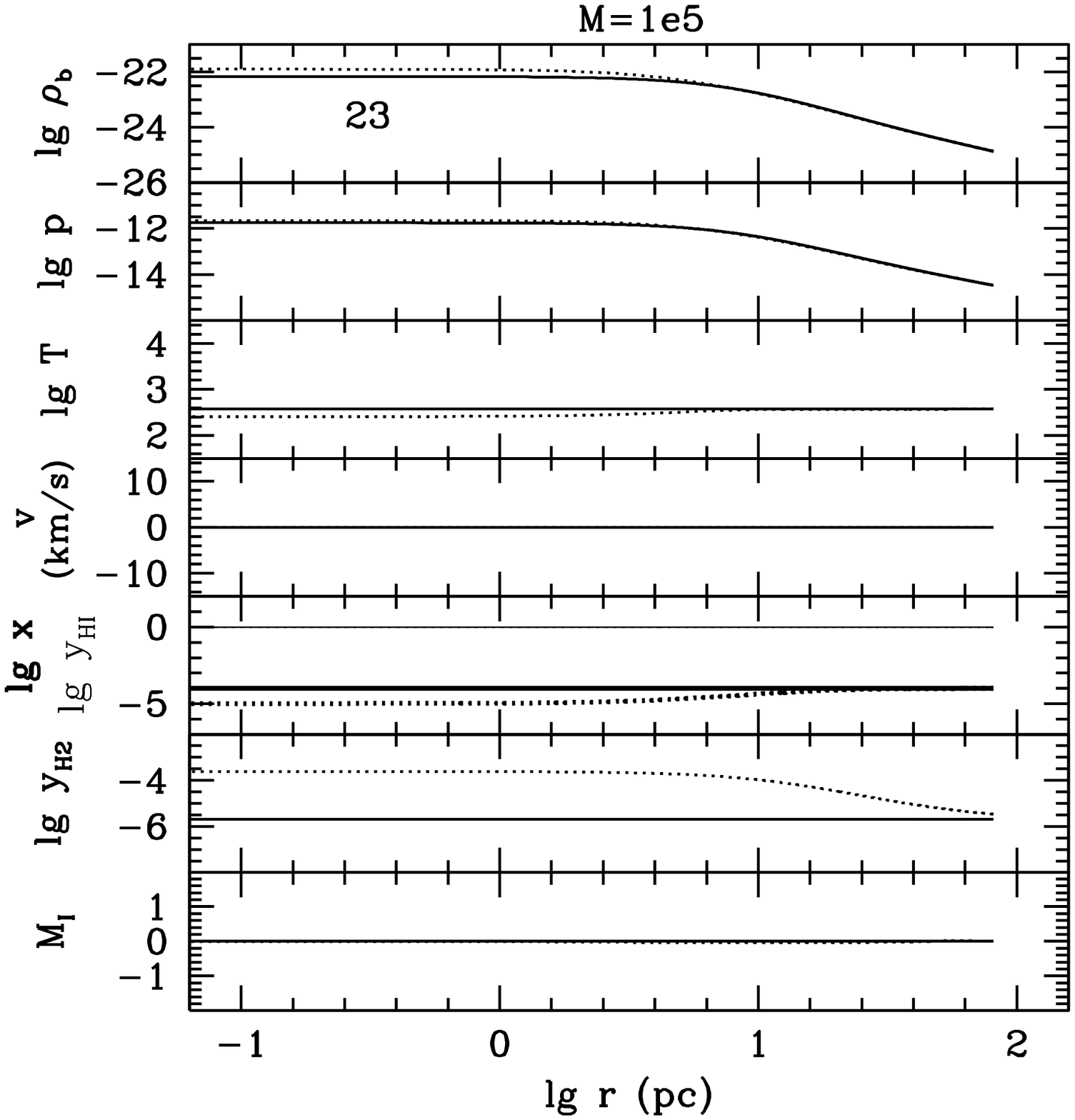}
\includegraphics[%
  width=68mm]{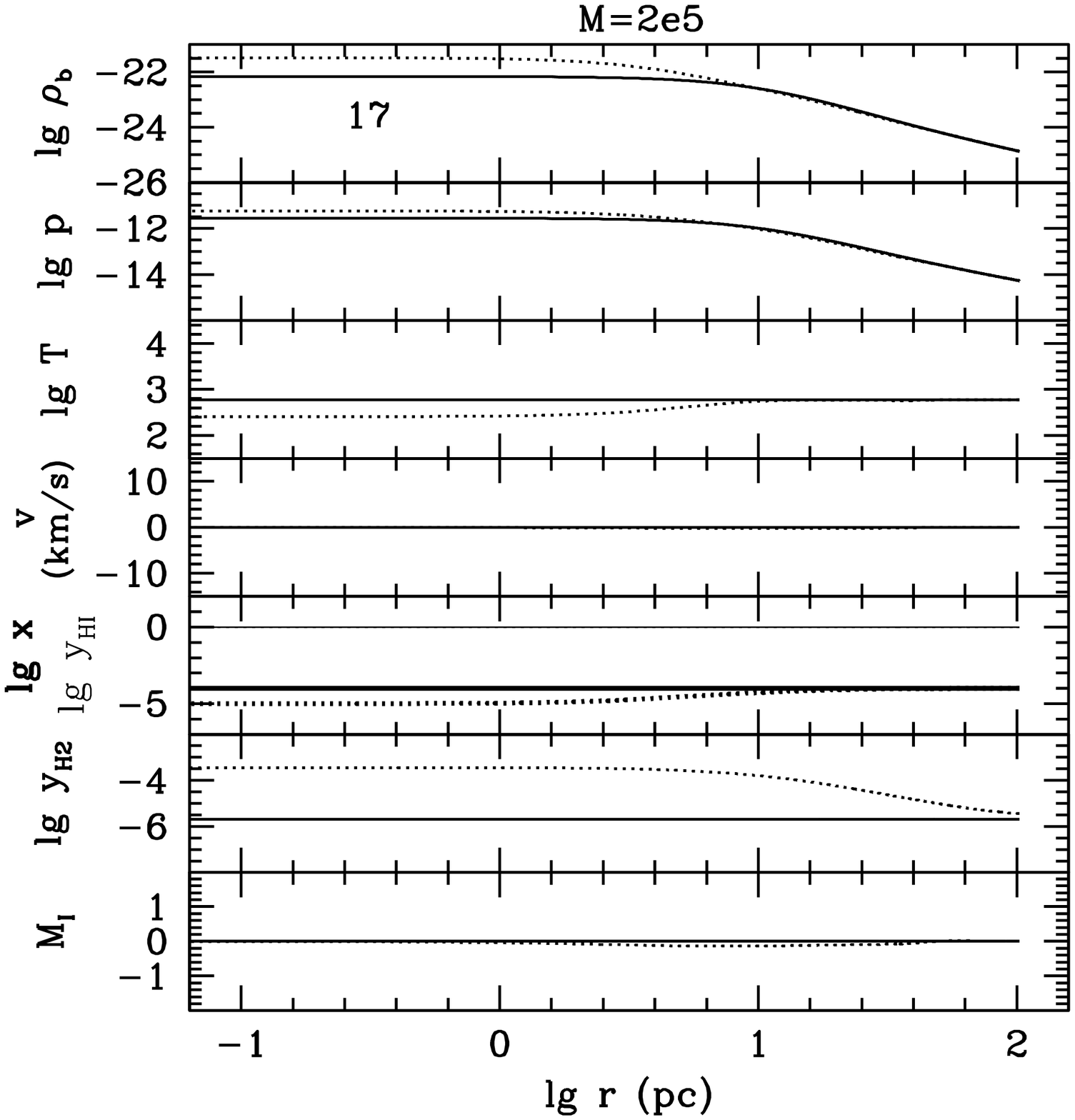}
\includegraphics[%
  width=68mm]{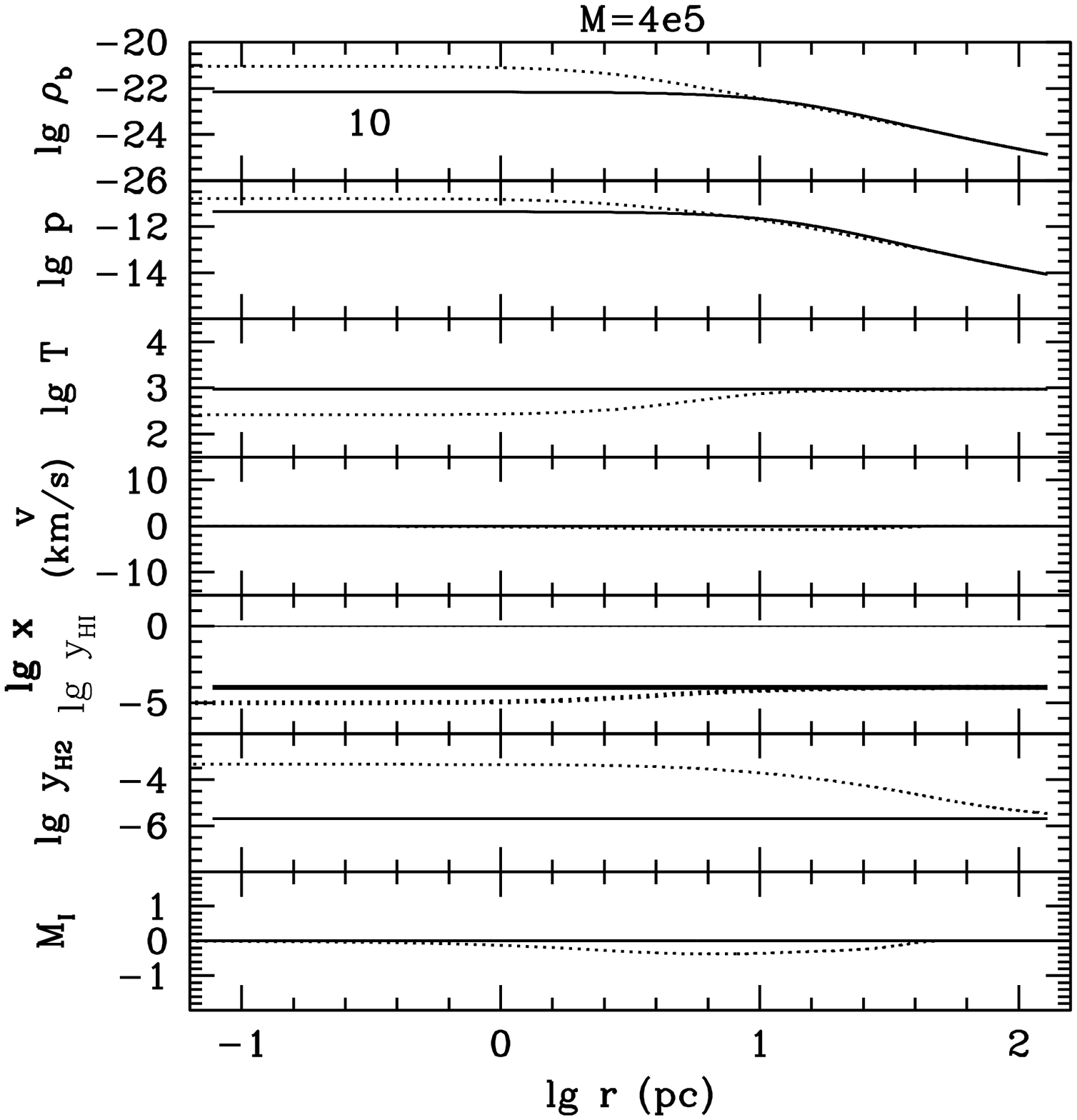}
\includegraphics[%
  width=68mm]{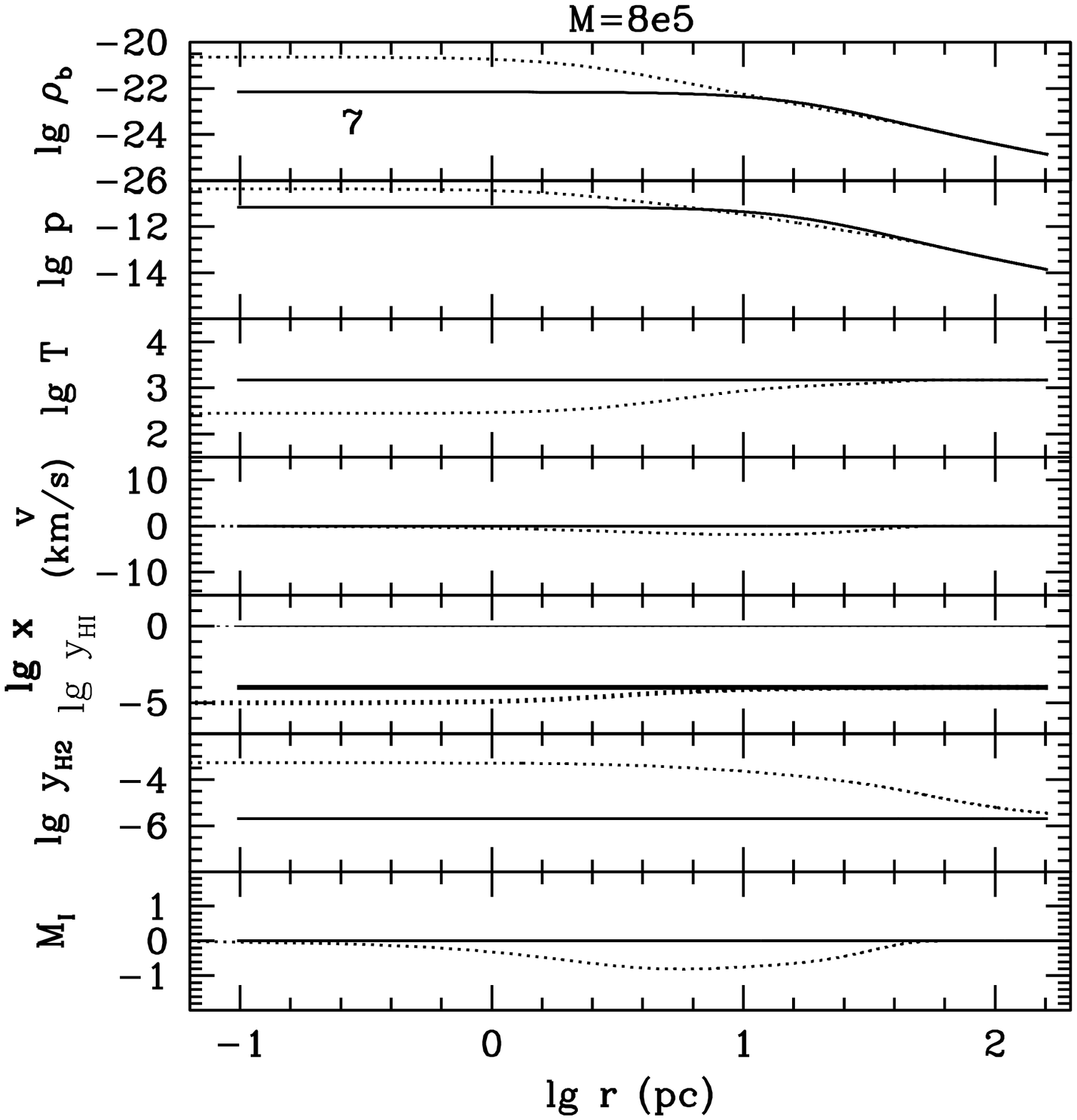}

\caption{Initial conditions for target haloes. We choose two different
  phases of TIS halo evolution as separate initial conditions. Phase I
  (unevolved; IGM abundance; solid) and Phase II (evolved from Phase I
  for a time $\Delta t_{\rm I,II}$ until $x=10^{-5}$
  at centre; dotted) 
  are plotted for each mass of target halo. 
Each panel is labelled with the value of $\Delta t_{\rm I,II}$ in Myrs.
Note that hydrodynamic
  difference between two phases is evident in haloes of mass $M\ga
  2\times 10^5 \, M_\odot$.
\label{fig-init}}
\end{figure*}

The time to
reach Phase II is different for different mass haloes because of different
gas properties. Initially,
as we start from the TIS density profile whose central density is
independent of the halo mass, the recombination rate is higher for
smaller mass haloes, because  hydrogen recombines according to
the following:
\begin{equation}
\label{eq:k2_dep}
\frac{dx}{dt} \propto n_{\rm H} n_{e^-} T^{-0.7}.
\end{equation}
The situation becomes complicated, however, once
evolution begins and density changes. 
The ${\rm H_2}$ cooling and collapse in the central region of the
haloes is increasingly effective as
halo mass increases, because of the increasingly large difference between the
virial temperature 
and the ${\rm H_2}$ cooling temperature plateau, $\sim 100\,{\rm
  K}$. The corresponding rapid collapse and cooling in massive haloes
can easily  
offset the initial temperature dependence by obtaining high density
and low temperature, as is
seen in equation (\ref{eq:k2_dep}).
Phase II for large mass haloes represents haloes that have already
started their cooling and collapse.

In Fig. \ref{fig-init}, we show halo profiles in Phase I and Phase
II for different halo masses. 
We also
show how much time it takes for the haloes to evolve from
Phase I to Phase II.
The times for gas at the halo centre to recombine to
$x=10^{-5}$ are in the range $7 \le \Delta t_{\rm I,\,II}(\rm Myrs)
\le 24$ for halo masses $0.25 \le M/(10^5\,M_\odot) \le 8$,
peaked at $\Delta t_{\rm I,\,II}=24 \,\rm Myrs$
for $M=5\times 10^4 \, M_\odot$.  In all cases, $\Delta t_{\rm I,\,II}
\ll t_{\rm H}=186\,\rm Myrs$, the age of the universe at $z=20$.

\section{Halo Evolution from Fully-Ionized Initial Conditions: The
  Consequences of Irradiation Without Optical Depth}
\label{sec:opt-thin}
Before describing the results of our full radiative transfer,
hydrodynamics calculation, we describe an experiment 
designed to show the effect of neglecting the optical depth of the
minihalo to ionizing radiation from the external star during the
star's lifetime on the minihalo's evolution after the star shuts
off. For this purpose, we assume the target minihalo is initially
fully-ionized and heated to the temperature of a photoionized gas as
it would be if it were instantaneously flash-ionized by starlight in
the optically-thin limit.
Such a setup is equivalent to that used by
\citet{2005ApJ...628L...5O}, where they find that 
second-generation star formation is triggered 
when the ionization of the minihalo caused by the nearby Pop III star
leads to
cooling by ${\rm
  H_{2}}$. The high initial electron fraction is present because of
the assumption of full ionization allows quick formation of ${\rm
  H_{2}}$, which then cools the central region before it reaches
the escape velocity.

For this experiment, we 
initialized ionized fractions as following:
$y_{\rm H I}=6.4\times 10^{-4}$,
$x=1.15$, $y_{\rm H II} = 1$, 
$y_{\rm He I}=6.8\times 10^{-6}$, $y_{\rm He II}=8.9\times 10^{-3}$,
$y_{\rm He III}=7\times 10^{-2}$, $y_i =0$ for other
species. Without disturbing the halo density profile -- we use the TIS halo
model, which is described in Section \ref{sub:phase1} --, we also assigned
a high initial temperature appropriate for photoionized gas,
$T=2\times 10^{4} {\rm K}$. These abundance and temperature values roughly
mimic the condition found in typical H II regions.

\begin{figure}
\includegraphics[%
  width=84mm]{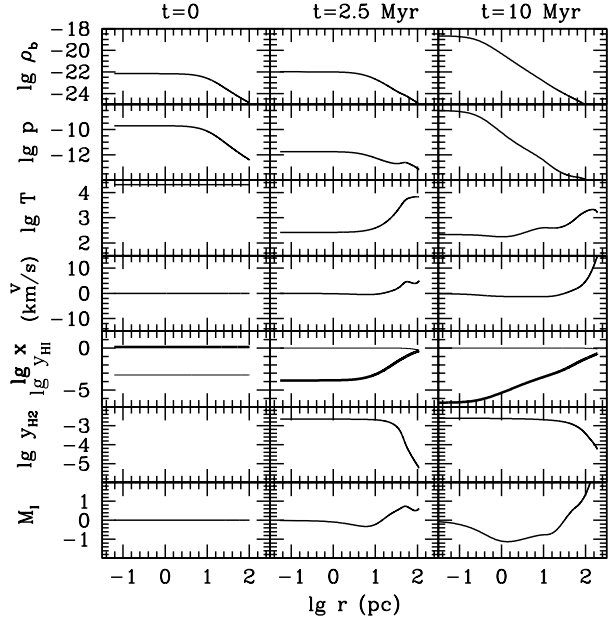}

\caption{Evolution of a ``flash-ionized''
  TIS halo -- i.e. initially fully-ionized at $T=2\times 10^4\,\rm K$
  of $M=2\times 10^5 M_{\odot}$.
  From left to
  right, each panel corresponds to $t$=0, 2.5, and 7.5 Myrs. Note that
  initially ($t$=0) absent $\rm H_2$ is quickly created and cools the
  central region,
  while the initially hot gas in the outskirts evaporates before it
  cools.
  In each panel, from top to bottom, baryon density $\rho_b$, pressure
  $p$,
  temperature $T$, velocity $v$, electron fraction($x$; thick)/neutral
  fraction($y_{\rm H I}$; thin), molecule fraction $y_{\rm H_2}$ and
  the isothermal Mach number $M_{\rm I}$ are plotted,
  respectively. Unless specified otherwise, the subsequent figures
  will follow the convention used in this figure.
\label{fig-zap}}
\end{figure}

We find that such an initial condition leads to the collapse of the
core region, when the
 formation of ${\rm H_{2}}$ stimulated by the high initial electron
fraction enables $\rm H_2$ cooling.
Gas in the outskirts 
evaporates from the halo, however,
because pressure forces accelerate
the gas to escape velocity before it can
form ${\rm H_{2}}$ and cool. 
The ${\rm H_{2}}$ cooling and adiabatic cooling which happen later 
in this outflowing gas do not reverse
the evaporation (Fig. \ref{fig-zap}). 

Our results for this case agree with the outcome of
\citet{2005ApJ...628L...5O}. 
This led those authors to suggest that the first stars exerted a positive
feedback effect on their surroundings, triggering a second generation of
star formation.
A question arises, however, as to whether this
fully-ionized initial condition of nearby minihaloes is actually
achieved
by the first Pop III star to form in their neighbourhood. 
As
already mentioned in Section~\ref{sec:Secondstar-Intro},
\citet{2006ApJ...639..621A} found that the I-front from this Pop III star
gets trapped in those minihaloes and cannot reach the central region
before the star dies.
In this paper, we will confirm that the fully-ionized initial
condition of \citet{2005ApJ...628L...5O} is never achieved when one
considers the coupled radiative and hydrodynamic processes more fully.
We will also show that, if any protostellar
region is to form in the target halo, 
it does so in the neutral core region
which the ionizing photons do not penetrate. 

\section{Minimum Halo Mass for Collapse: the case without radiative
  feedback} 
\label{sec:mcm}
When a minihalo forms as a nonlinear, virialized,
gravitationally-bound  structure out of the linearly
perturbed IGM, a change of chemical abundance occurs due to the change
of gas properties. Most importantly, the hydrogen molecule fraction
changes from the 
IGM equilibrium value, $y_{\rm H_2}\sim 2\times 10^{-6}$, to a new
equilibrium value, $y_{\rm H_2}\ga 10^{-4}$. Even with such a small
fraction, ${\rm H_2}$ can cool gas to $T_{\rm H_2}\simeq 100\,{\rm
  K}$, where $T_{\rm H_2}$ represents the temperature ``plateau'' that
gas in primordial composition can reach by $\rm H_2$ cooling.

There exists a minimum collapse mass of minihaloes, $M_{\rm c,min}$,
above which haloes, in the absence of external radiation, 
can form cooling and collapsing cores within the Hubble time at a
given redshift. 
The gap between the $\rm H_2$ cooling plateau temperature, $T_{\rm
  H_2}$, and the 
minihalo virial temperature, $T_{\rm vir}$, given by equation
(\ref{eq:tvir}) is a useful indicator of the success or failure of
collapse. For instance, at $z\approx 
20$, $T_{\rm vir}\sim 160\,{\rm K}$ for $M=2.5\times
10^4\,M_\odot$. As $T_{\rm vir}\simeq T_{\rm H_2}$ , even after gas
cools to $T_{\rm H_2}$, it cannot collapse fast enough to serve as a
site for star formation. 
On the other hand, $T_{\rm
  vir}\sim 10^3 \,\rm K $ for $M=4\times 10^5\,M_\odot$, and the
temperature's cooling down to $T_{\rm H_2}\approx 100\,\rm K$ will make
the gas gravitationally unstable, which will
lead to runaway 
collapse. This argument is supported by the results of
\citet*{1996ApJ...464..523H}, for example, that collapse can occur
only in haloes with $T_{\rm vir}\ga 100\,\rm K$.

We model the initial minihalo structure by the TIS model as described in Section
\ref{sub:phase1} and let it evolve in the absence of radiation,
starting from the IGM chemical abundance and
minihalo virial temperature (Phase I). We determine $M_{\rm c,min}$
by the criterion
\begin{equation}
t_{\rm coll}=t_{\rm H},
\end{equation} 
where $t_{\rm coll}$ is the time at which the central density reaches
$n_{\rm H}=10^8\,\rm cm^{-3}$ (the density suitable for initiating
three-body $\rm H_2$ formation; see
e.g. \citealt{2000ApJ...540...39A}), and 
$t_{\rm H}$ is the Hubble time at a given redshift.

\begin{figure*}
\includegraphics[%
  width=68mm]{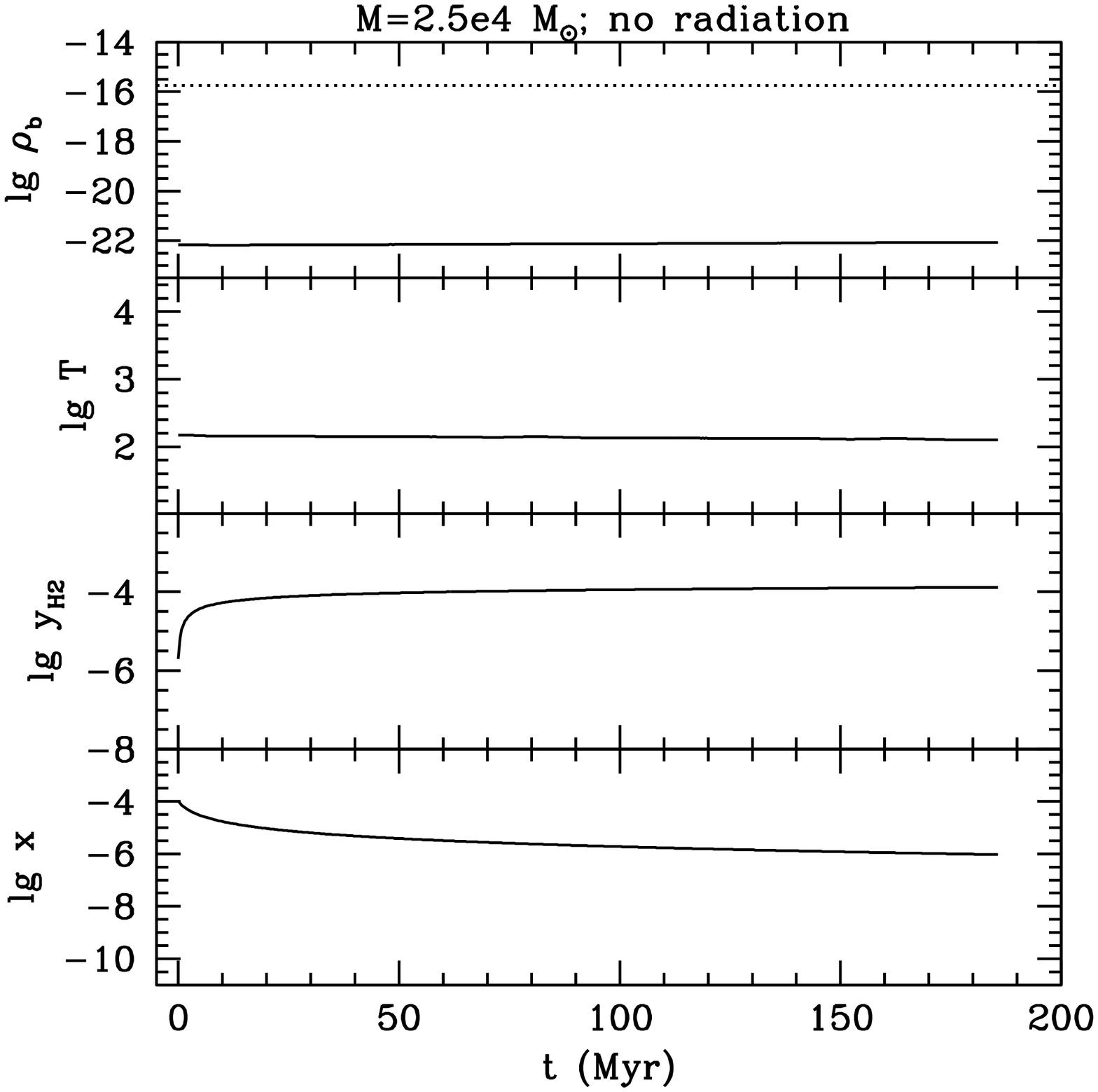}
\includegraphics[%
  width=68mm]{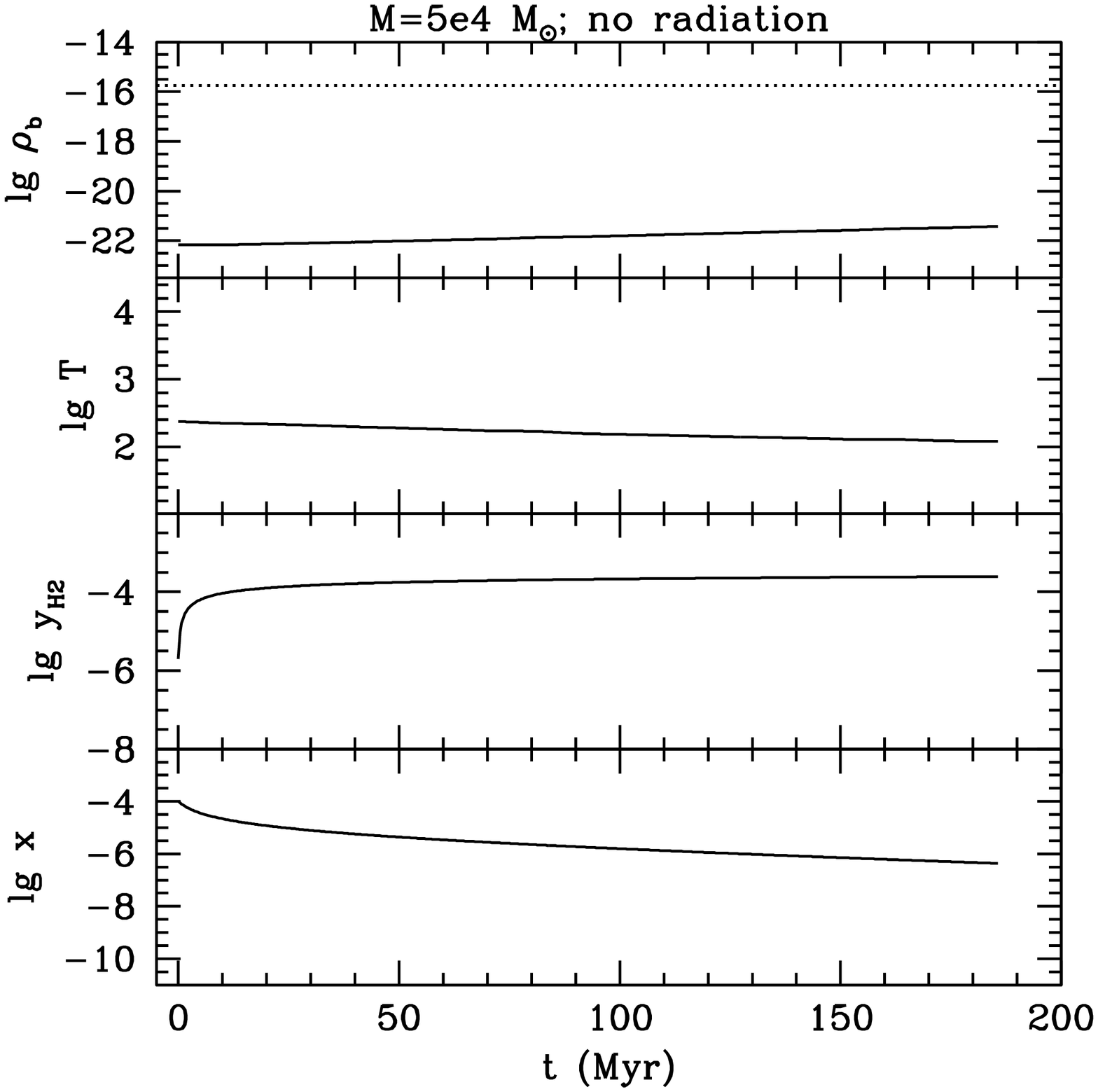}
\includegraphics[%
  width=68mm]{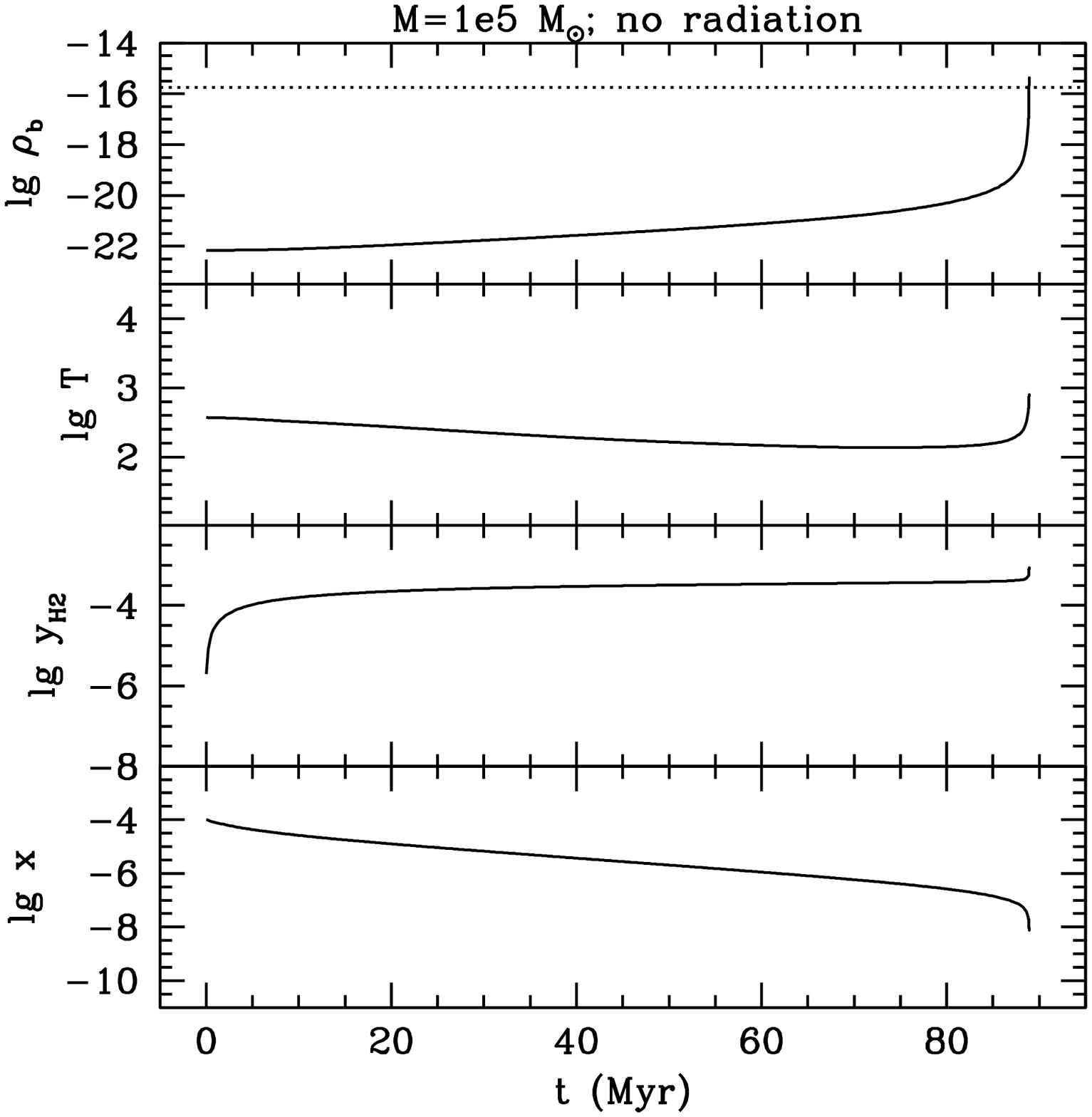}
\includegraphics[%
  width=68mm]{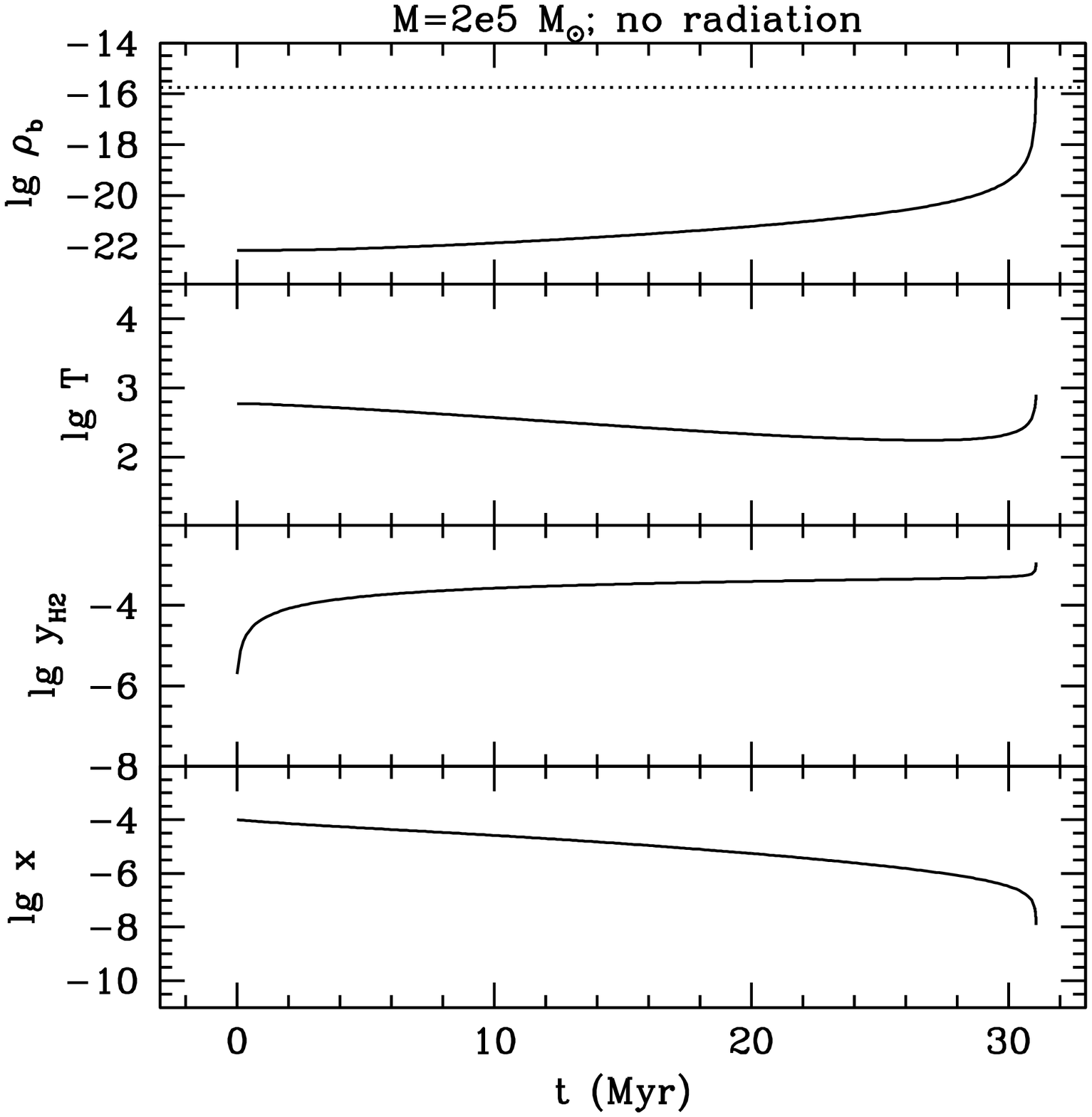}
\includegraphics[%
  width=68mm]{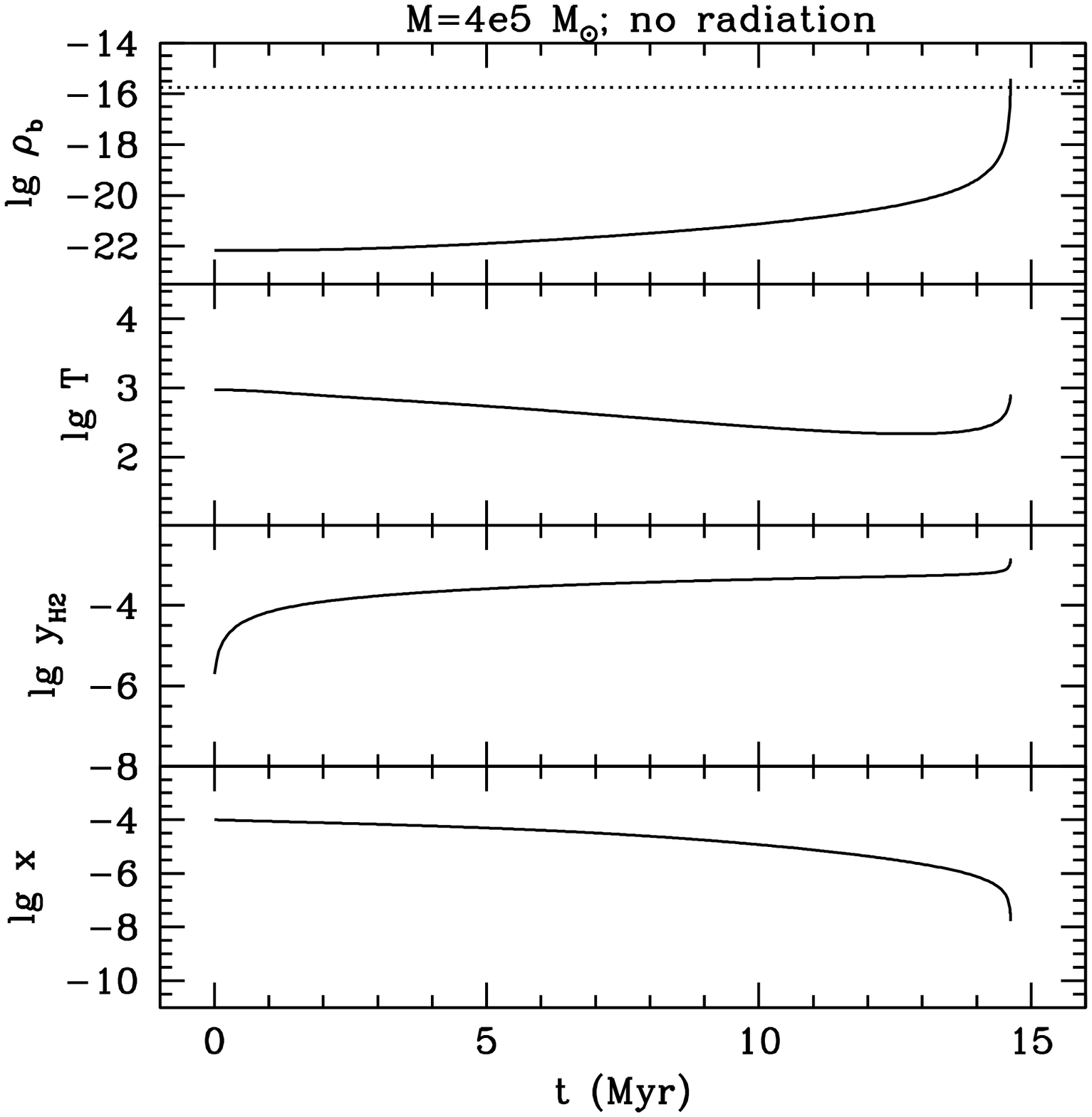}
\includegraphics[%
  width=68mm]{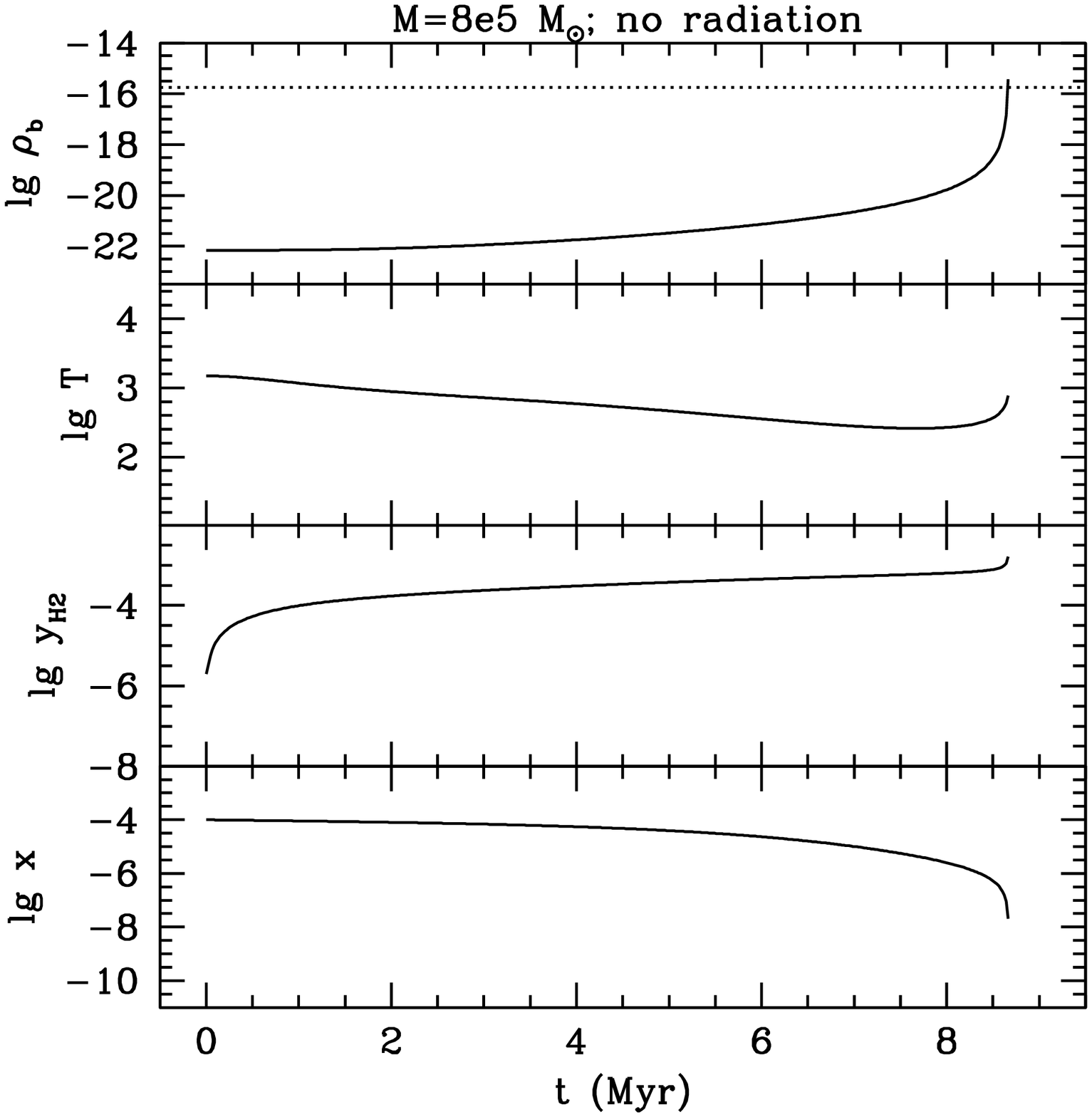}

\caption{Evolution of the centre of TIS minihaloes without radiation.
  Initially, minihaloes have structure described by the TIS model (see
  Phase I depicted by solid lines in Fig. \ref{fig-init}) with the equilibrium
  value of primordial IGM chemical abundances. We define $t_{\rm
  coll}$ as the time  to reach $n_{\rm H}=10^{8}\,\rm cm^{-3}$,
  represented by the horizontal dotted lines.
\label{fig-freeevol}}
\end{figure*}

We find that $M_{\rm c,min}\simeq 7\times 10^4\,M_\odot$ at $z=20$ (see
Fig. \ref{fig-freeevol}). We
have plotted the evolution of minihalo centres in the absence of
radiation, where each run starts from Phase I. This is in rough
agreement with $M_{\rm c,min}\simeq 1.25\times 10^5\,M_\odot$, the
value found by \citet{2001ApJ...548..509M}. The discrepancy is
larger with results by \citet{2000ApJ...544....6F} and
\citet{2003ApJ...592..645Y}, where they obtain $M_{\rm c,min}\simeq
7\times 10^5\,M_\odot$. The biggest contrast exists with
\citet{1997ApJ...474....1T},  where they find $M_{\rm c,min}\simeq 2\times
10^6\,M_\odot$ at $z\approx 20$, almost 30 times as large as our findings.

We argue that this discrepancy in minimum collapse mass results 
primarily from
how well the minihalo structure is resolved. Unless the centre, which
gains the highest molecule formation rate due to the highest density,
is fully resolved, one could be misled by a poor numerical resolution such that
certain low-mass haloes, which can cool and collapse in reality, are in 
hydrostatic equilibrium in the simulation.
The resolution becomes
poorer in the following sequence:
\citet{2001ApJ...548..509M}, which gives the best agreement with our
result, used an adaptive mesh refinement (AMR) scheme, resolving 
baryonic mass down to $M_b\sim 5\,M_\odot$. Such high resolution is
suitable to resolve even the central part of the smallest minihaloes
whose total baryonic 
mass content is roughly $ 2-3\times 10^3\,\rm M_\odot$.  \citet{2000ApJ...544....6F} and
\citet{2003ApJ...592..645Y}, on the other hand, used the smoothed
particle hydrodynamics (SPH) scheme, using SPH particles of mass $M_b \sim
40-140\times 10^2\,\rm M_\odot$. Finally, \citet{1997ApJ...474....1T}
used a uniform top-hat model, where there is no radial variation in
gas properties such as density and temperature, thus the central
region is, in effect, completely unresolved.
In addition, some of the rates used in \citet{1997ApJ...474....1T}
were not accurate \citep{2000ApJ...544....6F}.

We
believe that $M_{\rm c,min}\simeq 7\times 10^4\,M_\odot$ at $z=20$ is
close to reality, because
our 1-D spherical setup is based upon the TIS model which is  a highly
concentrated structure, and the
resolution of our code is superior to previous
calculations\footnote{After this paper was written a new preprint
  was posted which is consistent with our description here, finding
  $M_{\rm c,min}\approx 10^5\, M_\odot$ 
\citep{2006astro.ph..7013O}.}.
It is not our objective, however, to settle
the exact value of $M_{\rm c,min}$.
This estimate is based upon our
specific criterion described in this section, and is subject to change
under different criteria. This may also change if one adopts a more
realistic halo formation history to account, for instance, for
dynamical heating by accretion (see
\citealt{2003ApJ...592..645Y}). As the haloes we choose are rather
conservatively 
divided into successful collapse (for $M\ge 10^5\,M_\odot$) and
failure (for $M < 10^5\,M_\odot$), agreeing with AMR simulation
result by \citealt{2001ApJ...548..509M}, we shall proceed with our
choice of parameter space and {\em see how this fate of minihaloes
changes as a result of external radiation from a Pop III star.}

\section{Results: Radiative Feedback on Nearby Minihaloes by an
  External Pop III Star} 
\label{sec:2star-Result}
As described in Section \ref{sec:Initial-Setup}, we expose target haloes of
different mass
to the radiation from a Pop III star whose spectrum is approximated as
a $10^5\,\rm K$
 blackbody radiation field and whose flux is attenuated by the geometrical
factor $\left(\frac{D}{R_{*}}\right)^{-2}$ for different values of
$D$. 
In this section, we summarize the simulation results for both the
Phase I (early irradiation)
and the Phase II (late irradiation) initial conditions.

\subsection{I-front trapping and photo-evaporation}

In all cases, even in the presence of  evaporation, we find no
evidence of penetration of ionizing radiation into the halo
core. This is consistent with the results of
\citet{2006ApJ...639..621A} for the H II regions of the first Pop III
stars and of \citet{2004MNRAS.348..753S} and
\citet{2005MNRAS.361..405I} for the encounters between intergalactic
I-fronts and minihaloes during reionization.
There are two main reasons for this behaviour. First, 
the total intervening hydrogen
column density is initially high enough to trap the I-front outside
the core.
Second, the lifetime of the source is short compared to
the evaporation time. If the source 
lived
longer than the
evaporation time, 
the I-front would eventually have
reached the centre of the halo. 
In that case, \citet{2004MNRAS.348..753S} find that the
minihalo gas is completely evaporated.
In our
problem, however, the slow evaporation does not allow the I-front to reach the
centre within the lifetime of a Pop III star. 

\begin{figure}
\includegraphics[%
  width=84mm]{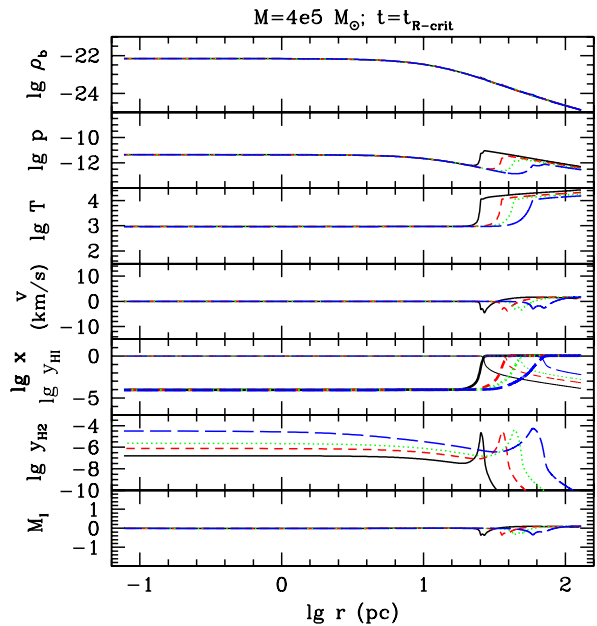}

\caption{Radial profiles of $4\times 10^5 M_\odot$ halo at $t=t_{\rm
  R-crit}$ in Phase I, for different fluxes 
  (distances). R-critical condition 
  is met by the condition $v_{\rm I}\approx 2 c_{\rm s,I,2}\approx
  20\,{\rm km\,s^{-1}}$, when the I-front
  makes a transition from R-type to D-type. Different distances to the source
  are represented by different line types: ${\rm D=180 pc}$
  ($F_0=46.3$; black, solid), ${\rm D=360 pc}$ ($F_0=11.6$; red,
  short-dashed), ${\rm D=540 
  pc}$ ($F_0=5.14$; green, dotted), ${\rm D=1000 pc}$ ($F_0=1.5$;
  blue, long-dashed). In the 
  fifth panel from top, electron fraction $x$ and neutral
  hydrogen fraction $y_{\rm H}$ (thin) are plotted. 
  As is shown in the density
  ($\rho_b$) plot, gas just starts to respond to the I-front
  hydrodynamically, as the initial R-type, supersonic I-front slows
  down to reach the R-critical phase.
  Also note that
  the shorter the distance, the deeper the I-front is (i.e. smaller r)
  at $t=t_{\rm R-crit}$.
\label{fig-Rcrit}}
\end{figure}

The I-front entering the minihaloes propagates as a
weak R-type 
front in the beginning. The I-front then makes the
transition to the D-type, after reaching the R-critical
state. This R-critical state is reached when the I-front velocity $v_{\rm I}$
satisfies the following condition:
\begin{equation}
v_{\rm I}=c_{\rm I,2}+(c_{\rm I,2}^2 - c_{\rm I,1}^2)^{0.5},
\end{equation}
where $c_{\rm I}$ is the isothermal sound speed, $c_{\rm I}\equiv
\sqrt{p/\rho}$, 
and subscripts 1 and 2 represent pre-front and post-front,
respectively. When the I-front propagates into a cold region ($T\ll
10^{4} \, {\rm K}$), as in our problem, this condition is approximately
$v_{\rm I}\approx 2\,c_{\rm I,2} \approx 20\,{\rm km\,s^{-1}}.$
In all
cases, we find that this transition occurs in times less than the
lifetime of the source star, 2.5 Myrs. After reaching the
R-critical state,
gas in front of the I-front forms a shock, which
then detaches from the slowed I-front. As an example, we plot in
Fig. \ref{fig-Rcrit} 
the profiles of Phase I, $4\times 10^5\,M_\odot$ halo at
$t=t_{\rm R-crit}$ under different fluxes.

\begin{figure}
\includegraphics[%
  width=84mm]{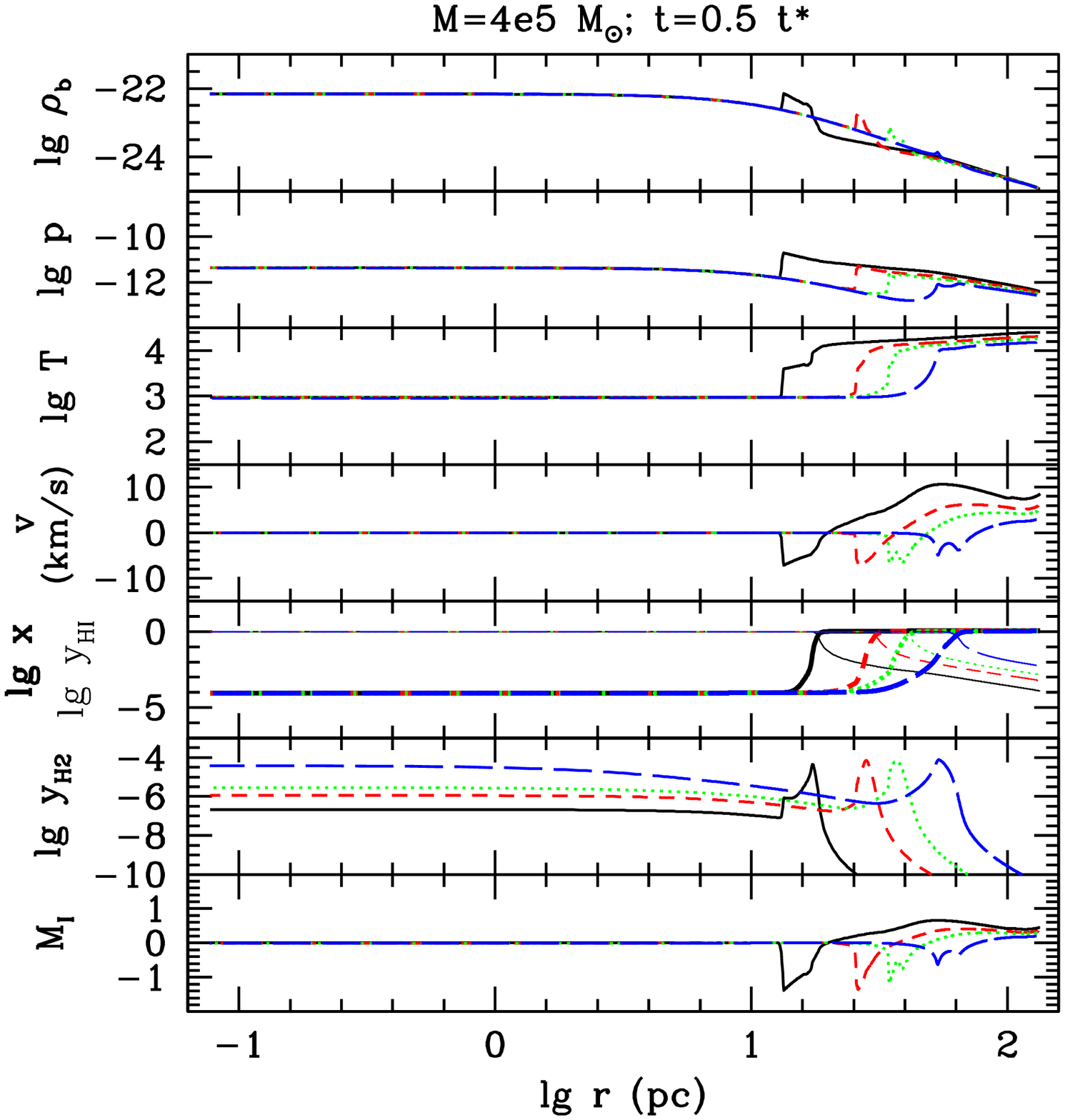}

\caption{Radial profiles of $4\times 10^5 M_\odot$ halo at $t=0.5 t_*$
  in Phase I, 
  for different fluxes  
  (distances). Same line types are used as in Fig. \ref{fig-Rcrit}.
\label{fig-midpoint}}
\end{figure}

\begin{figure}
\includegraphics[%
  width=84mm]{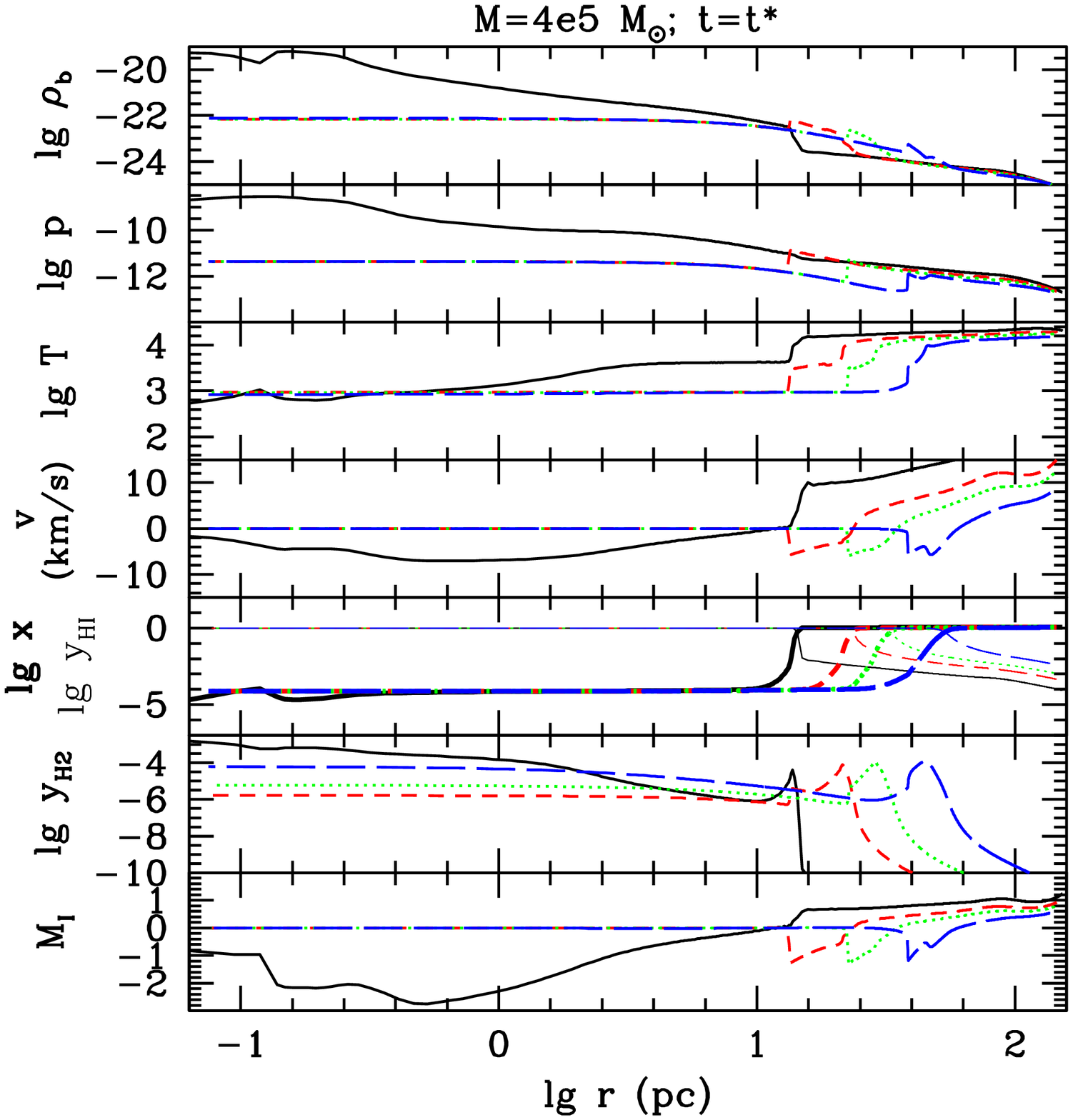}

\caption{Radial profiles of $4\times 10^5 M_\odot$ halo at $t=t_*$ in
  Phase I, 
  for different fluxes  
  (distances). Same line types are used as in Fig. \ref{fig-Rcrit}.
\label{fig-endpoint}}
\end{figure}

All of the post-front (ionized) gas, initially undisturbed, eventually
evaporates away, accelerated
outward by a large pressure gradient.
As the line-of-sight is cleared by this evaporation,
ionizing radiation penetrates deeper, until the source turns off. See
Figs \ref{fig-Rcrit}, \ref{fig-midpoint} and \ref{fig-endpoint} for
the evolution of the I-front.

This result invalidates the initial condition adopted by
\citet{2005ApJ...628L...5O} and \citet{2006astro.ph..4148M} which led
them to find that $\rm H_2$ 
formed in the core region after it was ionized and then cooled while
recombining, once the source turned off. 
As we show, the core remains neutral before and after the
source is turned off, so the mechanism explored by
\citet{2005ApJ...628L...5O} does not
work. This neutral core, therefore, must find a different way 
to cool and collapse if star formation is to happen in the target minihalo.

What happens to the initially ionized gas after the star turns off? This
gas recombines as it
cools radiatively and by adiabatic
expansion, even forming $\rm H_2$ molecules.
We find that this cooling cannot reverse the
evaporation, however. Gas is simply carried away with the initial momentum
given to it when it was in an ionized state. In Table \ref{table:ionized},
we list the 
fraction of the baryonic halo mass which is ionized during the
lifetime of the star. This mass 
serves
as a crude 
estimate of the mass lost from these haloes by evaporation. We found
no major difference 
between Phase I and Phase II in this matter, so we provide only one table.

\begin{table*}
\caption{Ionized mass fraction of baryons for different mass target haloes
  (columns) at different distances from a $120\,M_\odot$ Pop III star
  (rows; fluxes in units of $10^{50}\,\rm s^{-1}\,
  kpc^{-2}$, $F_0$,
in square brackets). The ratio shown here is the  mass ionized
  during the lifetime of the star to the total
  baryon mass.
  }
\label{table:ionized}
\begin{tabular}{ccccccc}
\hline
&\multicolumn{6}{c}{Total Halo Mass in $10^5\,M_\odot$ units} \\ 
&\multicolumn{6}{c}{(Halo Baryon Mass in $10^5\,M_\odot$ units)} \\ 
 \cline{2-7}
$D$ (pc) [$F_0$]&
$0.25 $&
$0.5 \cdot 10^{4} $&
$1 $&
$2 $&
$4 $&
$8 $\\
&
$(0.043 )$&
$(0.086 )$&
$(0.17 )$&
$(0.34 )$&
$(0.69 )$&
$(1.371 )$\\
\hline
\hline 
180 [46.3]&
0.95&
0.92&
0.88&
0.84&
0.82&
0.79\\
\hline 
360 [11.6]&
0.85&
0.81&
0.77&
0.74&
0.70&
0.67\\
\hline 
540 [5.14]&
0.78&
0.74&
0.70&
0.66&
0.62&
0.59\\
\hline 
1000 [1.5]&
0.66&
0.60&
0.55&
0.50&
0.47&
0.43\\
\hline
\end{tabular}
\end{table*}

\subsection{Formation of ${\rm H_2}$ precursor shell in Front of
  the I-Front} 
\label{sub:H2shell}

We find that a thin shell of ${\rm H_{2}}$ is formed just ahead of
the I-front, with peak abundance $y_{{\rm H_{2}}}\approx10^{-4}$.
It happens mainly because the increased electron fraction across the
I-front promotes the formation of $\rm H_2$.
More precisely,
the gas ahead of the I-front is ionized to the extent that the electron
abundance is large enough to form ${\rm H_{2}}$, but at the same time
too low to drive significant collisional dissociation of $\rm H_2$.
The width of this ${\rm H_{2}}$ shell and the amount of $\rm H_2$ in
this region is determined by the hardness of the energy spectrum of
the source: the width of the I-front is of the order of the mean free
path of the ionizing photons. Pop III stars, in general, produce a
large number of hard photons due to their high temperature,
which can penetrate deeper into the neutral region than soft photons.

\begin{figure}
\includegraphics[%
  width=84mm]{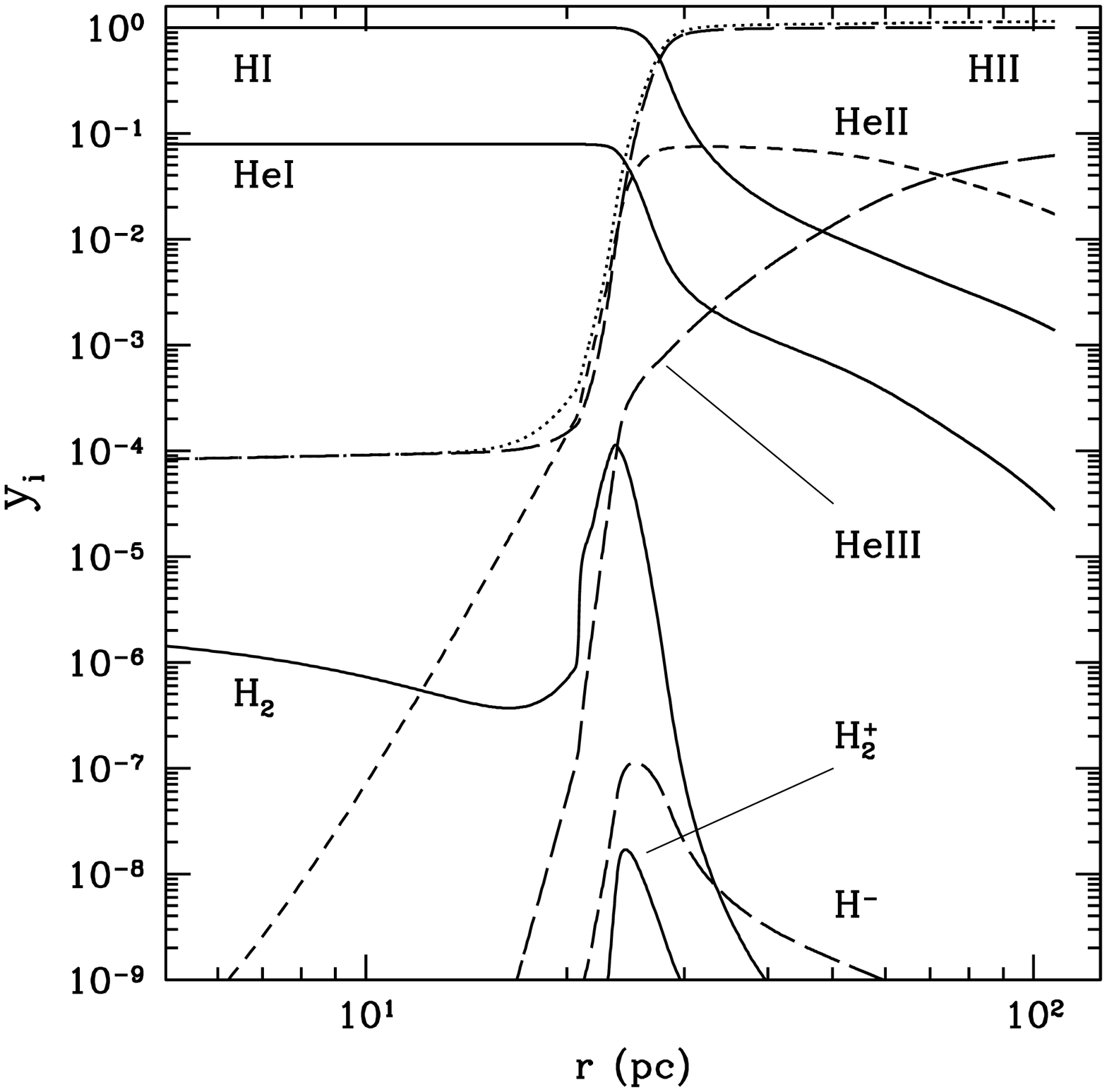}

\caption{Radial profile of abundance of primordial gas species at
  $t=0.6\, t_{*}$, for a 
  halo of $M=2\times 10^{5}\,M_\odot$ in Phase I, illuminated at
  $D=540\,\rm pc$ ($F_0 = 5.14$).
 Labels are self-explanatory; electron abundance
$x$ is represented by the dotted curve, which closely follows
the H II abundance.
The
flux is coming from the right hand side, so this figure can be compared
to the mirror image of Fig. 3 in \citet{2001ApJ...560..580R} for an
  I-front in a static, uniform IGM at the mean density.
\label{fig-midtrap}}
\end{figure}

This precursor ${\rm H_2}$ shell feature is evident in Figs
\ref{fig-Rcrit}, \ref{fig-midpoint}, and \ref{fig-endpoint}.
We show the detailed structure of these ${\rm H_{2}}$
shells in Fig. \ref{fig-midtrap}, where we plot the radial profile
of the abundance of different species for the case of $M=4\times 10^5
\,M_\odot$, Phase I, $D=540\,\rm pc$ ($F_0=5.14$) at
$t=0.5\, t_{*}$. 
We note the similarity between our results and those of
\citet*{2001ApJ...560..580R} for an I-front in a uniform, static IGM
at the mean density
(see Fig. 3 in \citealt{2001ApJ...560..580R}) which also show a
precursor $\rm H_2$ shell. A similar effect
was reported by \citet{2006ApJ...645L..93S}, as well.

What is the importance of this $\rm H_2$ shell in protecting the
central region of haloes from dissociating radiation?
The molecular column density obtained by this $\rm H_2$ shell sometimes reaches
$\sim 10^{16}\,\rm cm^{-2}$, which provides an appreciable amount of
self-shielding. The self-shielding due to the $\rm H_2$ shell,
however, is not the major factor that determines whether or not the $\rm H2$
in the core region is protected. A more important factor is which
evolutionary phase the target halo is in when it is
irradiated. Roughly speaking, when a target halo is irradiated early
in its evolution (Phase I), the precursor $\rm H_2$ shell dominates
the total $\rm H_2$ column density available to shield the central
region, but this shielding is not sufficient to prevent
photodissociation there anyway. On the other hand, if the halo is
irradiated later in its evolution (Phase II), the $\rm H_2$ column
density of the shell is only a small part of the total $\rm H_2$
column density, so shielding is successful independent of the
precursor shell. We describe this in more detail as follows.

\begin{figure}
\includegraphics[%
  width=84mm]{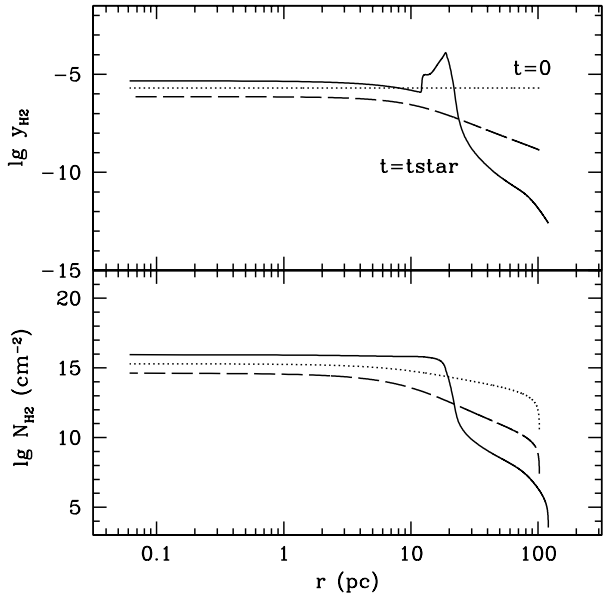}

\caption{Radial profiles of $\rm H_2$ fraction (top) and the $\rm H_2$
  column density (bottom) for $M=2\times 10^5 M_\odot$, Phase I
  initial conditions, at $t=0$
  (dotted) and at $t=t_{*}$ (solid). The source is at a distance
  $D=540\,\rm pc$ ($F_0=5.14$).
  Also plotted are those for a radiation composed only of dissociating
  photons (dashed) at $t=t_{*}$.
  Even though ${\rm H_2}$ shell
  provides self-shielding by contributing $N_{\rm H_2}\approx 10^{16}
  \, {\rm cm}^2$, the molecule fraction $y_{\rm H_2}$ is held at $\la
  10^{-5.3}$ due to strong dissociating radiation.
\label{fig-nh2-C1}}
\end{figure}

\begin{figure}
\includegraphics[%
  width=84mm]{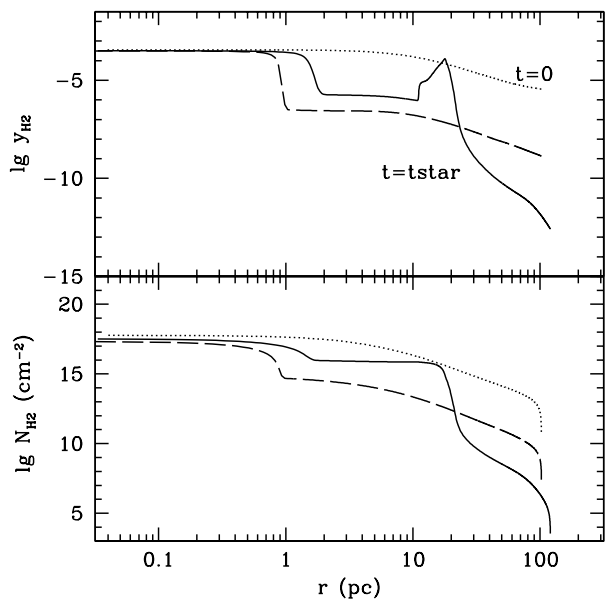}

\caption{Radial profiles of $\rm H_2$ fraction (top) and the $\rm H_2$
  column density (bottom) for $M=2\times 10^5 M_\odot$, Phase II
  initial conditions, at $t=0$
  (dotted) and at $t=0.5 t_{*}$ (solid). The source is at a distance
  $D=540\,\rm pc$ ($F_0=5.14$).
  Also plotted are those for a radiation composed only of dissociating
  photons (dashed) at $t=t_{*}$.
  Contrary to Fig. \ref{fig-nh2-C1}, the core ${\rm H_2}$ fraction is
  well protected to a high level, $y_{\rm H_2}\approx 10^{-3.5}$. The
  role of $\rm H_2$ shell is, however, not dramatic, because even without
  this shell, the core is protected from dissociating radiation
  (dashed). It rather offsets the loss to the molecular column density
  from collisional dissociation in the ionized region.
\label{fig-nh2-C4}}
\end{figure}

In order to understand quantitatively the importance of the $\rm
H_2$ shell in 
protecting the central $\rm H_2$ fraction, we have performed
simulations with a source SED that is identical to the Pop III SED
below 13.6 eV, but zero above 13.6 eV. As the radiation is now
incapable of ionizing the halo gas, the $\rm H_2$ shell formation by partial
ionization will not occur. This enables us to compare our results where
the $\rm H_2$ shell is present to those
cases without an $\rm H_2$ shell.
We describe a specific case of $M=2\times 10^{5} \,M_\odot$ as an
illustration. 
Roughly speaking, the ${\rm H_2}$ shell which forms only in the
presence of ionizing radiation compensates for the amount by which the
initial molecular column density, $N_{\rm H_2}$, is reduced when
molecules  in the ionized region are destroyed by collisional dissociation.
The nett column density in the case where the $\rm H_2$ shell is
present even
exceeds that in the case without the $\rm H_2$ shell
(Figs \ref{fig-nh2-C1} and \ref{fig-nh2-C4}). The  
nett effect is the increase of the self-shielding.
Such an increase of the self-shielding, however, is not too dramatic. 
In the case of
$M=2\times 10^{5} \,M_\odot$ with Phase I initial conditions, $y_{\rm
  H_2}\approx 10^{-5.3}$ at the centre, 
about an order of magnitude higher than the central $y_{\rm H_2}$ of
the case without ionizing photons (Fig. \ref{fig-nh2-C1}). This
molecule fraction is
still too low, however, to cool the 
gas. On the other hand, in the case of $M=2\times 10^{5} \,M_\odot$ with
Phase II initial conditions, $y_{\rm H_2}\approx 10^{-3.5}$ at the
centre throughout the lifetime of the Pop III source,
whether or not the
${\rm H_2}$ shell is formed. The depth (radius) of penetration of dissociating
  photons differs by a factor of 2 if the shell is included, but the
  central ${\rm H_2}$ is still
  protected because of the high ${\rm H_2}$ column density {\em apart} from
  the precursor shell (Fig. \ref{fig-nh2-C4}).
The major factor that determines the fate of the central $\rm H_2$
fraction is instead the evolutionary phase of a target halo when
it is irradiated. The short lifetime of a Pop III star plays an
important role of either reconstituting or protecting molecules in
the core, depending upon the evolutionary phase of the halo, as will
be described in Section \ref{sub:Collapse}. 

Note that in all cases, we use equation (\ref{eq:DB_shield_factor}),
the shielding function for thermally-broadened lines 
with $T=10^{4}\,{\rm K}$. This is justified by the fact that the
${\rm H_2}$ shell moves inward with $v\approx 2-5 \,{\rm km\,s^{-1}}$
and the shell achieves $T=T_{\rm sh}\approx 10^{3}-5\times
10^{3}\,{\rm K}$, where $T_{\rm sh}$ denotes the temperature of the shell.
If we take this peculiar velocity as sound speed, $v\approx 2-5 \,{\rm
  km\,s^{-1}}$ corresponds to $T=T_p\equiv v^2 \mu m_{\rm H}/k=6\times 10^2
- 3.7\times 10^3\,{\rm K}$, where the subscript $p$ denotes the
peculiar velocity. A crude way to imitate both effects by thermal
broadening is to use the sum of these two temperatures ($T_{\rm sh}$ and
$T_p$). We take the most conservative stand -- the least
self-shielding effect --
in order not to overestimate the self-shielding, and use
$T=10^{4}\,{\rm K}$ as the temperature responsible for the nett 
thermal broadening of the molecular LW bands.

\subsection{Formation of shock and Evolution of core}
\label{sub:shock}

After the I-front decelerates as it enters the target halo,
transforming from R-type to D-type, a shock front forms to lead the
D-type front. 
The neutral gas in the core is strongly affected by this shock front
as it
propagates.
This shock plays an important role
in providing both positive and negative feedback effects. 
By identifying successive evolutionary stages of the shock, we
now describe how the core responds to the shock and evolves accordingly.

\subsubsection{Stage I: Formation and acceleration of Shock}
\label{sub:shockstageI}

A shock starts to form as the I-front, initially moving supersonically
as an R-type, slows down
and turns into a D-type. The pre-front gas -- neutral gas ahead of the
I-front -- can respond to the I-front before it is swept by the
I-front, because the D-type front moves subsonically into the neutral
gas. It is easier
to understand the formation of the shock by using the I-front jump
conditions: the pre-front gas speed in the rest frame of the I-front, $v_1$,
derived from the I-front jump conditions,
should satisfy either $v_1\ge v_R \equiv c_{\rm I,2}+(c_{\rm
  I,2}^2 - c_{\rm I,1}^2)^{0.5},$ or $v_1\le v_D \equiv c_{\rm I,2}-(c_{\rm
  I,2}^2 - c_{\rm I,1}^2)^{0.5}$, where 
$c_{\rm I,1}$ and $c_{\rm I,2}$ are the isothermal sound speeds of the
pre-front and post-front gas, respectively.
$v_R$ and $v_D$ have a gap of $2(c_{\rm
  I,2}^2 - c_{\rm I,1}^2)^{0.5}$, which is nonzero in general. As the
I-front slows down and $v_1$ starts to cross $v_R$, $v_1$ encounters a value
which is not allowed mathematically. This paradox is resolved, however,
because the pre-front gas now ``prepares'' a new hydrodynamic
condition by forming a shock. 
The shock wave increases $\rho_1$ and thereby reduces $v_1$ and
increases $v_D$, making it
possible to satisfy the D-type condition, $v_1\le v_D$.
 
\begin{figure*}
\includegraphics[%
  width=68mm]{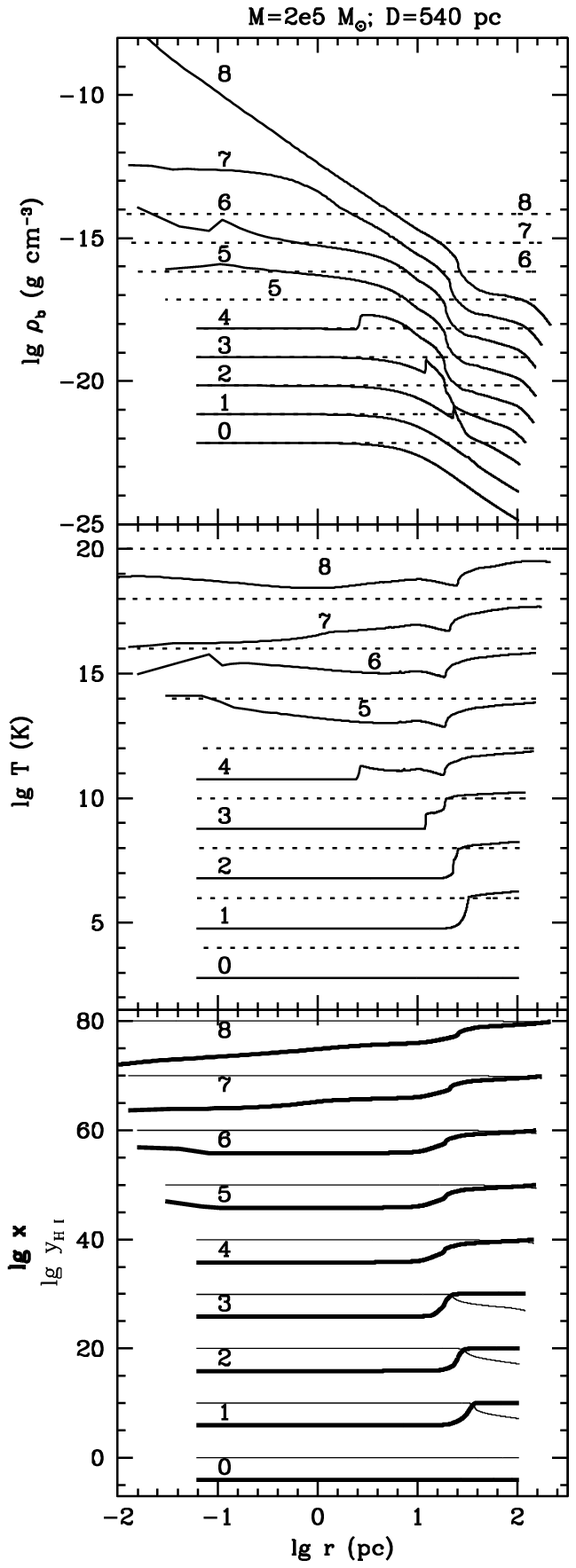}
\includegraphics[%
  width=68mm]{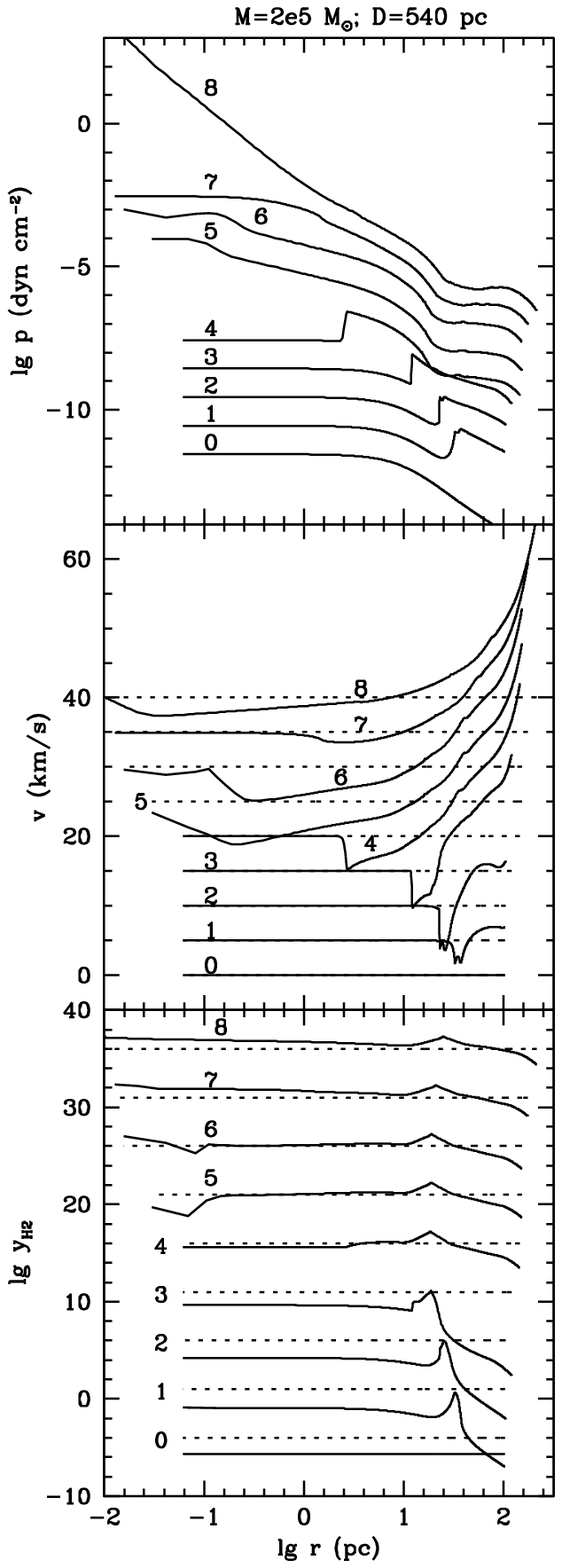}

\caption{Success of shock-induced molecule formation (SIMF): evolution
  of $2\times 10^5 M_\odot$ halo irradiated early (Phase I) by
  a Pop III star at $D=540 {\rm pc}$ ($F_0=5.14$). 
  Radial profiles of fluid parameters --
  baryon gas density ($\rho_b$), pressure ($p$), temperature ($T$),
  radial velocity ($v$), 
  electron fraction 
  ($x$; thick line),
  neutral fraction ($y_{\rm H I}$; thin line), and molecule fraction ($y_{\rm
  H_2}$) -- are
  labelled by different time frames as following: 
  0 - $t=0$; 1 - $t=t_{\rm R-crit}=0.2\,t_*$; 2 -
  $t=0.5\,t_*$; 3 - $t=t_*$; 4 - $t=1.5\,t_*$, 5 - $t=t_{\rm
  shock\,bounce}=1.611 \, t_*$;  
  6 - $t=t_{\rm
  shock\,bounce}+\varepsilon=1.617\,t_*$; 7 - $t=2\,t_*$; 8 -
  $t=t_{\rm coll}=2.6 \,t_*$. These time frames are
  shifted along the y-axis for clarity, with equal displacements as
  following: $\Delta \,{\rm lg}\, \rho=1$; $\Delta \,{\rm lg}\, p=1 $;
  $\Delta \,{\rm lg}\, T=2$; $\Delta v = 5$ km/s; 
  $\Delta \,{\rm lg}\, x=10$; $\Delta \,{\rm lg}\, y_{\rm H I}=10$;
  $\Delta \,{\rm lg}\,y_{\rm H_2}=5$. Dotted lines represent 
  the initial central density, $T=10^4 \,\rm K$, $v=0 \,\rm km/s$, and
  $y_{\rm H_2}=10^{-4}$ in the  
  $\rho_b$, $T$, $v$, and $y_{\rm H_2}$ plot, respectively.
  $t_{\rm shock-bounce}$ is the time
  when the shock front reaches the centre. Note that at this moment
  the shock-front accelerates to heat the gas up to $T \ga 10000
  \rm K$ at the centre. 
  This temperature is high enough to cause
  collisional ionization, which leads to rapid formation of $\rm H_2$
  and cooling at the centre afterwards ($t \ga t_{\rm
  shock-bounce}+\varepsilon$). 
  Thus the thermal energy delivered is dissipated very easily, and the core
  collapses in a runaway fashion. We show here the fast evolution of $\rm
  H_2$ around the time of shock-bouncing, using $\varepsilon\approx
  1.5\times 10^4$ years.
\label{fig-evol}}
\end{figure*}

This shock-front then propagates inward, separating from the
I-front, due to the discrepancy between the speed of the
shock-front and the
speed of the I-front. As the shock-front enters the flat-density
core, the shock front starts to accelerate, leaving behind the
post-shock gas 
with ever increasing temperature (e.g. see time steps 4 and 5 in
Fig. \ref{fig-evol}, where the post-shock temperature increases as the
radius $r$ decreases).

As the shock
boosts the density and temperature in the neutral, post-shock gas, the ${\rm
  H_2}$ formation rate there increases, boosting the $\rm H_2$ column
density even further.
We can understand the evolution of $y_{\rm H_2}$ in the
presence of this shock quantitatively by using its equilibrium value, $y_{\rm
  H_2,eq}$. The increase of density and temperature due to this shock
promotes $\rm H_2$ formation, as follows. When there is no 
significant $\rm H^-$ destruction mechanism, the dominant
$\rm H_2$ formation mechanism is through $\rm H^-$
(equation \ref{eq:solomon}), and the $\rm H_2$ formation rate becomes
equivalent to the $\rm H^-$ formation rate. 
Photo-dissociation dominates over collisional dissociation in
destroying  $\rm H_2$, which
occurs when $x\la 4\times 10^{-3}\,T_{\rm 
  K}^{1/2}$ and $n_{\rm H}\ga 0.045 \times (F_{\rm LW}/10^{-21}\,\rm
erg\,s^{-1}\,cm^{-2}\,Hz^{-1})$ 
(e.g. \citealt{2001MNRAS.321..385G}). Using the $\rm H^-$
formation rate coefficient \citep{1972AA....20..263D}
\begin{equation}
k_{\rm H^-}=10^{-18}\,T_{\rm K}\,{\rm cm}^{3}\,{\rm s}^{-1},
\end{equation}
and the photo-dissociation rate coefficient $k_{\rm H_2}$ given by
equation (\ref{eq:DB_rate}), we obtain
\begin{eqnarray}
y_{\rm H_2, eq}&=&4.1\times 10^{-5} \left( \frac{T}{5000\,{\rm K}}\right)\left(
\frac{x}{10^{-4}}\right)\nonumber \\ 
&& \times \left(
\frac{n_{\rm H}}{30\,{\rm cm}^{-3}}\right) \left( F_0\cdot F_{\rm
  shield}\right)^{-1},
\label{eq:yh2}
\end{eqnarray}
where we have used the fact that one can scale $F_{\rm LW}$ by $F_0$
according to the following:
\begin{equation}
F_{\rm LW}\approx 3.25\times 10^{-21} \,{\rm erg
  \,s^{-1}\,cm^{-2}\,Hz^{-1}}\, F_0,
\label{eq:flw}
\end{equation}
if one adopts a black-body spectrum with $T=10^{5}\,\rm K$.
As seen in equation (\ref{eq:yh2}), both the high temperature ($\sim
1000-5000\,\rm K$) and increased density ($\times 4$ in the case
of strong shock) of the post-shock gas contributes to boosting the $\rm H_2$
fraction. As $y_{\rm H_2}\propto F_{\rm shield}^{-1}$, molecular
self-shielding also plays an important role in determining $y_{\rm H_2}$.
If the shock boosts the formation rate of $\rm H_2$ and $y_{\rm H_2}$
increases, so will $N_{\rm H_2}$, and with it the shielding. These two
effects, therefore, amplify each other.

There is an additional mechanism to create molecules: the
shock-induced molecule formation (SIMF). The
acceleration of the shock-front accompanied by an increasing
post-shock 
temperature, leads to a partial
ionization of the post-shock gas in many cases, when the right
condition ($T\ga 10^4\,\rm K$) is met to trigger
collisional ionization -- see, for example, step 5 in
Fig. \ref{fig-evol}: the centre is shock-heated above $10^4\,\rm K$,
with a boost in $x$. The electron
fraction $x$ now reaches $\sim 10^{-4} - 10^{-2}$, which
promotes
further ${\rm H_2}$ formation.
This mechanism is indeed identical to the $\rm H_2$ formation
mechanism in a gas 
that has been shock-heated to temperatures above $10^4\,\rm K$
\citep{1987ApJ...318...32S,1992ApJ...386..432K}. 
When a gas cools
radiatively from a temperature well above $10^4\,\rm K$, it cools
faster than it recombines. As a result, the recombination is out of
equilibrium, and an enhanced electron fraction exists at temperatures
even below $10^4\,\rm K$ compared to the equilibrium value. This
electron fraction triggers the formation of $\rm H_2$ through the
gas-phase reactions (equations \ref{eq:solomon} and
\ref{eq:solomon2}).

\begin{figure*}
\includegraphics[%
  width=68mm]{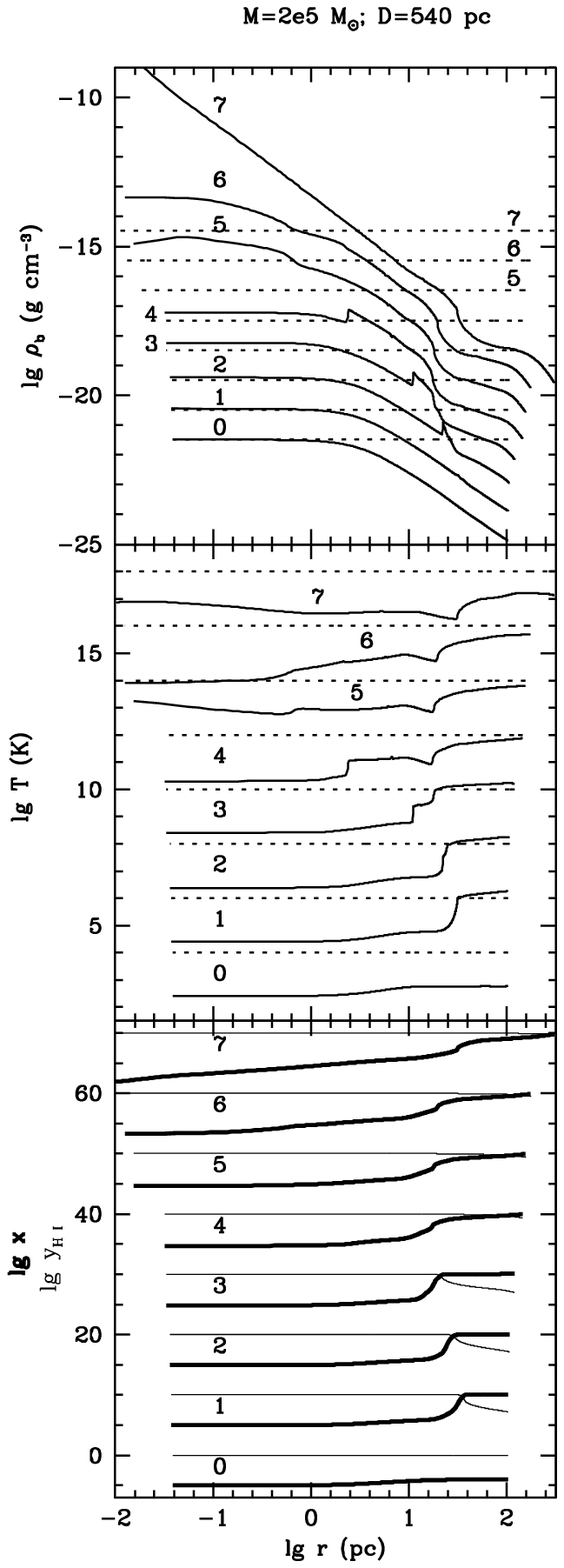}
\includegraphics[%
  width=68mm]{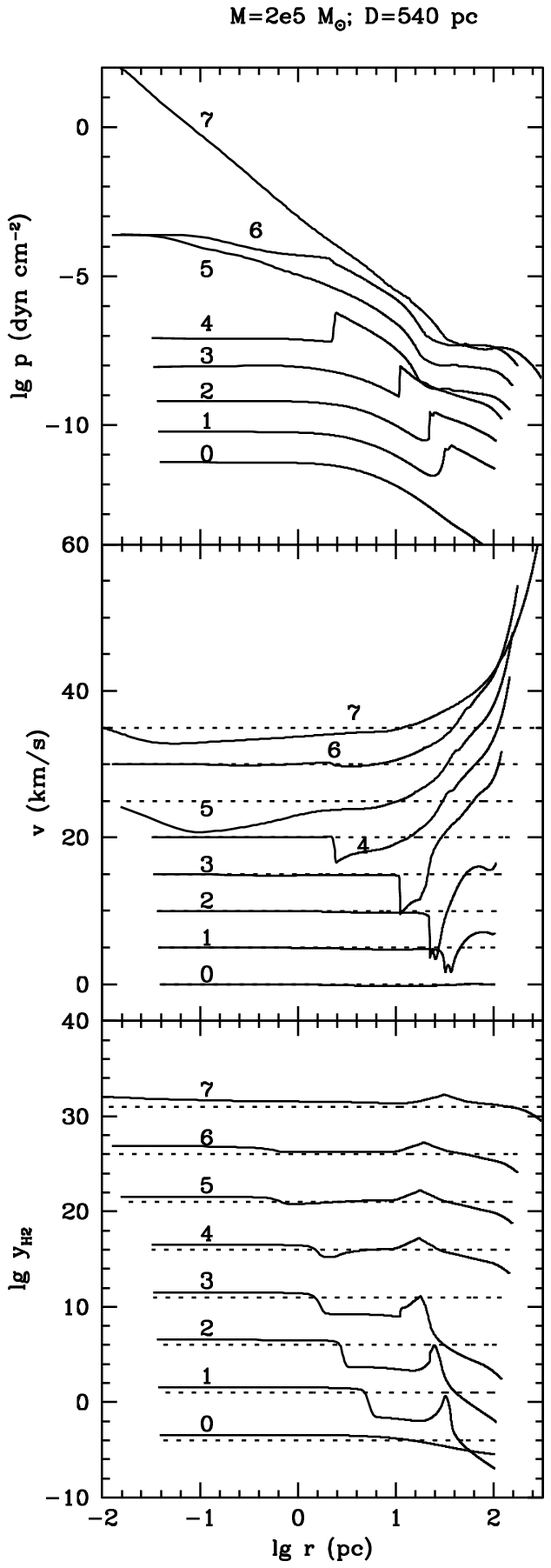}

\caption{Failure of shock-induced molecule formation (SIMF): evolution
  of $2\times 10^5 M_\odot$ halo irradiated late (Phase II) by
  a Pop III star at $D=540 {\rm pc}$ ($F_0=5.14$). 
  Contrary to the case with the Phase I
  initial conditions with the same mass and flux
  (Fig. \ref{fig-evol}), SIMF does   not occur in this case.
  Gas profiles are
  labelled by different time frames as following: 
  0 - $t=0$; 1 - $t=t_{\rm R-crit}=0.2\,t_*$; 2 -
  $t=0.5\,t_*$; 3 - $t=t_*$; 4 - $t=1.5\,t_*$; 5 - $t=t_{\rm
  shock\,bounce}=1.67\,t_*$; 6 - $t=2\,t_*$; 7 - $t=t_{\rm
  coll}$. These time frames are 
  shifted in the same way as in Fig. \ref{fig-evol}. Dotted lines have
  the same meaning as those in Fig. \ref{fig-evol}.
  Note that even at $t=t_{\rm shock-bounce}$, the shock-front velocity
  is not high enough to heat the gas up to $T \ga 10000
  \rm K$ at the centre. The SIMF, therefore, does not occur.
  The thermal energy delivered, however, is dissipated anyway by
  radiative cooling, 
  because the core is well protected from the dissociating radiation
  and the high $\rm H_2$ fraction is maintained throughout the evolution.
  The core  collapses in a runaway fashion afterwards.
\label{fig-evol2e5C4}}
\end{figure*}

SIMF does not always occur, however.
The shock-front can accelerate
when the pre-shock density remains almost constant (e.g. Fig. \ref{fig-evol}). If the
density increases faster than the shock propagates, on the other hand,
the shock-front
will encounter an ever increasing density ``hill'' and it
will never accelerate to generate post-shock temperature above
$10^4\,\rm K$ (e.g. Fig. \ref{fig-evol2e5C4}). The dependence of SIMF
on the halo mass, source flux, and the initial phase  will be
described in Section \ref{sub:Collapse}.

\subsubsection{Stage II: Cooling and Compression of Core}

As the shock-front approaches the centre of the halo, the
post-shock gas there becomes more concentrated and denser than the pre-shock
gas. This shock-induced compression leads to a very fast
molecular cooling in the core and further compression in almost a runaway
fashion, as follows.

\begin{figure*}
\includegraphics[%
  width=80mm]{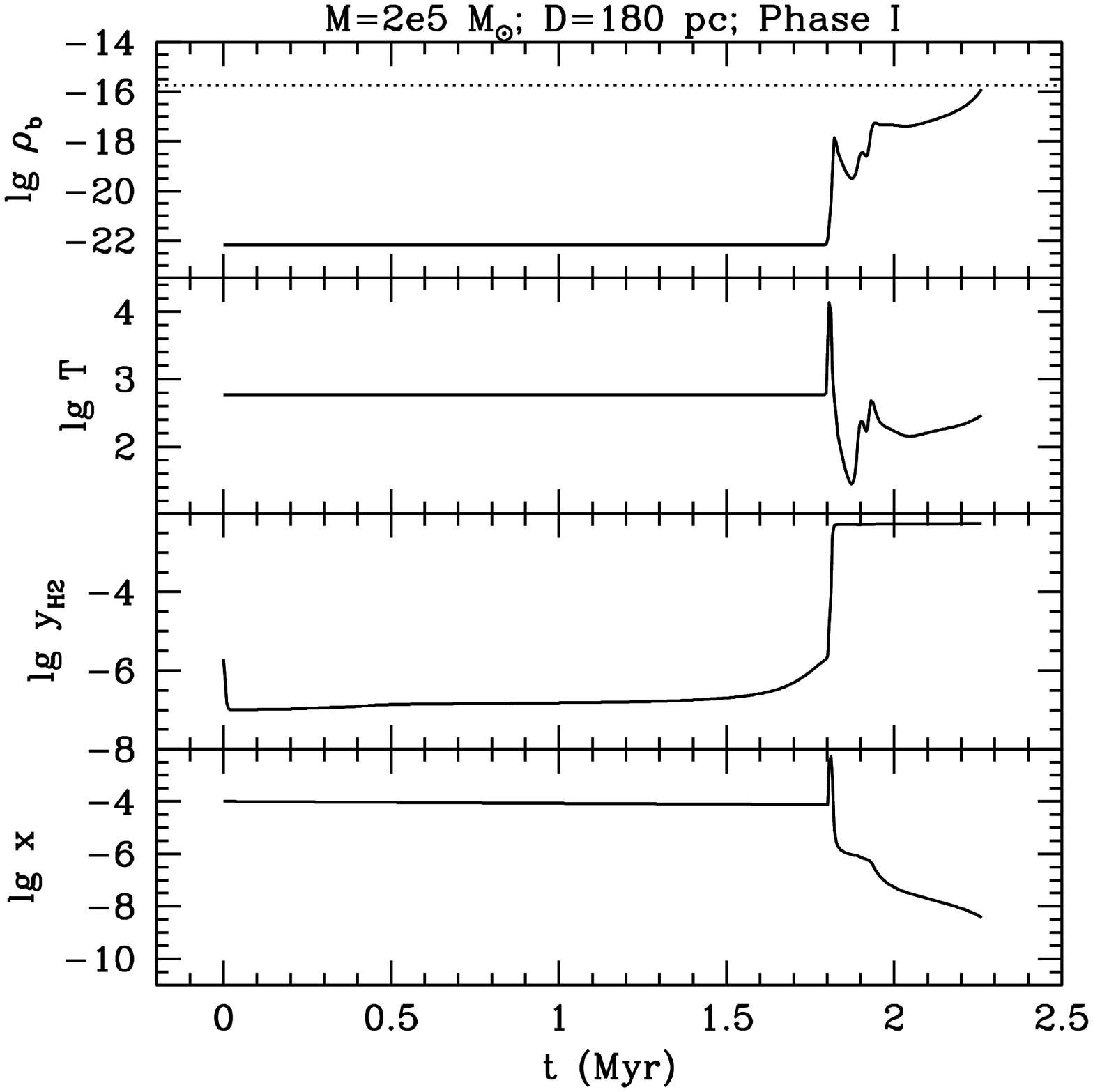}
\includegraphics[%
  width=80mm]{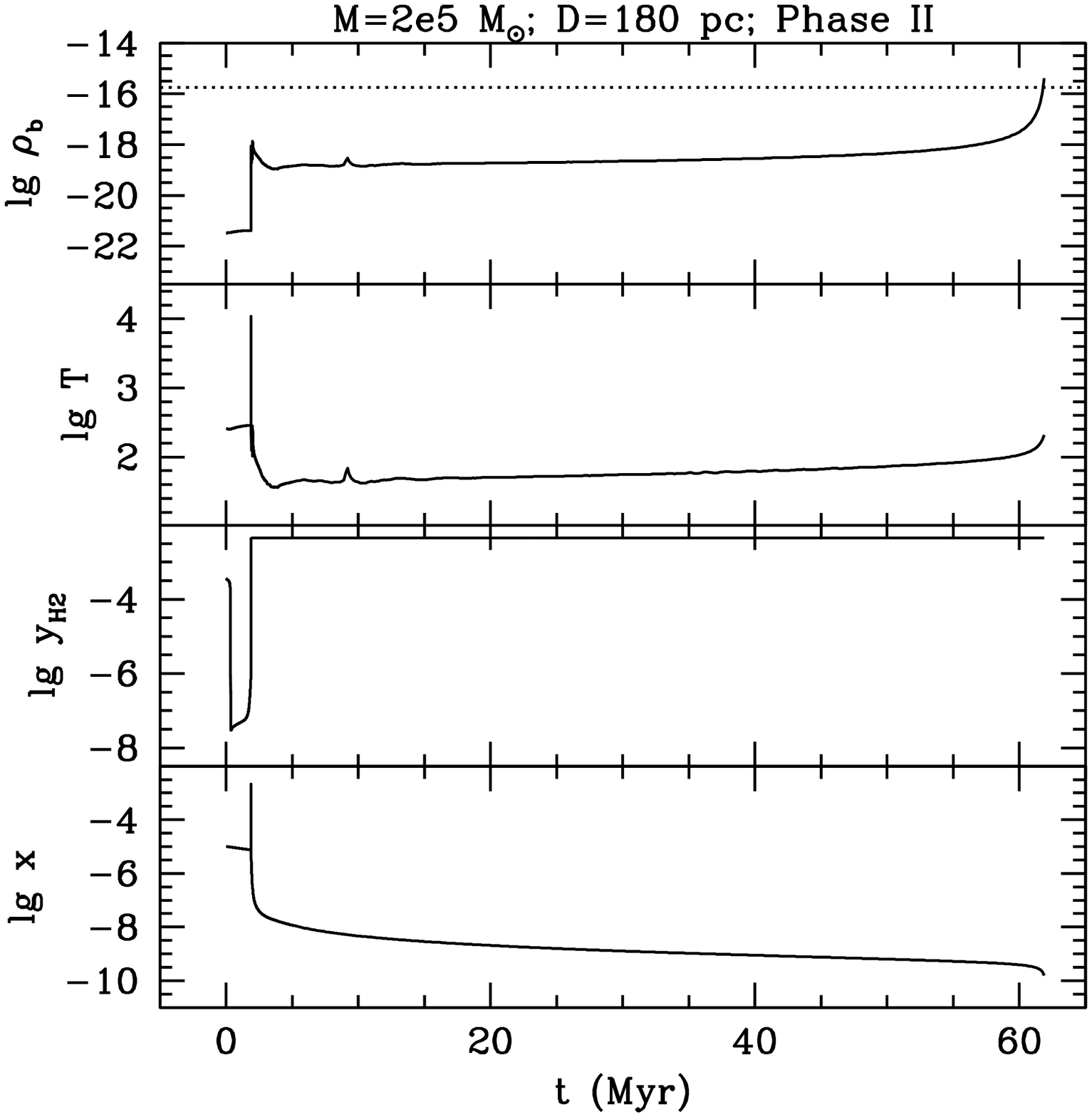}
\includegraphics[%
  width=80mm]{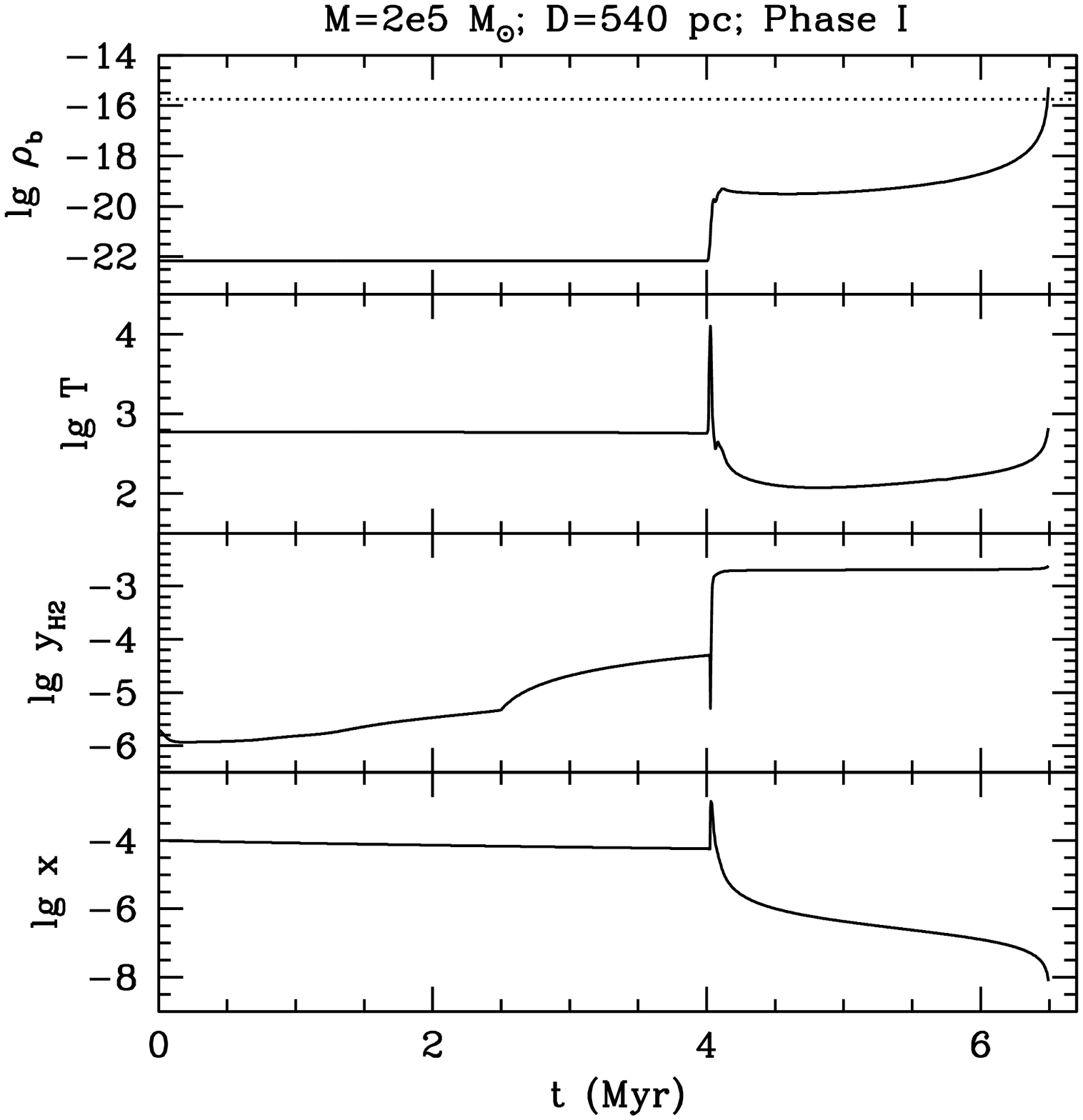}
\includegraphics[%
  width=80mm]{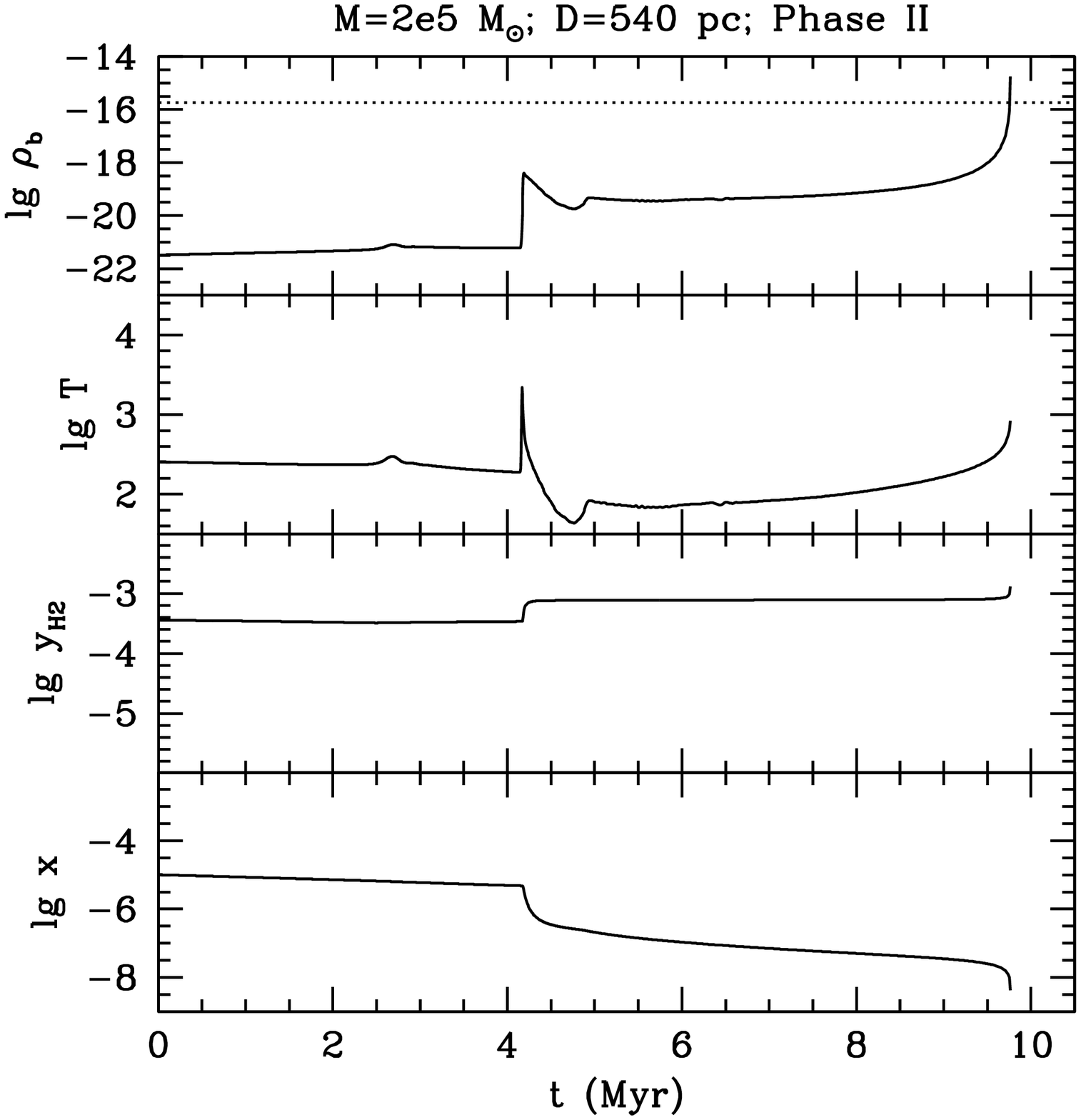}

\caption{Evolution of the central part of a $M=2\times
  10^{5}\,M_\odot$ halo, for Phase I (left) and Phase II (right) at
  $D=180\,{\rm pc}$ (top; $F_0=46.3$) and $D=540\,{\rm pc}$ (bottom;
  $F_0=5.14$). In Phase I, 
  expediting core collapse is observed for both distances. In Phase II,
  mixed results occur: delayed collapse for $D=180\,{\rm pc}$ while
  expedited collapse for $D=540\,{\rm pc}$. Another notable feature is
  the shock-ionization (electron) molecule formation in all cases
  except for the case of $D=540\,{\rm pc}$, Phase II. The increase of
  molecule fraction in the latter case is due to the increase of
  temperature and density due to shock compression, while in other
  cases, shock-induced electron formation promotes 
  further molecule formation.
\label{fig-M2e5-ing}}
\end{figure*}

Molecular cooling occurs very rapidly at a high density and temperature
condition. Assuming that the pre-shock gas of the halo core remains unchanged
before the shock-front arrives -- as is usually the case in Phase I --
and the shock is strong, the
post-shock density of the core becomes 4 times higher than that of the
pre-shock, namely $n_{\rm HI}\approx
4\times 30\,\rm cm^{-3}=120\,\rm cm^{-3}$ in a TIS halo core at
$z=20$. At the same time, post-shock temperature can be as high as
$10^4\,\rm K$. The molecular cooling time, $t_{\rm cool,H_2}\equiv
T/(dT/dt)$, is
\begin{equation}
t_{\rm cool,H_2}=\frac{kT}{X\mu(\gamma-1)y_{\rm H_2}n_{\rm
    HI}\Lambda_{\rm H_2}},
\label{eq:tcool}
\end{equation}
where $X=0.75$ is the hydrogen mass fraction, and $\Lambda_{\rm H_2}$
is the molecular cooling rate. For a gas with $n_{\rm HI}=120\,\rm
cm^{-3}$ and $T=10^4\,\rm K$, $\Lambda_{\rm H_2}\approx 3.4\times
10^{-22} \,\rm erg\,cm^{-3}\,s^{-1}$, and thus 
\begin{equation}
t_{\rm
  cool,H_2}\approx 1.8\times 10^3\,{\rm yr}\,
\left(\frac{y_{\rm H_2}}{10^{-3}}\right)^{-1}.
\label{eq:tool-num}
\end{equation}
With such a rapid cooling, the isothermal shock jump condition
($T_2=T_1$) is a good approximation, and the post-shock density
becomes even higher than that of the adiabatic strong shock, because
$\rho_{b,2}/\rho_{b_1}\approx M_{I,1}^2$ now. Such a strong
compression of the core is observed very frequently in our parameter
space of different halo masses and source fluxes. For example,
Fig. \ref{fig-M2e5-ing} shows how the centre of a halo with $M=2\times
10^4\,M_\odot$ evolves in response to the shock. As the shock hits the
centre, density increases by many orders of magnitude.

Does this compression eventually lead to the core collapse? 
As the shock carries the kinetic energy as well as the thermal energy,
the shock will bounce off the centre after it hits the centre.
In the following section, we
describe this
final stage of the shock propagation and show how it will affect the
core collapse.

\subsubsection{Stage III: Bounce of Shock and Collapse of Core}

After the shock hits the centre, the shock wave will be reflected and
propagate outward. In
our 1D calculation, this reflection will mimic the transmission of the
shock wave through the centre. This bouncing shock will try to disrupt
the gas. The core that is undergoing cooling and
compression due to the positive feedback effects mentioned so far will
be affected by this negative feedback effect, as well.

\begin{figure*}
\includegraphics[%
  width=80mm]{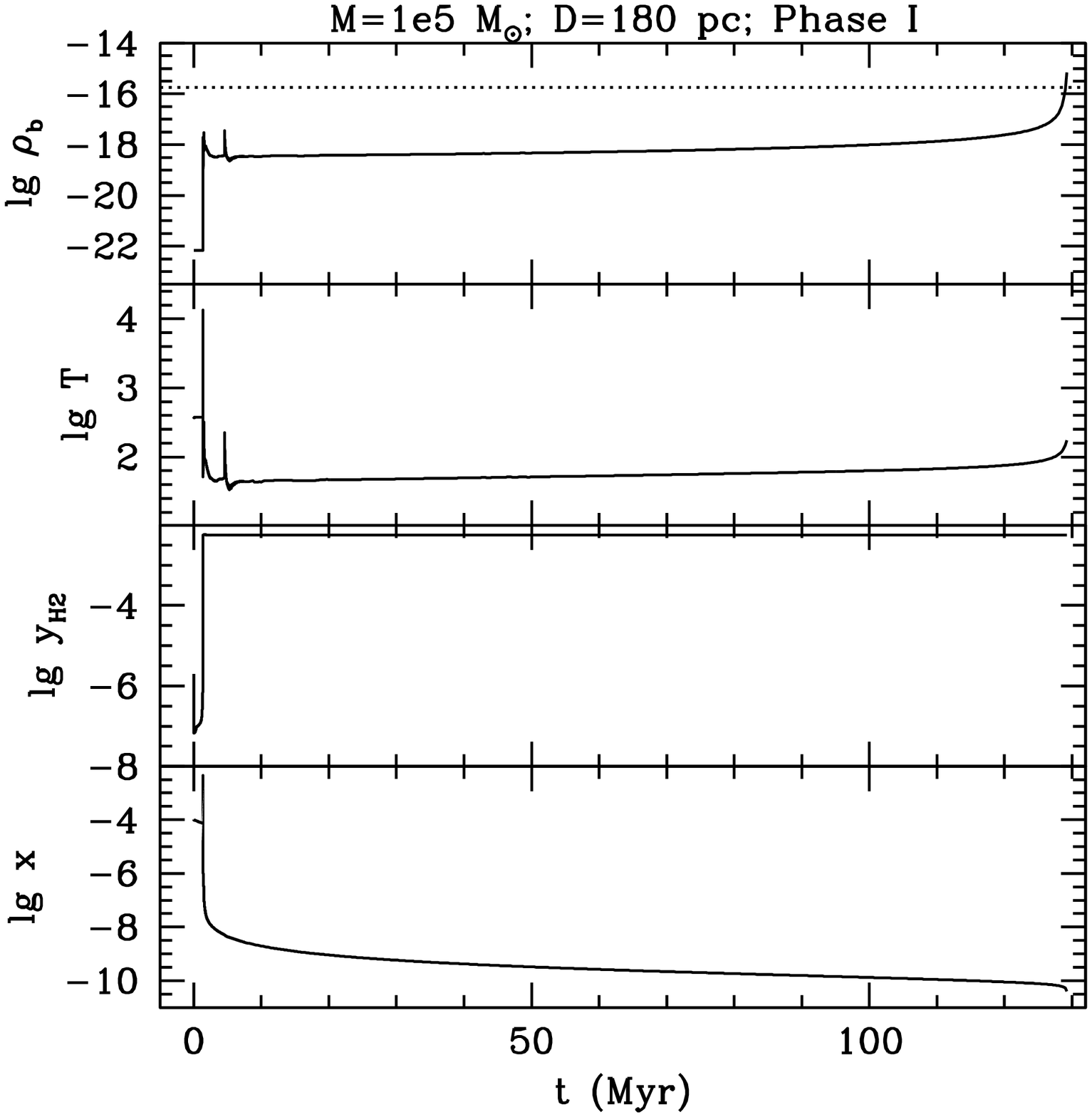}
\includegraphics[%
  width=80mm]{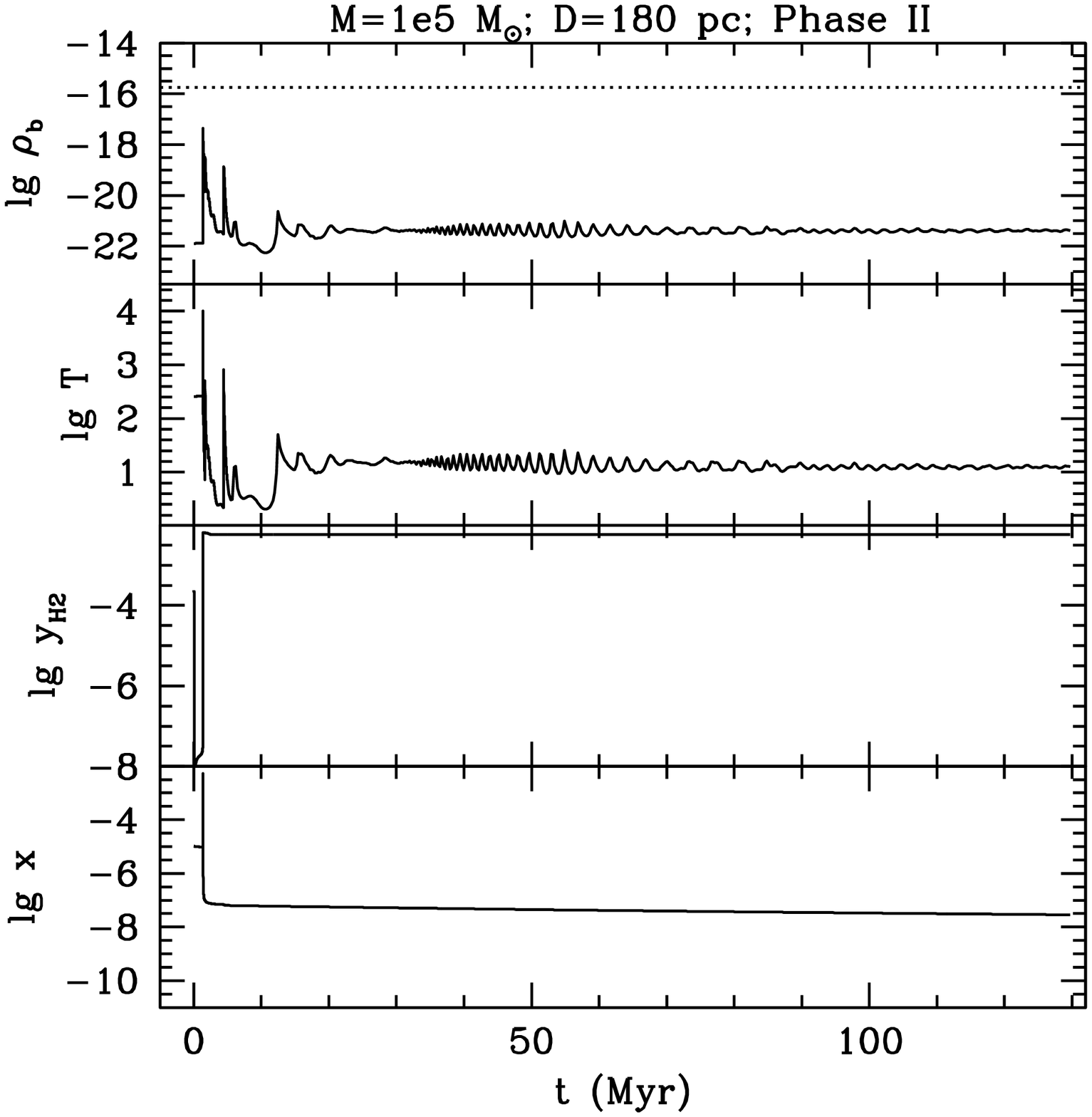}
\includegraphics[%
  width=80mm]{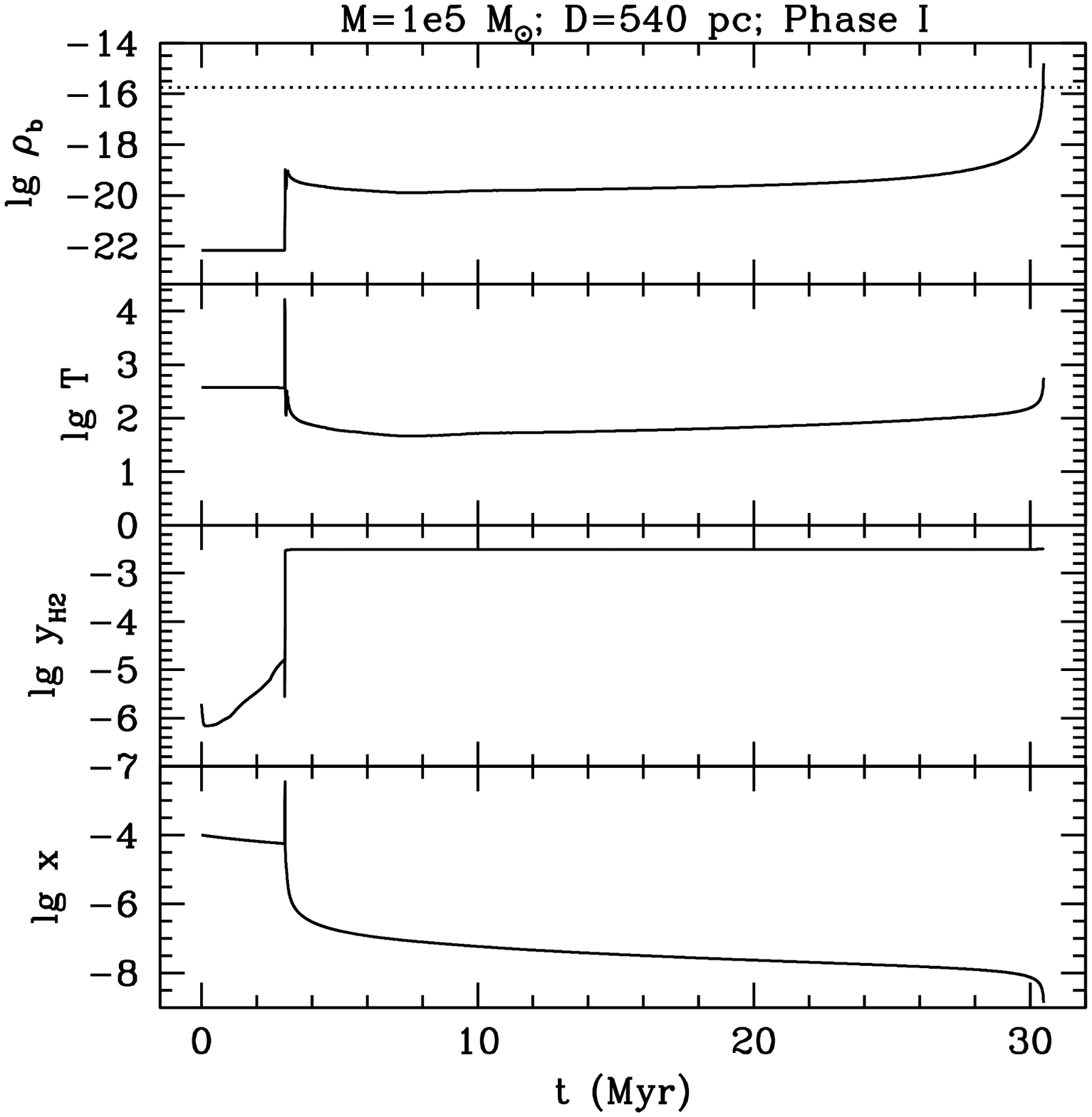}
\includegraphics[%
  width=80mm]{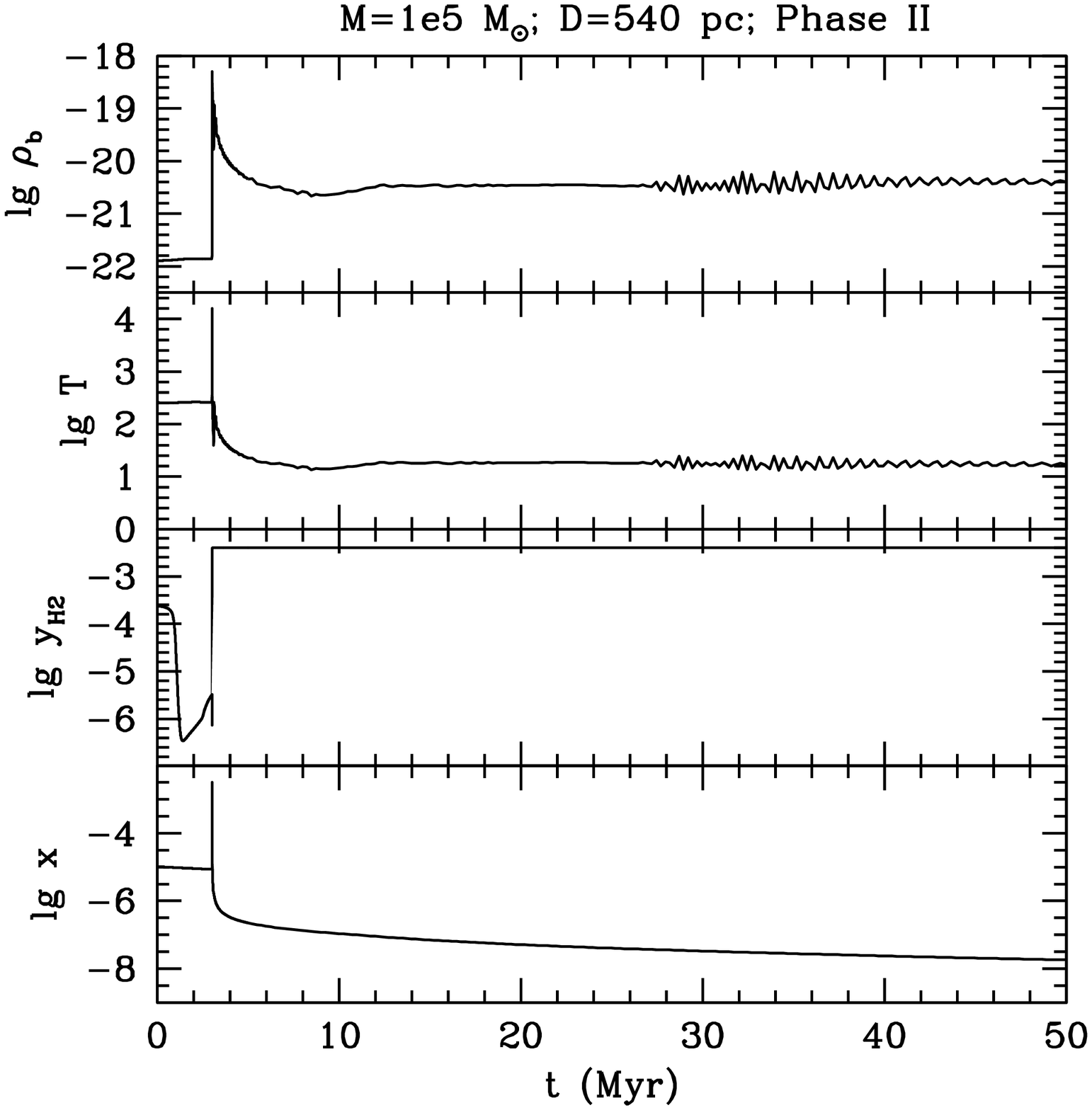}

\caption{Evolution of the central part of a $M=1\times
  10^{5}\,M_\odot$ halo, for Phase I (left) and Phase II (right) at
  $D=180\,{\rm pc}$ (top; $F_0=46.3$) and $D=540\,{\rm pc}$ (bottom;
  $F_0=5.14$). In Phase I, 
  high flux results in the delayed collapse ($D=180\,{\rm pc}$), while
  low flux results in the expedited collapse ($D=540\,{\rm pc}$).
  In Phase II, core collapse is completely halted at any flux.
  The shock-induced molecule formation occurs in all cases, but the
  negative feedback is stronger than the case of higher masses.
\label{fig-M1e5-ing}}
\end{figure*}

\begin{figure*}
\includegraphics[%
  width=80mm]{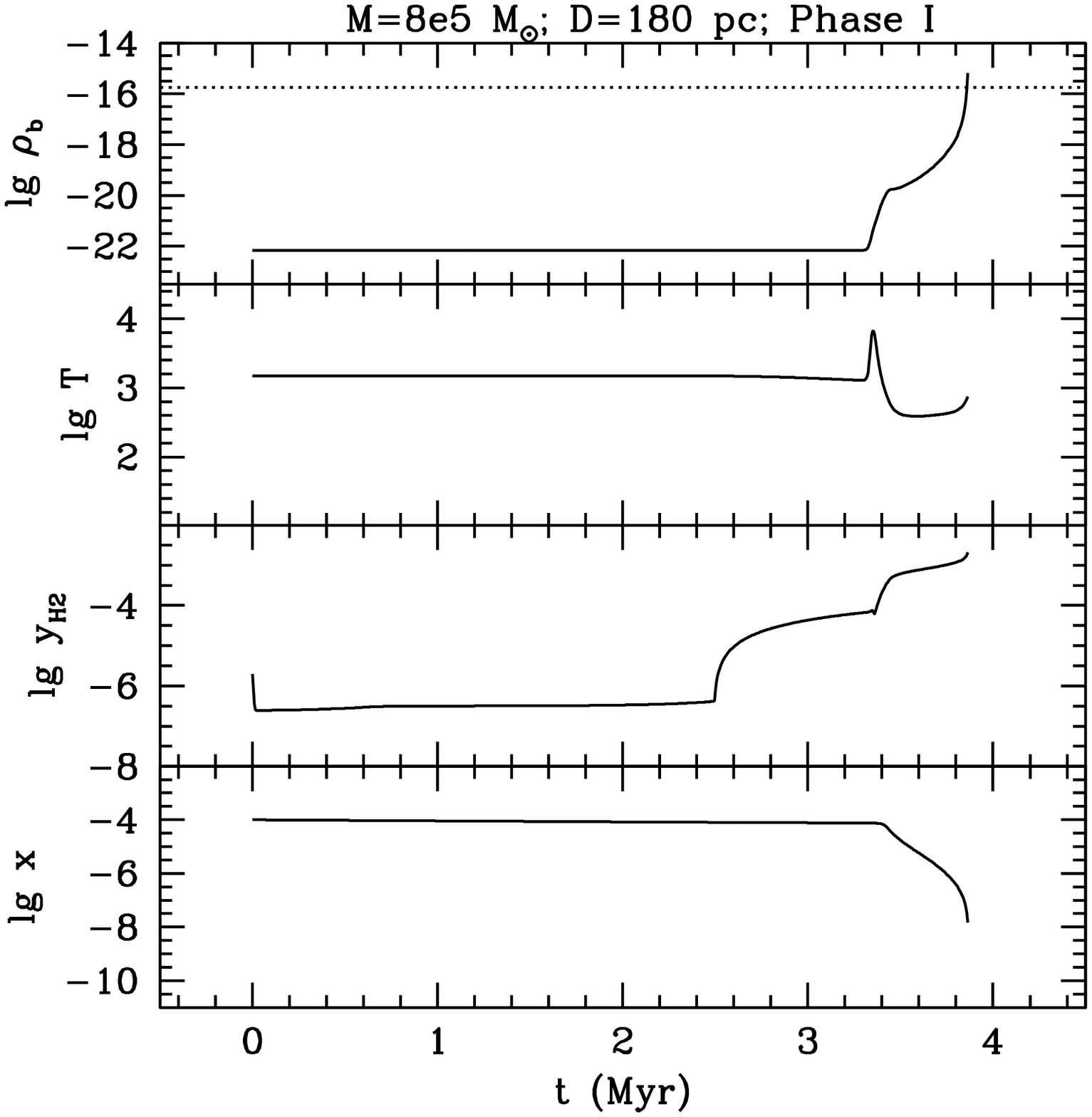}
\includegraphics[%
  width=80mm]{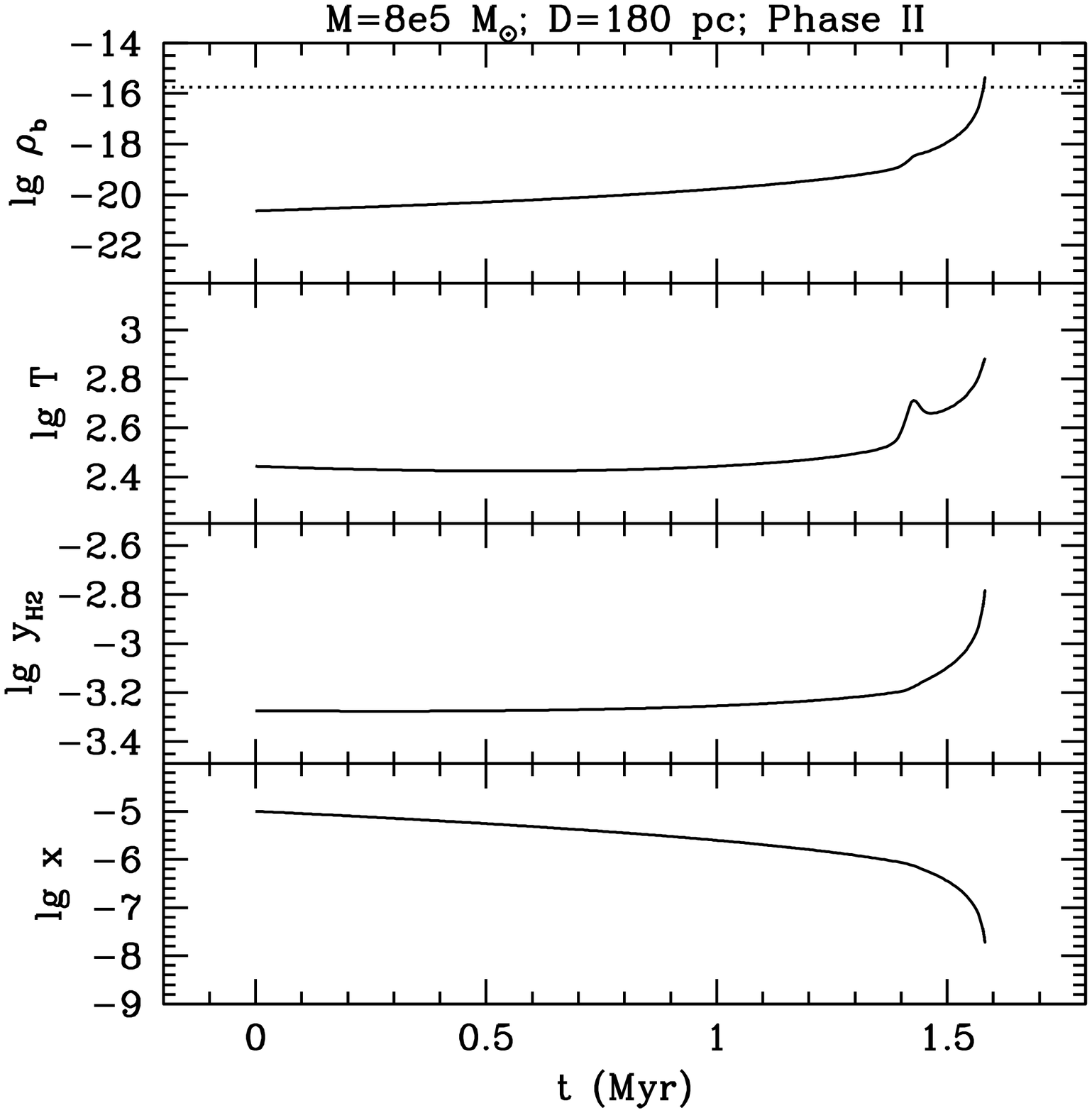}
\includegraphics[%
  width=80mm]{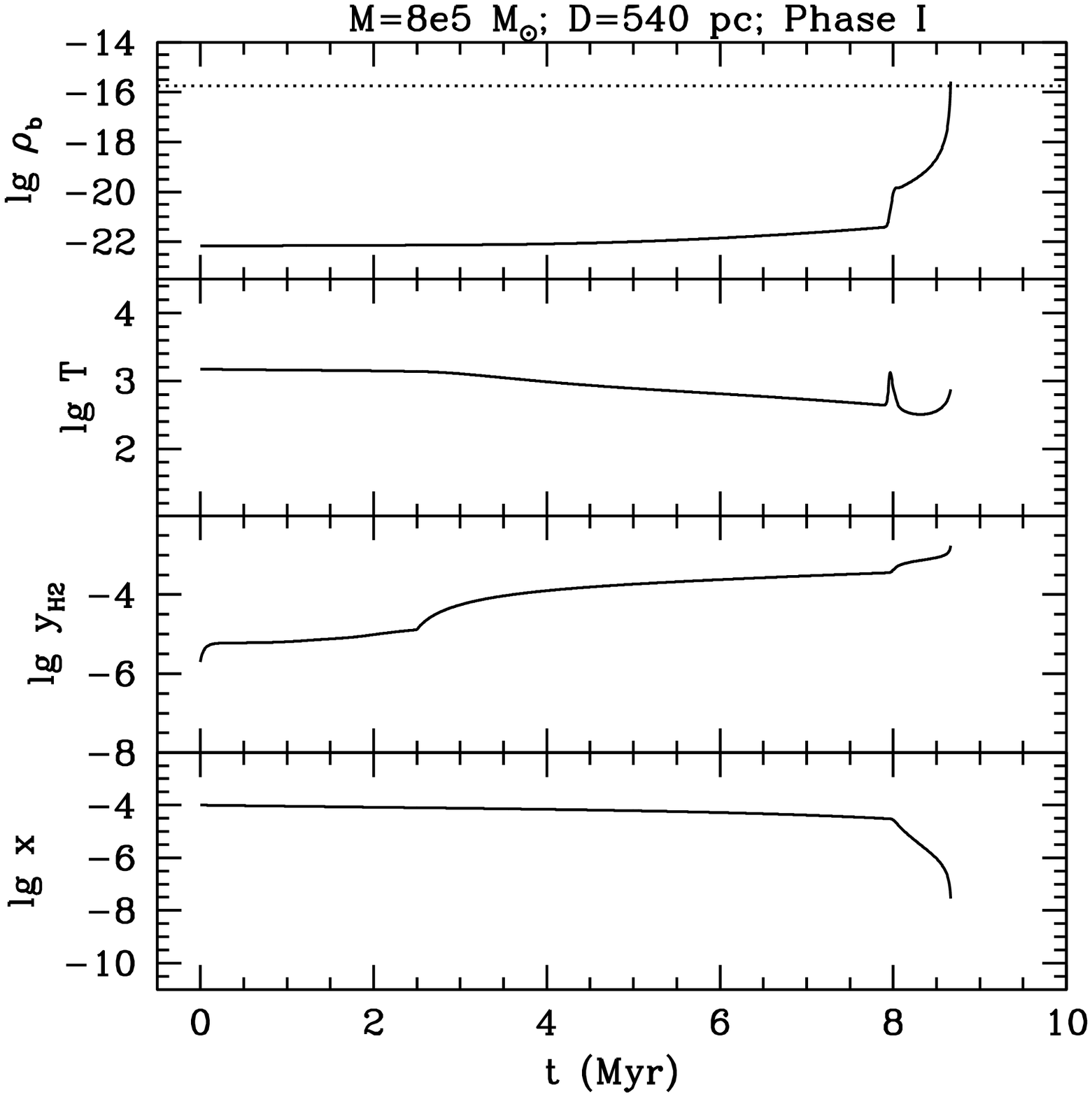}
\includegraphics[%
  width=80mm]{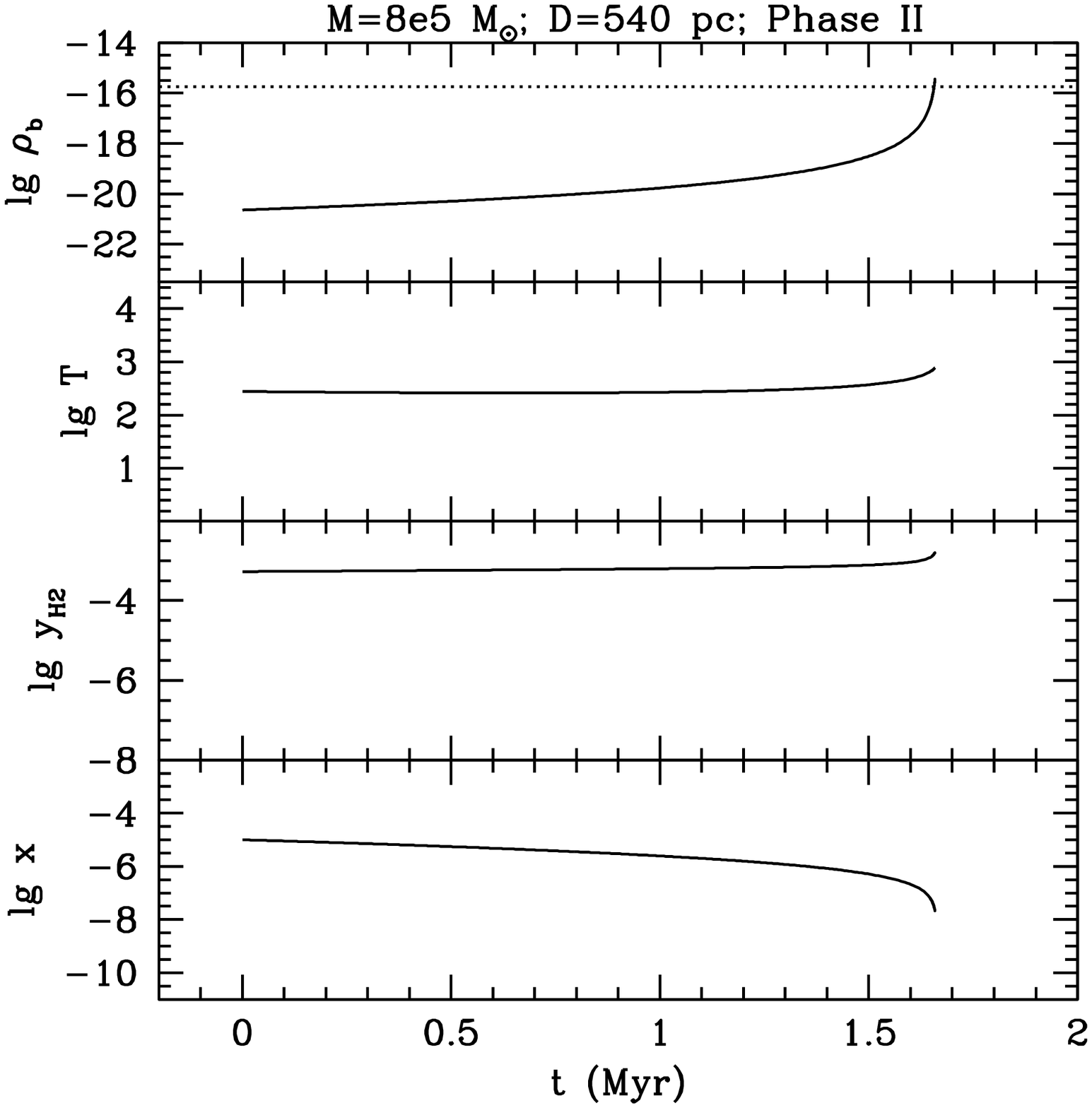}

\caption{Evolution of the central part of a $M=8\times
  10^{5}\,M_\odot$ halo, the highest mass end of our parameter space,
  for Phase I (left) and Phase II (right) at 
  $D=180\,{\rm pc}$ (top; $F_0=46.3$) and $D=540\,{\rm pc}$ (bottom;
  $F_0=5.14$).  
  An expedited collapse occurs for high flux ($D=180\,{\rm
  pc}$). Otherwise, collapse is unchanged in time. 
\label{fig-M8e5-ing}}
\end{figure*}

The final fate of the core depends on how well the core
  endures such a disruption. 
As the shock bounces off the centre, density starts to
decrease. If this bounce is weak, the core quickly reassembles, cools,
  and finally collapses. If
  this bounce is strong, the core will take a longer time to
  collapse and, in some cases, the core will never collapse within the
  Hubble time. Haloes of smaller mass seem to be more susceptible to
  this shock-bounce than those of larger mass (see Figs \ref{fig-M1e5-ing} and
  \ref{fig-M8e5-ing} for comparison).

If the core finally takes the collapse route, the central hydrogen
number density
increases to $\sim 10^{4}\,\rm cm^{-3}$, at which point the ro-vibrational
levels of $\rm H_2$ 
are populated at their equilibrium values and the molecular cooling
time becomes independent of density
(e.g. \citealt{2002Sci...295...93A}). Since then, 
adiabatic heating 
dominates over the molecular cooling, and the temperature increases as
collapse proceeds. Finally, when $n_{\rm HI}$ reaches $\sim 10^{8}\,\rm
cm^{-3}$, the three-body hydrogen reaction ensues and
converts most hydrogen atoms into the molecules, which will undergo a
further collapse and form a proto-star.  

\subsection{Feedback of Pop III starlight on Nearby Minihaloes:
  parameter dependence of core collapse}
\label{sub:Collapse}

We now summarize the outcome of our full parameter study of radiative
feedback effects of Pop III starlight on nearby minihaloes.
As we have described in the previous section, positive and negative feedback
effects of the shock compete and produce a nett effect which can be
either 1) an expedited collapse, 2) delayed collapse, 3) neutral (unaffected)
collapse, or 4) a disruption. 

\begin{figure*}
\includegraphics[%
  width=84mm]{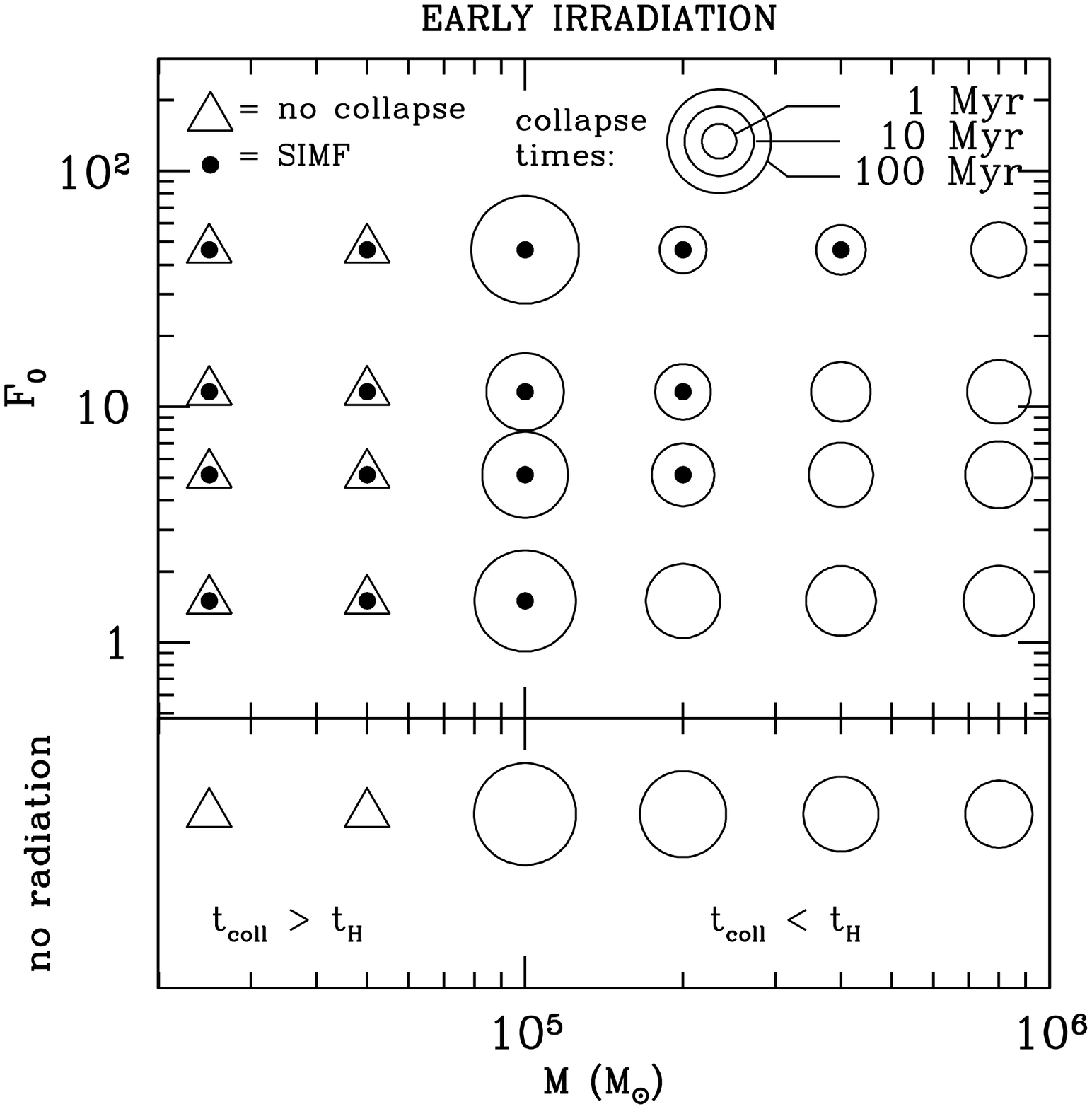}
\includegraphics[%
  width=84mm]{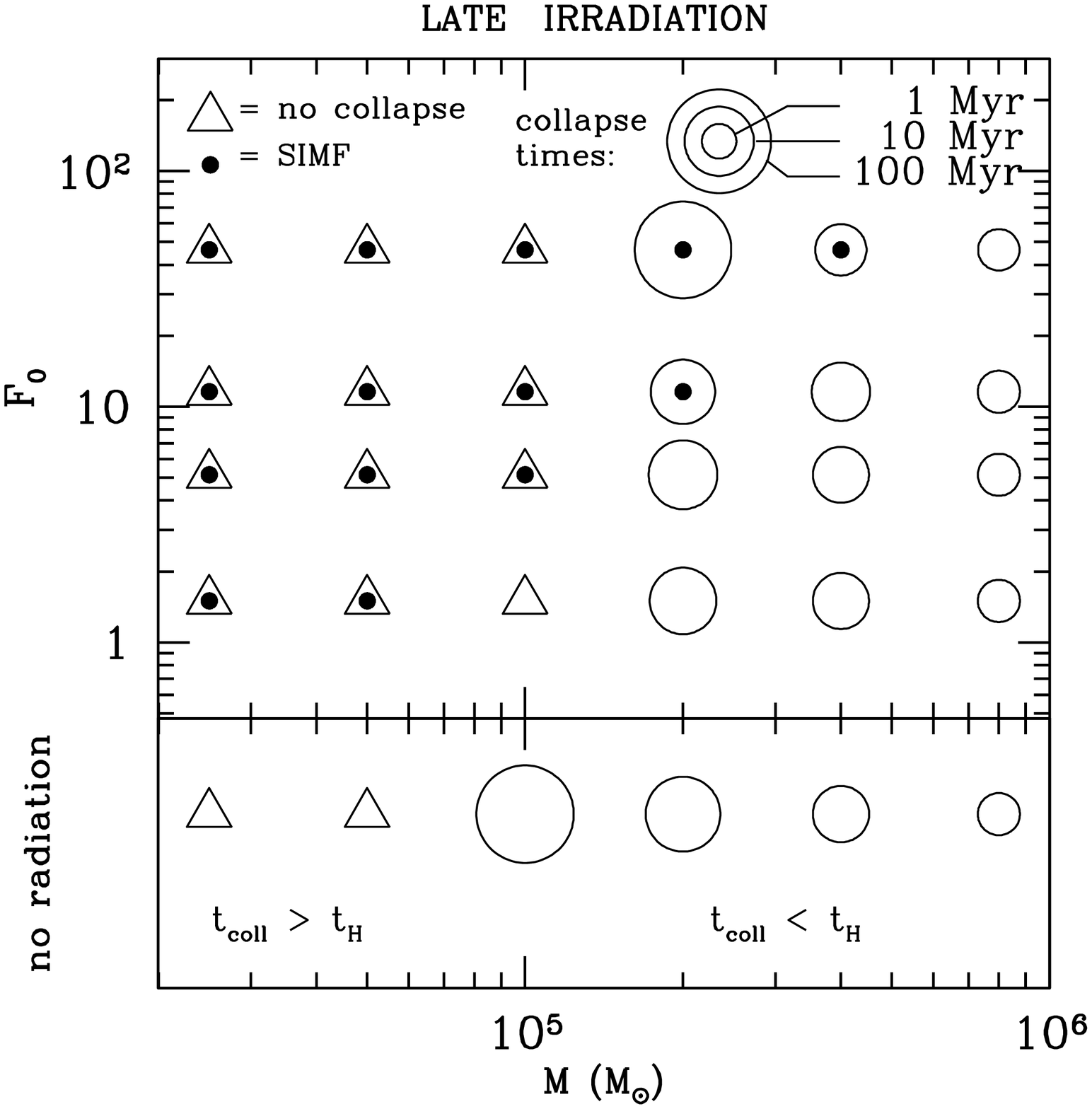}

\caption{
Feedback of Pop III starlight on Nearby Minihaloes: To collapse or not
to collapse? Outcome depends on minihalo mass and stellar flux
$F_0\equiv F/(10^{50}\,\rm s^{-1}\,kpc^{-2})$
($\propto$ distance$^{-2}$) 
as plotted and on the timing of the feedback. 
The flux (distance) is 46.3
(180 pc), 11.6 (360 pc), 5.14 (540 pc) and 1.5 (1000 pc), 
from top
to bottom in each panel, for a 120
$M_{\odot}$ Pop III star located at each distance. The two panels
correspond to different initial conditions when starlight arrives:
(left) (Early = Phase I), virialized halo in
hydrostatic equilibrium with IGM primordial chemical abundances and
(right) (Late = Phase II), halo evolved chemically and hydrodynamically
without radiation 
until $x=10^{-5}$ at centre. The outcome of
the radiative feedback is marked by a circle (collapse) or
a triangle (no collapse), as well as the logarithmic collapse
time (size of circle). Solid dots represent those cases in which
shock-induced molecule formation (SIMF) occurs.
Compared to the {}``no radiation'' cases on the bottom of each panel,
feedback makes collapse either (1) expedited (smaller circle), (2) delayed
(larger circle), (3) unchanged (same sized circle), (4) reversed/failed
(triangle), or (5) unchanged/no collapse (triangle).
Expedited or unchanged collapse occurs widely for
$10^{5}\la M/M_\odot \la 8\times10^{5}$ in the left
panel (Phase I), with exception of $M=10^{5}\, M_{\odot}$ and
$F_0=46.3$($D=180\,{\rm pc}$). 
For right panel (Phase II),
reversed cases
occur for $M=10^{5}\, M_{\odot}$, delayed collapse for $M=2\times10^{5}\, M_{\odot}$
and $F_0=46.3$($D=180\,{\rm pc}$), and expedited or unchanged
collapse for the rest.
\label{fig-coll}}
\end{figure*}

Overall, the radiative feedback effect of a Pop III star is not as
  destructive as naively 
  expected. Minihaloes with $M\ga [1-2]\times 10^5\,M_\odot$, which
  can cool and collapse without radiation, are still
  able to form 
  cooling and collapsing clouds at their centre even in the presence of
  Pop III starlight. 
  The quantitative results are
  summarized in Tables \ref{table:case1}, \ref{table:case2} and
  Fig. \ref{fig-coll}. 

The relatively short lifetime of a Pop III star, compared to the
recombination timescale in the core, is a key to
understanding this behaviour. One of the necessary conditions for the
core collapse is that $\rm H_2$ molecular cooling should occur in the core.
As this requires a sufficient molecular fraction, namely $y_{\rm H_2}\ga
10^{-4}$, it is crucial to understand how molecules are created at
such a level. 
  In Phase I (low $y_{\rm H_2}$ and high $x$), radiation can easily
  dissociate $\rm H_2$ while the source is on, but after the source dies,
  the high electron fraction stimulates $\rm H_2$ formation. 
This is possible because the recombination time in the TIS core is
longer than the lifetime of the source Pop III star.
On the contrary,
  in Phase II (high $y_{\rm H_2}$ and low $x$), $\rm H_2$ is more
  easily protected against the dissociating radiation because the
  higher $\rm H_2$ column density provides self-shielding and
  compression increases the formation rate. Because the source
  irradiates these haloes for a short period of time, the {\em
    dissociation front} does not reach the centre, and its high
  molecule fraction is preserved throughout the Pop III stellar
  lifetime. 

\subsubsection{Phase I}

When haloes start their evolution from Phase I -- IGM chemical
abundance and the TIS structure --, other than the change of collapse
times, there is no reversal of collapse. 
In other words, haloes that were destined to cool and
collapse would do so even when exposed to the first Pop III star in
the neighbourhood.
Minihaloes with
$M\ga 10^5 \,M_\odot$ are able to collapse without radiation, while those with 
$M < 10^5 \,M_\odot$ are not. In the presence of radiation, haloes
with $M\ga 10^5 \,M_\odot$ are still able to collapse, while those
with $M < 10^5 \,M_\odot$ are still unable to do so, even with the help of
shock-induced molecule formation (Fig. \ref{fig-coll}; Table
\ref{table:case1}). 

The core collapse in Phase I occurs mostly as an expedited collapse (Table
\ref{table:case1}). The shock plays a major role in driving such an
expedited collapse: the $\rm H_2$ fraction becomes boosted by the higher
density and high temperature delivered by the shock. Whether or not
SIMF has occurred, such a boost in $y_{\rm H_2}$ is sufficient to 
expedite the core collapse.

There is one delayed collapse case at the low mass and the high flux end. For
$M=10^5\,\rm M_\odot$ at $F_0=46.3$, the boosted molecule formation is
not sufficient to bring the core to an immediate collapse. As the
shock bounces, the momentum carries gas away from the centre until it
cools and recollapses.

The unchanged collapses occur at the high mass and the low flux
  end. For $M=8\times 
  10^5\,M_\odot$ at $F_0=[1.5,\,5.14]$, the shock propagates into the
  already collapsing core. The shock energy delivered in these cases
  is not significant enough to change the course of collapse. 

\begin{table*}
\caption{Collapse times of Phase I for different target haloes (columns) at
  different locations (rows). Each element represents the ratio
  $t_{\rm coll,R}/t_{\rm coll,NR} $, where $t_{\rm coll,R}$ is the
  collapse time (time the halo core takes to reach $n_{\rm crit}=10^8
  {\rm cm^{-3}}$) under
  radiation, and $t_{\rm coll,NR}$ the collapse time without radiation.
  $t_{\rm coll,NR}$ is denoted by values in parentheses. Dot
  represents the case where the core collapse never occurs during the Hubble
  time at $z=20$, or 186 million years.}
\label{table:case1}
\begin{tabular}{ccccccc}
\hline 
&\multicolumn{6}{c}{Total Halo Mass in $10^5\,M_\odot$ units} \\ 
&\multicolumn{6}{c}{(Collapse Time without Radiation in Myrs units)} \\ 
 \cline{2-7}
$D$ (pc) [$F_0$]&
$0.25 $&
$0.5 $&
$1 $&
$2 $&
$4 $&
$8 $\\
&
$(\cdot)$&
$(\cdot)$&
$(88.82)$&
$(31.02)$&
$(14.61)$&
$(8.66)$\\
\hline
\hline 
180 pc [46.3]&
$\cdot$&
$\cdot$&
$1.455$&
$7.288 \cdot 10^{-2}$&
$1.838 \cdot 10^{-1}$&
$4.712 \cdot 10^{-1}$\\
\hline 
360 pc [11.6]&
$\cdot$&
$\cdot$&
$1.935 \cdot 10^{-1}$&
$1.308 \cdot 10^{-1}$&
$3.597 \cdot 10^{-1}$&
$8.177 \cdot 10^{-1}$\\
\hline 
540 pc [5.14]&
$\cdot$&
$\cdot$&
$3.427 \cdot 10^{-1}$&
$2.093 \cdot 10^{-1}$&
$4.919 \cdot 10^{-1}$&
$1.000$\\
\hline 
1000 pc [1.5]&
$\cdot$&
$\cdot$&
$9.497 \cdot 10^{-1}$&
$4.525 \cdot 10^{-1}$&
$7.144 \cdot 10^{-1}$&
$1.241$\\
\hline
\end{tabular}
\end{table*}

\subsubsection{Phase II}

The overall effect of radiation from a Pop
III star on neighbouring minihaloes in Phase II is similar
to the effect on the minihaloes in Phase I: haloes that were
destined to cool and 
collapse would do so even when exposed to the first Pop III star in
the neighbourhood. A slight shift of the trend exists, however, in
Phase II (Fig. \ref{fig-coll}; Table \ref{table:case2}). When haloes
start their evolution from 
Phase II,  
those with
$M\ga 10^5 \,M_\odot$ are able to collapse without radiation, while those with 
$M \la 2 \times 10^5 \,M_\odot$ are not. 
The collapse in Phase II is
reversed (halted) for the low mass end: for $M=10^5\,M_\odot$, the shock
disrupts the core and it never recollapses. SIMF occurs at $F_0>1.5$
for $M=10^5\,M_\odot$, but this does not prevent such a destructive
process from happening.

As haloes start their evolution from Phase II, in which the halo cores
are already 
cooling and collapsing, the neutral (unaffected) collapse cases occur
more frequently 
than in Phase I. At high and intermediate masses, the collapse time
hardly changes from the case without radiation.
Haloes with $M=8\times 10^5 \,M_\odot$ collapse {\em before} the source
dies, as they do without radiation, simply because the shock wave
does not affect the core. In this case, shock
propagates into the centre after collapse has advanced significantly.

There is one delayed collapse case: compared to the delayed collapse
in Phase I, which occurred at low mass/high flux end ($M=10^5\,M_\odot$
at $F_0=46.3$), this now occurs at an intermediate mass/high flux end 
($M=2\times 10^5\,M_\odot$ at $F_0=46.3$).
Otherwise, for intermediate mass, collapse is either neutral or
expedited.

\begin{table*}
\caption{Collapse times of Phase II for different target haloes (columns) at
  different locations (rows). Each element represents the ratio
  $t_{\rm coll,R}/t_{\rm coll,NR} $, where $t_{\rm coll,R}$ is the
  collapse time (time the halo core takes to reach $n_{\rm crit}=10^8
  {\rm cm^{-3}}$) under
  radiation, and $t_{\rm coll,NR}$ the collapse time without radiation.
  $t_{\rm coll,NR}$ is denoted by values in parentheses. Dot
  represents the case where the core collapse never occurs during the Hubble
  time at $z=20$, or 186 million years.}
\label{table:case2}
\begin{tabular}{ccccccc}
\hline 
&\multicolumn{6}{c}{Total Halo Mass in $10^5\,M_\odot$ units} \\ 
&\multicolumn{6}{c}{(Collapse Time without Radiation in Myrs units)} \\ 
 \cline{2-7}
$D$ (pc) [$F_0$]&
$0.25 $&
$0.5 $&
$1 $&
$2 $&
$4 $&
$8 $\\
&
$(\cdot)$&
$(\cdot)$&
$(65.66)$&
$(14.49)$&
$(4.23)$&
$(1.65)$\\
\hline
\hline 
180 pc [46.3]&
$\cdot$&
$\cdot$&
$\cdot$&
$4.269 $&
$7.151 \cdot 10^{-1}$&
$9.541 \cdot 10^{-1}$\\
\hline 
360 pc [11.6]&
$\cdot$&
$\cdot$&
$\cdot$&
$4.997 \cdot 10^{-1}$&
$1.155 $&
$1.002 $\\
\hline 
540 pc [5.14]&
$\cdot$&
$\cdot$&
$\cdot$&
$6.740 \cdot 10^{-1}$&
$9.794 \cdot 10^{-1}$&
$9.964 \cdot 10^{-1}$\\
\hline 
1000 pc [1.5]&
$\cdot$&
$\cdot$&
$\cdot$&
$5.794 \cdot 10^{-1}$&
$9.926 \cdot 10^{-1}$&
$9.994 \cdot 10^{-1} $\\
\hline
\end{tabular}
\end{table*}

\subsection{The structure of haloes at the moment of collapse}
\label{sub:onset}

The structure of halo at collapse determines how a protostar evolves into
a star and
how the starlight will later propagate through the host halo. We first show how
halo profiles at collapse vary for different mass without radiation. We then 
describe how halo structure is affected by the Pop III starlight.
 
\begin{figure*}
\includegraphics[%
  width=148mm]{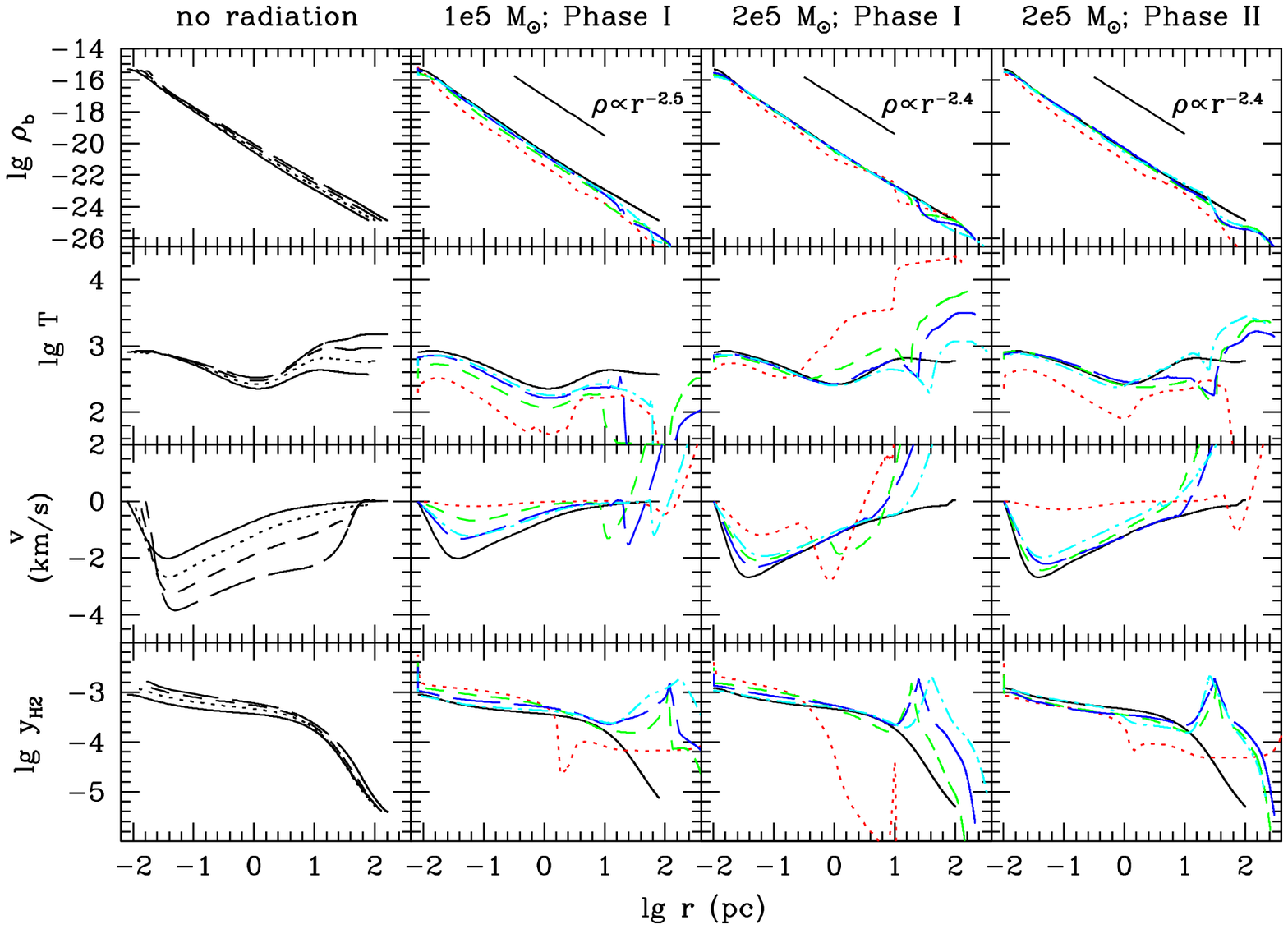}
\includegraphics[%
  width=148mm]{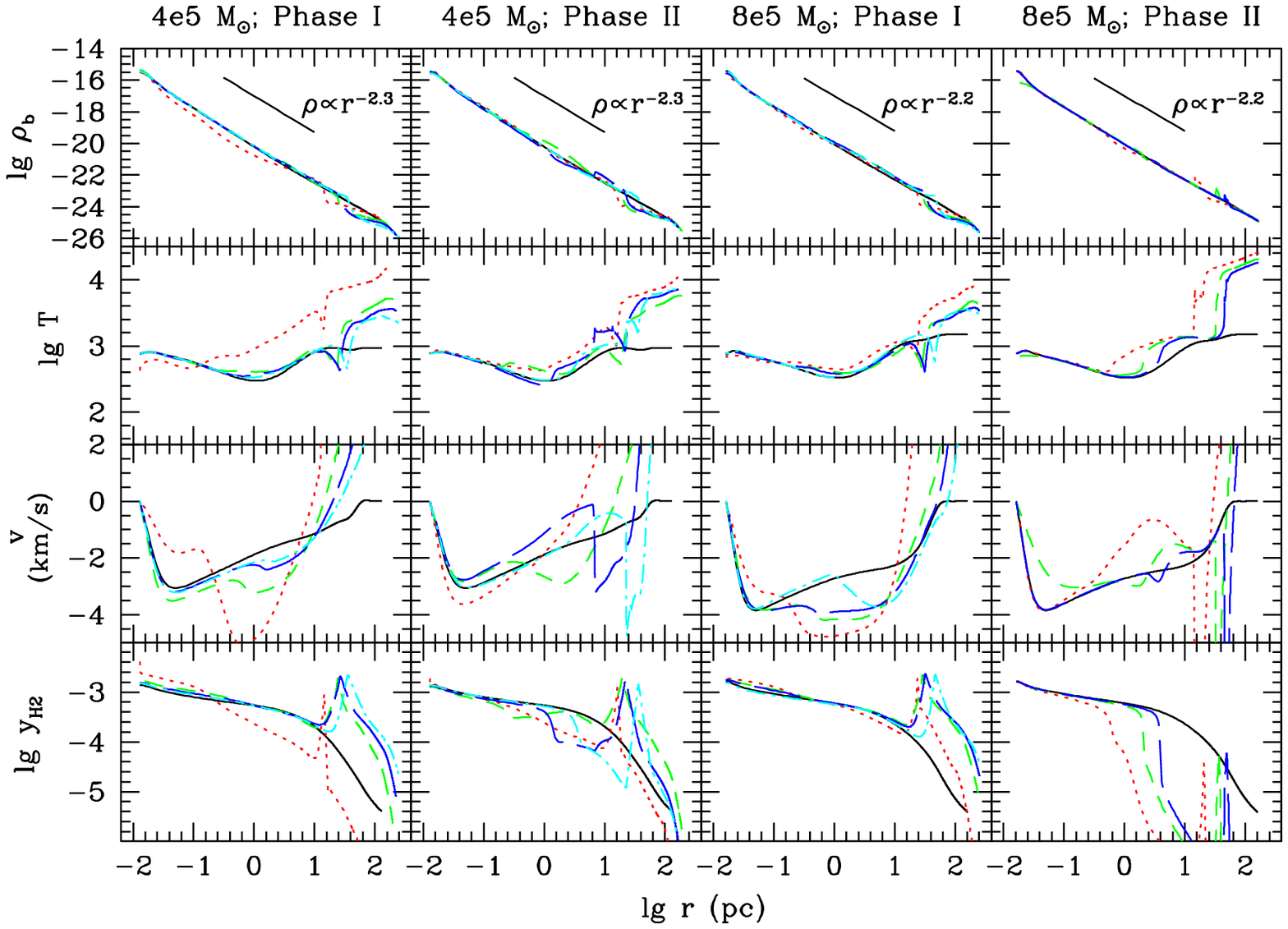}

\caption{Halo profiles at the onset of collapse. The top left panel shows 
  profiles of different mass haloes that collapse without radiation, with mass 
  $M=10^5 \,M_\odot$ (solid), $M=2\times 10^5 \,M_\odot$ (dotted), 
  $M=4\times 10^5 \,M_\odot$ (short-dashed), and $M=8\times 10^5 \,M_\odot$
  (long-dashed). Other panels show profiles of haloes of different masses and 
  phases.
  In each of these panels (except for the top-left panel), no radiation (black;
  solid),
  $D=180\,\rm pc$ ($F_0=46.3$; red; dotted), 
  $D=360\,\rm pc$ ($F_0=11.6$; green; short-dashed), 
  $D=540\,\rm pc$ ($F_0=5.14$; blue; long-dashed),  and
  $D=1000\,\rm pc$ ($F_0=1.5$; cyan; dot-dashed) cases are plotted.
  Note that even though the region ionized during the stellar lifetime
  is heavily affected, the final structure of the neutral
  core remains hardly changed at the onset of collapse in most cases. 
  Some variation is
  seen at the low-mass end, $M=10^5\,M_\odot$, 
  or the high-flux end, $D=180\,\rm pc$ ($F_0=46.3$).
 \label{fig-onset}}
\end{figure*}

We note that halo structure shows a strong dependence on the halo mass.
For radius $r\ga 10^{-2}\,\rm pc$, density profiles of haloes without
radiation are well fit 
by a power law, $\rho \propto r^{-w}$. The value of $w$, however, is dependent
upon the mass of the halo. We find that $w=2.5$, 2.4, 2.3, and 2.2 for 
haloes of mass $M=10^5$, $M=2\times 10^5$, 
$M=4\times 10^5$, and $M=8\times 10^5\,M_\odot$, respectively.
In all cases, the temperature is somewhat flat with 
$T\sim 10^{2.5} - 10^3\,\rm K$. The temperature at $r\approx 10^{-2}\,\rm pc$, 
where $\rho\approx 3\times 10^{-16} \rm g\,cm^{-3}$ 
(or $n_{\rm H}\approx 10^8\,\rm cm^{-3}$), is about $800 \,\rm K$ in all cases.
The universality of these core properties seems to  originate from
the fact that the dominant process, $\rm H_2$ cooling, causes loss memory of 
the
initial condition (e.g. different virial temperatures for different
virial masses). The outer part of these haloes, however, still retain
the memory 
virial equilibrium because radiative cooling is negligible. Overall, 
as mass decreases, density slope increases (see
Fig. \ref{fig-onset}). 

The radiative feedback effect of the starlight on final halo profiles is 
found to be negligible
in most cases. The region that has been photo-ionized during the stellar
lifetime is obviously strongly affected. The neutral region, however,
is almost  
indistinguishable from the case without radiation in most cases. The variance
of temperature profile exists only at the low-mass end, $M=10^5\,M_\odot$, or
the high-flux end, $F_0=46.3$ ($D=180\,\rm pc$). Such variance completely 
disappears at the high-mass end, $M=8\times 10^5\,M_\odot$, because collapse
is mostly unaffected (Fig. \ref{fig-onset}).

This result indicates that the mass of secondary Pop III stars would be
almost identical to that of the Pop III stars which form without radiative
feedback effect. A more fundamental variance may exist, however, due to
the environmental variance of star forming regions:
\citet{2006astro.ph..7013O} show that temperature variance of 
different regions result in the
variance of protostellar masses, due to the corresponding variance of
mass infall rate. As our simulation does not advance 
beyond $n_{\rm H}=10^8\,\rm cm^{-3}$, where three-body collision can
produce copious amount of $\rm H_2$ molecules and change the adiabatic
index of the gas, we are unable to quantify the final mass of the 
protostar at this stage. 

\subsection{Feedback of Pop III Starlight on Merging Haloes and Subclumps}
\label{sub:abel}

While we were preparing this manuscript, two preprints were posted
describing simulations of the radiative feedback of the first Pop III
star on dense gas clumps even closer to the star than the external
minihaloes we have considered so far, for the case of subclumps
\citep{2006ApJ...645L..93S} and the case of a second minihalo
undergoing a major merger with the minihalo that hosts the first star
\citep{2006astro.ph..6019A}.
The centre of the target halo or clumps in this case 
is well within the virial radius of the halo which hosts the first
star, and, these authors find that secondary star formation occurs in
these subhaloes.
\citet{2006astro.ph..6019A}, for
instance, report that the first star forms inside a minihalo of mass $M=4\times
10^5\,M_\odot$ as it merges with a 
second minihalo of mass $M=5.5\times
10^5\,M_\odot$ (the target halo). The centre of this target halo is at a
distance of only 50 parsecs from the first star. Cooling and collapse
leading to the formation of a protostar is found to occur inside the
target halo
about 6 Myrs {\em after} the first star has died.

We ask the same question that whether or not a halo would collapse to
form a secondary Pop III star if a nearby Pop III star 
irradiates the halo at a distance of 50 pc. Note
that the target halo we consider now would collapse anyway if there
were no
radiation, in $\sim 11$ Myrs for Phase I and $\sim 3$ Myrs for Phase
II (see Table \ref{table:collsub}). This problem requires us to extend our
parameter space beyond what 
has been considered so far, because of the short distance (high flux)
between the source and the target.

We have attempted to reproduce the result of \citet{2006astro.ph..6019A}
using our code for a target halo of mass $M=5.5\times 10^5\,M_\odot$
and $D=50\,\rm pc$, corresponding to the ionizing flux $F_0=600$. Note
that the LW band flux is very high: $F_{\rm LW}\sim
2000\times 10^{-21}\,\rm erg\,s^{-1}\,cm^{-2}\,Hz^{-1}$ (equation
\ref{eq:flw}).
As 
$D$ is smaller than the virial radius of the target halo, we truncated
the halo profile at 50 pc. 
To be consistent with our previous calculations, we neglect the
geometrical variation of 
the flux with position inside the target halo.

Surprisingly
enough, contrary to the outcome of \citet{2006astro.ph..6019A},  we
find that collapse is expedited, occurring {\em within the 
  lifetime of the first star}, for both Phase I and Phase II initial
conditions. The main mechanism was SIMF:
initially, $\rm H_2$ is completely wiped out by a strong dissociating
radiation, but as the SIMF occurs, newly created molecules lead to
cooling and collapsing. This result is in 
disagreement with the result of \citet{2006astro.ph..6019A}, which
shows that the second star forms {\em after the star has died}.

This puzzling result shows the importance of $\rm H_2$
self-shielding. \citet{2006astro.ph..6019A} performed an
optically-thin calculation for Lyman-Werner bands, neglecting the  $\rm H_2$
self-shielding, while our calculation took the self-shielding into
account. In order to mimic their calculation more consistently,
we artificially performed an optically-thin calculation for
Lyman-Werner bands. We found that, if the target halo is irradiated 
without $\rm H_2$ self-shielding, 
the core collapse is delayed and occurs {\em after the star
  dies} both in Phase I and Phase II. In our simulations without $\rm
H2$ self-shielding, the core bounced
and recollapsed in $\sim$44 Myrs and $\sim$111 Myrs after the star has
turned off 
in Phase I and Phase II, respectively (Table \ref{table:collsub}).

Qualitatively, our calculation without $\rm H_2$ self-shielding agrees
with the result of 
\citet{2006astro.ph..6019A}, that collapse 
in the target halo occurs
after the source dies. We find that SIMF is the main mechanism for the
formation of $\rm H_2$. Initially, the strong LW band photons destroy
molecules in the core. As the shock propagates inward, however,
boosted density and temperature of the post-shock gas enhances the
molecule fraction (equation \ref{eq:yh2}), and increases the $\rm H_2$
column density. As the shock front accelerates, SIMF occurs, and 
newly created $\rm H_2$ is protected from the LW band photons because
of increased self-shielding. If self-shielding is not accounted for, however,
this $\rm H_2$ is destroyed and never restored, so collapse does not
proceed during the lifetime of the source.

We conclude, therefore, that neglecting $\rm H_2$ self-shielding
in calculation explains why \citet{2006astro.ph..6019A} observes a
delayed collapse. The quantitative disagreement between
our collapse times (when we neglect self-shielding) and theirs may
originate from the difference in the structure and 
chemical abundances of the target halo when the source irradiates it.

How do our results compare with those of \citet{2006ApJ...645L..93S}?
A fundamental 
difference exists other than the fact that
their work is limited to 
subclumps of a halo that hosts a Pop III star. 
They interpret the shock only as a carrier of negative feedback
effect, while 
the shock, in our case, delivers both
the positive and negative feedback effects. In
their shock-driven evaporation (Model C) case, the collapsing core
eventually fails to collapse, because the shock heats the core before
it finishes collapse. 
Their successful
collapse case (Model B) is simply an unaltered collapse: an already
collapsing core finishes collapse before the shock front reaches the
centre. On the other hand, we have observed expedited collapses as
well as delayed or failed collapse. Such expedited collapses we observe
are truly positive feedback effects. Quantitatively, because of their
limited interpretation of the role of the shock, they argue that
only regions with hydrogen number density $n_{\rm H}\ga 10^{2-3} \, \rm
cm^{-3}$, high enough to finish collapse before the shock front
reaches the centre, can collapse under the influence of Pop III
starlight. On the contrary, we find, 
for instance,
that regions with  $n_{\rm H}\sim 30 \, \rm cm^{-3}$ -- core density
of TIS haloes in Phase I -- can cool and collapse even
after the shock front has reached the centre.
As the shock-front accelerates and delivers strong positive
feedback effects in the small core region, high resolution is required
to produce this mechanism in simulations. The relatively poor
resolution of SPH simulations by \citet{2006ApJ...645L..93S} might
have prevented them from fully resolving the shock structure in the
core, and potentially producing the positive feedback effects.

Our result indicates that secondary star formation may occur even in
subclumps of the host halo, which are subject to much stronger
radiative feedback than isolated, nearby minihaloes. We have shown in
this section that $\rm H_2$ self-shielding is important even at
this high level of ionizing ($F_0=600$) and dissociating ($F_{\rm LW}=
2\times 10^{-18}\,\rm erg\,s^{-1}\,cm^{-2}\,Hz^{-1}$) fluxes.
It is even more surprising because
the collapse is expedited and {\em coeval formation} of Pop III stars
in the same neighbourhood is possible. The naive
expectation of negative feedback effect of a Pop III star in its
neighbourhood,
therefore, should be revisited.

\begin{table}
\begin{tabular}{llll}
\hline 
& no radiation & self-shielding & no self-shielding\\
\hline
\hline  
Phase I  & 11.2 & 1.1  & 47 \\
Phase II & 2.7  & 1.3 & 114 \\
\hline

\end{tabular}
\caption{Collapse time (in units of Myrs) of a subclump with
  $M=5.5\times 10^5\,M_\odot$ irradiated by a Pop III star at distance
  $D=50\,\rm pc$ ($F_0=600$). For both Phase
  I and Phase II, 
  we show how a case with a proper treatment of $\rm H_2$
  self-shielding (2nd 
  column) differs in collapse time from a case without self-shielding
  (3rd column) and a case without radiation. When $\rm H_2$
  self-shielding is properly treated, collapse occurs in
  $\sim 1\,\rm Myr$, {\em before} the neighbouring Pop III turns off,
  while when $\rm H_2$ self-shielding is neglected, collapse occurs
  {\em after} the star turns off, which is qualitatively consistent
  with the simulation results by \citet{2006astro.ph..6019A}.
  }
\label{table:collsub}
\end{table}

\section{Summary/Discussion}
\label{sec:2star-Discussion}

We have studied the radiative feedback
effects of the first stars (i.e. Pop III stars) on their nearby
minihaloes, by solving radiative transfer and hydrodynamics
self-consistently using the 1-D spherical, radiation-hydrodynamics
code we have developed.
The results can be summarized as follows:
\begin{itemize}
\item We identified the minimum collapse mass, namely the mass
  of minihaloes which are able to have a core which cools and collapses
  in the absence of external radiation.
  We find that $M_{\rm c,min}\sim 7\times
  10^4\,M_\odot$ at $z=20$. In determining $M_{\rm c,min}$, we applied two criteria.
  First, the collapsing region should reach $n_{\rm
  H}=10^8\,\rm cm^{-3}$ to be considered as a collapse. Second, this
  should occur within the Hubble time. The minimum collapse mass we
  find roughly agrees with that of \citet{2001ApJ...548..509M}, where
  the AMR scheme they used seems to have resolved the
  inner structure of minihaloes.
\item Minihaloes could have been in very different stages of their
  evolution when they were 
  irradiated by a Pop III star. We used two different initial
  conditions to represent such phase differences.
  In Phase I,
  chemical abundances have not yet evolved away from their IGM
  equilibrium values. This stage is characterized by low $\rm H_2$
  fraction, $y_{\rm 
  H_2}\sim 2\times 10^{-6}$ and high electron fraction, $x\sim
  10^{-4}$ at the centre.
  Haloes can be irradiated in Phase II, which is the state of these haloes
  evolved from Phase I, where $x$ has dropped to $10^{-5}$ by recombination.
  Phase II is characterized by high $\rm H_2$
  fraction $y_{\rm 
  H_2}\sim 10^{-4} - 10^{-3}$, low electron fraction  $x=
  10^{-5}$, and core density higher than that of Phase I.
\item Within our parameter space, the I-front is trapped before reaching
  the core in all cases. Ionized gas evaporates, and a shock-front
  develops ahead of the I-front and travels into the core. The shock front
  leads to both positive and negative feedback effects. 
  A boost in density and temperature by a shock increases the $\rm
  H_2$ formation 
  rate. In some cases, the shock accelerates and obtains a temperature
  above  $10^4\,\rm K$, which is high enough to drive collisional
  ionization, which then leads to a further boost in $\rm H_2$ fraction. The
  high temperature and kinetic energy delivered by the shock, on the
  other hand, tries to 
  disrupt the gas. The nett effect is either
  1) an expedited collapse, 2) delayed collapse, 3) neutral
  (unaffected) collapse, or 4) a disruption,
  depending upon the flux, halo mass, and the initial condition when irradiated.
\item At the moment of collapse, halo profiles under radiation are almost 
  identical to those without radiation. Density profiles of different mass
  haloes are well fit by different power-law profiles, $\rho\propto r^{-w}$, 
  where $w=2.5$, 2.4, 2.3, and 2.2 for
  $M=10^5$, $2\times 10^5$, $4\times 10^5$, 
  and $8\times 10^5\,M_\odot$, respectively. Some variation in temperature
  profile exists at the low-mass end, $M=10^5\,M_\odot$, and the high-flux
  end $F_0=46.3$ ($D=180\,\rm pc$).
\item Overall, the radiative feedback effect of Pop III stars is not as
  destructive as naively 
  expected. Minihaloes with $M\ga [1-2]\times 10^5\,M_\odot$ are still
  able to form 
  cooling and collapsing clouds at their centres even in the presence of
  radiation. A simple explanation is possible for such behaviour.
  In Phase I (low $y_{\rm H_2}$ and high $x$), radiation can easily
  dissociate $\rm H_2$ while the source is on, but after the source dies,
  high electron fraction allows $\rm H_2$ formation. On the contrary,
  in Phase II (high $y_{\rm H_2}$ and low $x$), $\rm H_2$ is more
  easily protected against the dissociating radiation because the
  higher $\rm H_2$ column density provides self-shielding and
  compression increases the formation rate.
  The situation becomes more
  complicated, however, by other feedback effects which will be
  described in the following bullets.
\item Within our parameter space, haloes that are irradiated at
  Phase I experience expedited collapse predominantly for $10^5 \la
  M/M_\odot \la 8\times 10^5$, except for the delayed or neutral
  collapses occurring at the low mass/high flux and the high mass/low
  flux extremes (e.g.
  for $M=10^5\,M_\odot$ at $F_0=46.3$ and for $M=8\times
  10^5\,M_\odot$ at $F_0=[1.5,\,5.14]$). 
\item Haloes that are irradiated at
  Phase II show a more complicated behaviour. In this case, unaffected
  collapse is more frequent, in general, at high and intermediate
  masses, while for $M=10^5\,M_\odot$,
  core collapse is now reversed at any $F_0$. Delayed collapse occurs
  for $M=2\times 10^5\,M_\odot$ at $F_0=46.3$. Unaffected collapse
  occurs for $M=8\times 10^5\,M_\odot$ for any $F_0$, and for
  $M=4\times 10^5\,M_\odot$ at $F_0\la 11.6$. Otherwise, for
  intermediate mass, collapse is either neutral or expedited.
\item We first find in this paper that  coeval formation of Pop
  III stars is possible even
  under the influence of ionizing and dissociating radiation from a
  first star. This occurs
  either as an expedited collapse or an unaffected collapse.
  Among those parameters explored in this paper, expedited collapse
  occurs during the 
  lifetime of the source star 
  when a halo of mass $M=2\times 10^5 \,M_\odot$ in Phase I is
  irradiated by a
  Pop III star at a distance $D=180 \,\rm pc$ ($F_0=46.3$). Unaffected
  collapse occurs for haloes of mass $M=8\times 10^5$ in Phase II
  during the lifetime of the source star for all different distances
  (fluxes). 
\item Extending our parameter space to include a specific case
  studied by 
  \citet{2006astro.ph..6019A}, a minihalo merging with a halo hosting
  a Pop III star,
  we find that the coeval formation of Pop III stars is 
  possible even in this high ionizing ($F_0\approx 600$) 
  and dissociating ($F_{\rm LW}\sim
  2\times 10^{-18}\,\rm erg\,s^{-1}\,cm^{-2}\,Hz^{-1}$) flux case.
  While \citet{2006astro.ph..6019A} find that
  the secondary star formation in this target halo occurs after the
  first star dies because of $\rm H_2$ destruction by
  photodissociation, we find 
  that the minihalo core collapse is expedited to form a star in  $\sim 1\,\rm
  Myr$, long before the first star dies, due to the SIMF and
  $\rm H_2$ self-shielding. This discrepancy comes from 
  the fact that we account for the effect of $\rm H_2$ self-shielding,
  while they do not. A proper treatment of $\rm H_2$ self-shielding is
  important even for such a high flux regime, 
  because the central $\rm H_2$ fraction can reach $y_{\rm H_2}\ga
  10^-3$ due to the SIMF and strong $\rm H_2$ self-shielding is
  possible due to newly  created $\rm H_2$.
\end{itemize}

We find the minimum collapse mass $M_{\rm c,min}\sim 7\times
  10^4\,M_\odot$ at $z=20$ without radiation. While our result agrees
  roughly with that 
  of the
  3D AMR simulation by \citet{2001ApJ...548..509M},
  discrepancy becomes larger with those of 3D SPH simulation results
  (e.g. \citealt{2000ApJ...544....6F}; \citealt{2003ApJ...592..645Y})
  and a semi-analytical calculation using a uniform-sphere model
  \citep{1997ApJ...474....1T}. This implies that the central region of
  haloes should be resolved well in order to quantify the minimum
  collapse mass exactly. 

What does the result of our paper imply for the ``first'' H II region
created by Pop III stars? Because a significant fraction of nearby
minihaloes can host second generation stars within the first H II
region, it is possible that such a subsequent star formation may at
least keep 
the first H II regions ionized. It may even be possible that individual
H II regions grow and overlap, thus finishing the first cosmological
reionization. 
A semi-analytic calculation of minihalo clustering around high
density peaks, for example, might allow us to
quantify how fast and how big such bubbles can grow. 
Without secondary star formation, this
would simply be a relic H II region in which gas recombines and
cools after the source star
dies, possibly with metal enrichment from supernova explosion
(e.g. \citealt{2003ApJ...596L.135B}). 

We found  that the minimum collapse mass is $\sim
1-2\times 10^5 \,M_\odot$ even in the presence of Pop III
starlight. Such a low value may affect the reionization history
significantly. \citet{2006ApJ...639..621A} estimates that the
instantaneous ionized mass fraction at $z=20$ is $\sim 0.1$, if individual
$\sim 10^6 
\,M_\odot$ haloes host one $\sim 100 \,M_\odot$ Pop III star each. 
If
the typical mass scale of host haloes is $\sim 10^5 \,M_\odot$ instead,
as the number density of haloes would be roughly 10 times as big as
that for $M\sim 10^6 \,M_\odot$, Pop III stars alone would be able to finish
cosmological reionization at $z \sim 20$\footnote{This argument is
  based upon the fact that the comoving number density of haloes,
  $M\,dn/dM$, is roughly proportional to $M^{-1}$. 
The minihalo population, however, might have been severely reduced by
the ``Jeans-mass filtering'' inside ionized bubbles created around
rare, but more massive objects (e.g. \citealt{2006astro.ph..7517I}),
in which case sources hosted by minihaloes would make negligible
contribution to cosmic reionization.}. New reionization sources will
form later in more massive haloes with $T_{\rm vir}\ga 10^4 \,\rm K$,
which will 
host a region cooling by the hydrogen atomic cooling. Depending
upon how fast such transition occurs, the global reionization history
will have different characteristics (e.g. monotonic growth of
ionization fraction vs. double reionization).

In this paper, we have considered only the radiative feedback
effect. 
Pop III stars, however, may exert additional feedback
effects. 
The H II region developed by a Pop III star inside the host halo
breaks out as a ``champagne flow'' inside the
host minihalo, where the I-front separates from the shock-front and
runs ahead, transforming from D-type to R-type.
The shock front left behind also expands into the IGM
and nearby minihaloes would be encountered by this shock-front
ultimately. Other 
feedback effects will come from supernova explosions. If the
first star dies and explodes as a supernova, both dynamical and
chemical feedback effects would alter the fate of nearby minihaloes,
as well. 

How would the additional presence of $\rm H_2$ dissociating background
radiation affect our results? 
In this paper, we have considered the effect of the radiation from
an individual nearby 
Pop III star, whose SED takes a black body form for a short lifetime
($\sim 2.5$ Myrs). 
This is the case appropriate to the earliest star formation.
It is valid whenever a minihalo resides
in a place and time where the background from other, more distant stars
is negligible. 
On average, however, the mean free path to $\rm H_2$ dissociating
radiation is greater than that for ionizing radiation prior to
reionization, so the situation can arise in which the ionizing
radiation from distant sources is filtered out but the UV radiation in
the LW bands is not.
Suppose a minihalo is under the influence of both  Pop
III starlight from a nearby star
and a persistent background radiation field in the LW bands. 
In the absence of the nearby star, the dissociating background can
only hinder the formation of $\rm H_2$ and its cooling. As such, the
$\rm H_2$ fraction inside the minihalo when
the nearby Pop III star starts to
irradiate it would be lower than it would have been without the
background. 
In this case, even if the background were intense enough on its own to
prevent the minihalo from cooling and collapsing,
the minihalo could still host a cooling core if $\rm H_2$
formed by the positive feedback from the Pop III star, despite the
presence of the background.
Indeed, this could occur
frequently, because we find that
a high electron fraction -- and, thus a high $\rm H_2$ fraction -- can be
achieved by collisional ionization in the postshock region in many
cases (SIMF; see Section \ref{sub:shockstageI}). 
This newly created $\rm H_2$ will then be easily protected from the
dissociating background by self-shielding, since our simulation results show
that this SIMF $\rm H_2$ survives  
even the much larger -- albeit short-lived
-- flux of $\rm H_2$ dissociating 
radiation from a nearby star in our most extreme case, $F_{\rm
  LW}\approx 2000 \times 
10^{-21} \rm erg \, s^{-1}\,cm^{-2}\, Hz^{-1}$, as has been shown in
Section~\ref{sub:abel}. 
Thus, the background would then only prevent
those haloes that cannot ``host'' this SIMF mechanism from cooling and
forming stars. We will
address this issue further in the future.

As the focus of our paper is the fate of neutral cores of target haloes,
in which the ionized fraction never exceeds  $\sim
10^{-2}$, we neglected processes which are relevant only when gas achieves
high ionized fraction, such as 
HD cooling and charge exchange between $\rm He^+$ ($\rm He$)
and $\rm H$
($\rm H^+$) (see e.g. \citealt{2006astro.ph..6106Y}). These processes
may be important, however, in the relic H II 
region outside the target minihalos. For instance, HD cooling may cool
gas down below the $\rm H_2$ cooling 
temperature plateau, $T_{\rm H_2}\sim 100\,\rm K$,
if $\rm H_2$ formation and cooling start from a highly ionized initial
state (e.g. \citealt{2006MNRAS.366..247J}). 

We chose two different evolutionary phases of nearby minihaloes as our initial
conditions. A more natural way to address this problem is to use the
structure and chemical composition of minihaloes and IGM from 3-D,
chemistry-hydrodynamics calculation. 
We intend to
extend our study in a more consistent manner by combining a 3-D,
chemistry-hydrodynamics simulation and the 1-D,
radiation-hydrodynamics simulation in the future.
In this paper, we simply adopted a model for
  virialized haloes (TIS profile). In the future, we will also implement a more
  realistic growth history of haloes
  (e.g. \citealt{2002ApJ...568...52W}) to account for the dynamical
  effect of mass accretion.

\section*{\sc Acknowledgments}
We thank M. Alvarez, T. Abel, S. Glover, I. Iliev, B. O'shea, H. Susa,
and D. Whalen for helpful discussions.
We also acknowledge the Institute for Nuclear Theory at the University
of Washington for their support and hospitality.
This work was supported by NASA 
Astrophysical Theory Program grants NAG5-10825, NAG5-10826,
NNG04G177G.

\appendix
\section{Numerical Method and Code Tests}
\label{finite_appendix}

Here we describe the finite-difference scheme used for our 1-D
spherical, radiation-hydrodynamics code. The subscript, unless noted
otherwise, denotes the position of a shell. The superscript denotes
the time. For instance, $\rho_{j+1/2}^{n+1}$ is the zone-centred
density of shell $j+1$ at time $t^{n+1}$, and $r_{j}^{n}$ is the
zone-edge-centred radius of shell $j$ at time $t^{n}$.

\subsection{The Gas Dynamical Conservation Equations}
Hydrodynamic conservation equations for the baryonic component
(eqs. [\ref{eq:realhydro_mass}] - 
[\ref{eq:realhydro_energy}]) are solved following the
finite-difference scheme by \citet{1995ApJ...442..480T}.
We first update
the velocity and position using the so-called ``leap-frog'' 
scheme, so that the velocity and the position are staggered in time:
\begin{equation}
v_{j}^{n+1/2} = v_{j}^{n-1/2} -\left[
  4\pi(r_{j}^{n})^{2}\frac{p_{j+1/2}^{n} -p_{j-1/2}^{n}}{dm_{j}}
  +\frac{m_{j}^{n}}{(r_{j}^{n})^{2}}\right] dt^{n},
\label{eq:leapfrog1}
\end{equation}
and
\begin{equation}
r_{j}^{n+1} = r_{j}^{n} + v_{j}^{n+1/2}dt^{n+1/2},
\label{eq:leapfrog2}
\end{equation}
which are second-order accurate. As the mass of each shell is
conserved for such a Lagrangian scheme, density is updated
following 
\begin{equation}
\rho_{j+1/2}^{n+1} = \frac{dm_{j+1/2}}{(4/3) \pi[(r_{j+1}^{n+1})^{3}
    -(r_{j}^{n+1})^{3}]}. 
\end{equation}
In these equations,
\begin{equation}
dt^{n}=\frac{1}{2}(dt^{n-1/2}+dt^{n+1/2}),
\end{equation}
and
\begin{equation}
dm_{j}=\frac{1}{2}(dm_{j-1/2}+dm_{j+1/2}).
\end{equation}
We then advance the energy by
\begin{eqnarray}
e_{i+1/2}^{n+1} &=& e_{i+1/2}^{n} -
p_{i+1/2}^{n}\left(\frac{1}{\rho_{i+1/2}^{n+1}} -
\frac{1}{\rho_{i+1/2}^{n}}\right) \nonumber \\
&& +\frac{(\Gamma-\Lambda)_{i+1/2}^{n}}{\rho_{i+1/2}^{n+1}}dt^{n+1/2}.
\label{eq:energy_tw}
\end{eqnarray}

Shocks are treated with the usual artificial viscosity technique. The
pressure in the momentum and energy conservation equations is replaced
by $P=p+q$, where
\begin{eqnarray}
q_{i+1/2}^{n+1} = -c_{q}\frac{2}{1/\rho_{i+1/2}^{n+1}
  -1/\rho_{i+1/2}^{n}} \left|v_{i+1}^{n+1/2}-v_{i}^{n+1/2} \right|
\nonumber \\
\times (v_{i+1}^{n+1/2}-v_{i}^{n+1/2}),
\label{eq:arti_viscos}
\end{eqnarray}
if $v_{i+1}^{n+1/2}-v_{i}^{n+1/2}<0$, and $q=0$ otherwise. We use
$c_{q}=4$, which spreads the shock fronts over four or five cells.

Dark matter shells are also updated according to equations
(\ref{eq:leapfrog1}) - (\ref{eq:arti_viscos}) -- note that we use
fluid approximation as described in Section \ref{sub:fluid-approx} --, 
except that the heating/cooling term is zero in equation
(\ref{eq:energy_tw}). Note that the dark matter
shells are allowed to have effective shock in our fluid approximation,
and therefore we need to compute the artificial viscosity when dark
matter shells
are converging (equation \ref{eq:arti_viscos}), as in the case of the
baryonic gas component.

\subsection{Time Steps}
Time step for the finite-differencing is chosen such that important
fluid variables do not change abruptly. The relevant time scales are
the dynamical, sound-crossing (Courant), cooling(heating), and
species-change time scales. In addition, to ensure that the fluid
shells do not cross, we also adopt a shell-crossing time.
\begin{equation}
dt=\min\{ dt_{{\rm dyn}},\, dt_{{\rm Cour}},\, dt_{{\rm cool}},\,
dt_{{\rm spec}},\, dt_{vel}\}
\end{equation}
\begin{equation}
dt_{{\rm dyn}}=\min\left\{
c_{d}\sqrt{\frac{\pi^{2}r_{j}^{3}}{4m_{j}}}\right\} ,
\end{equation}
\begin{equation}
dt_{{\rm Cour}}=\min\left\{ c_{{\rm
    C}}\left|\frac{r_{j}-r_{j-1}}{\sqrt{\gamma(\gamma-1)u_{j}}}\right|\right\} ,
\end{equation}
\begin{equation}
dt_{{\rm cool}}=\min\left\{
c_{c}\left|\frac{u_{j}\rho_{j}}{(\Gamma-\Lambda)_{j}}\right|\right\} ,
\end{equation}
\begin{equation}
dt_{{\rm spec}}=\min\left\{ c_{{\rm sp}}\left|\frac{x_{j}}{dx_{j}/dt}\right|,\, c_{{\rm sp}}\left|\frac{y_{{\rm
      H I},\, j}}{dy_{{\rm H I,}\, j}/dt}\right|\right\} 
\end{equation}
\begin{equation}
dt_{{\rm vel}}=\min\left\{
c_{v}\left|\frac{r_{j}-r_{j-1}}{v_{j}-v_{j-1}}\right|\right\} ,
\end{equation}
where $c_{d}$, $c_{\rm C}$, $c_{c}$, $c_{\rm sp}$, and $c_{v}$ are
coefficients that ensure accurate 
calculation of the finite difference equations. We use $c_{d}=0.1$,
$c_{\rm C}=0.1$, $c_{c}=0.1$, $c_{\rm sp}=0.1$, and $c_{v}=0.05$. 

In
practice, we frequently find that $dt_{\rm dyn}$ can be very small compared to
other time scales. We sometimes disregard $dt_{\rm dyn}$ in order to
achieve computational efficiency. We confirmed, especially in our
problem, that such a treatment does not produce any significant
discrepancy from a calculation with $dt_{\rm dyn}$ considered. When
the virial temperature of a halo is close to the cooling temperature
plateau, for instance, $dt_{\rm dyn}$ must be irrelevant because gas
would be almost hydrostatic.

\subsection{Radiative Transfer}
For the radiation field generated
from a point source at the centre, the radiative rate coefficient
of species $i$ at radius $r$ is given by equation
(\ref{eq:rad_rate_ext_body}). 
Finite-differencing this rate coefficient, however, requires some caution.
For the baryonic shell at position $j$ (smaller $j$ means closer
to the centre) whose inner edge and outer edge have radii $r_{j-1/2}$
and $r_{j+1/2}$, respectively, the incident differential flux at
the outer edge is $F_{\nu}^{{\rm int}}(r_{j+1/2})$, and one could
naively calculate the rate coefficient of species $i$ by 
\begin{equation}
k_{i}(r_{j})=\int_{0}^{\infty}d\nu\frac{\sigma_{i,\nu}F_{\nu}^{{\rm
      ext}}(r_{j+1/2})}{h\nu}.
\label{nonconserving_k_ext2}
\end{equation}

As mentioned already in Section \ref{sub:External-source} and Section
 \ref{sub:photo-heating}, however, this 
 expression may not yield an accurate result when the
shell $k$ is optically thick. In this case, $F_{\nu}$ may change
substantially over the shell width, and equation
(\ref{nonconserving_k_ext2}) 
might overpredict the ionization rate by applying a constant flux
over the shell width ($\Delta r_{j}\equiv r_{j+1/2}-r_{j-1/2}$).
One may, in principle, choose to set up the initial condition such
that all shells are optically thin. However, such a scheme can be very
expensive computationally, especially when collapsed haloes are treated.
In order to resolve this problem, we use the {}``photon-conserving
scheme'' by \citet{1999MNRAS.309..287R} and
 \citet{1999ApJ...523...66A}. In this 
treatment, the number of photons that are absorbed in a shell is the
same as the number of ionization events. Equation
(\ref{nonconserving_k_ext2})
can then be re-written as 
\begin{eqnarray}
k_{i}(r_{j}) &=& \int_{0}^{\infty}d\nu\frac{L_{\nu}^{{\rm
      ext}}(r_{j+1/2})-L_{\nu}^{{\rm
      ext}}(r_{j-1/2})}{h\nu}\cdot\frac{1}{n_{i}V_{{\rm shell},j}}
\nonumber \\
&\simeq& \int_{0}^{\infty}d\nu\frac{F_{\nu}^{{\rm
      ext}}(r_{j+1/2})}{h\nu}
\cdot\frac{1-e^{-\Delta\tau_{i,\nu}(r_{j})}}{n_{i}\Delta r_{j}},
\label{eq:conserving_k_ext2}
\end{eqnarray}
 where $L_{\nu}^{{\rm ext}}(r)=4\pi r^{2}F_{\nu}^{{\rm ext}}(r)$,
$\Delta\tau_{i,\nu}(r_{j})\equiv n_{i}\Delta r_{j}\sigma_{i,\nu}$
is the optical depth of a shell $k$ on a species $i$, and $V_{{\rm
    shell},j}\simeq4\pi r_{j}^{2}\Delta r_{j}$ 
is the volume of the shell. Note that when $\Delta\tau_{\nu}\ll1$,
equation (\ref{eq:conserving_k_ext2}) becomes equivalent to equation
(\ref{nonconserving_k_ext2}).
For each species,
the corresponding radiative reaction rate is calculated by quadrature,
by summing the integrand in equation (\ref{eq:conserving_k_ext2}), 
then summing
over the frequency to obtain the nett radiative reaction rate.

\subsection{Nonequilibrium Chemistry}
As described in Section~\ref{sub:noneq_chem}, in order to update the
abundance of species $i$, we adopt the finite
difference scheme by \citet{1997NewA....2..181A}.
Based upon equation (\ref{eq:verygeneric_rate_eq}), each species $i$
is updated by  
\begin{equation}
n_{i}^{n+1}=\frac{C_{i}^{n+1}(T,\{ n_{j}\}) dt^{n+1/2} +
  n_{i}^{n}}{1+D_{i}^{n+1}(T,\{ n_{j}\})dt^{n+1/2}},
\label{eq:backward-diff}
\end{equation}
 where the species $\{ n_{j}\}$ is the previously updated value in
the order given by \citet{1997NewA....2..181A} (note that the letter
$n$ ($n+1/2$, $n+1$) in superscript denotes the time $t^{n}$ ($t^{n+1/2}$,
$t^{n+1}$). The order they find 
to be optimal is H, ${\rm H^{+}}$, He, ${\rm He^{+}}$, ${\rm He^{++}}$
and ${\rm e^{-}}$, followed by the algebraic equilibrium expressions
for ${\rm H^{-}}$ and ${\rm H^{+}}$, and finally ${\rm H_{2}}$,
again by equation (\ref{eq:backward-diff}).

\subsection{Numerical resolution}
\label{sub:resolution}

In practice, we use $500$ dark matter and $1000$ fluid shells sampled uniformly
(in radius) from the centre to the truncation radius $r_{{\rm tr}}$.
We put a small reflecting core at the centre with negligible size,
namely $r_{{\rm core}}=10^{-4}r_{{\rm tr}}$. Such a core is
found to be useful in reducing undesirable numerical instability at
the centre. Our choice is conservative enough not to affect the overall
answer.

A wide range of radiation frequency (energy),
$h\nu\sim[0.7\,-\,7000]\,{\rm eV}$,
is covered by $100$, logarithmically spaced bins, 
$\Delta E/E \approx 0.04$,
together
with additional, linearly-spaced bins where radiative cross sections
change rapidly as frequency changes. About a dozen linearly spaced
bins at each of those rapidly changing points turned out to produce
reliable results.

\section{Rate coefficients}
\label{sub:rates}

In Table \ref{table:rates}, we list the chemical reaction rates we
implemented in 
our code and the corresponding references. The rate coefficients
(1-19) and
radiative cross sections (20-26) are mostly from the fit by
\citet{1987ApJ...318...32S}, except for a few updates. 

\begin{table*}
\begin{tabular}{lll}
\hline 
&Reactions & Reference\\
\hline
\hline  
1 & $\rm H + e^- \rightarrow H^+ + 2e^- $ & \citet{1987ephh.book.....J}\\
2 & $\rm H^+ + e^- \rightarrow H + \gamma $ & Case B; \citet{1989agna.book.....O}\\
3 & $\rm He + e^- \rightarrow  He^+ + 2e^- $ & \citet{1987ephh.book.....J}\\
4 & $\rm He^+ + e^-  \rightarrow He + \gamma $ & \citet{1973AA....25..137A}\\
5 & $\rm He^+ + e^-  \rightarrow He^{++} + 2e^- $ & AMDIS Database; \citet{1997NewA....2..181A}\\
6 & $\rm He^{++} + e^- \rightarrow He^+ + \gamma $ & \citet{1978ppim.book.....S}\\
7 & $\rm H + e^-  \rightarrow H^- + \gamma           $ & \citet{1972AA....20..263D,1987ApJ...318...32S}\\
8 & $\rm H^- + H  \rightarrow H_2 + e^-              $ & \citet{bieniek}\\
9 & $\rm H + H^+  \rightarrow H_{2}^{+} + \gamma     $ & \citet{1976PhRvA..13...58R}\\
10& $\rm H_{2}^{+} + H  \rightarrow H_2 + H^+        $ & \citet{1979JChPh..70.2877K}\\
11& $\rm H_2 + H           \rightarrow  3H          $ & \citet{1986ApJ...311L..93D}\\
12& $\rm H_2 + H^+         \rightarrow H_{2}^{+} + H $ & \citet{2004ApJ...606L.167S}\\
13& $\rm H_2 + e^-         \rightarrow 2H + e^-      $ & \citet{1983ApJ...266..646M}\\
14& $\rm H^- + e^-         \rightarrow H + 2e^-      $ & \citet{1987ephh.book.....J}\\
15& $\rm H^- + H           \rightarrow 2H + e^-      $ & \citet{1984SvA....28...15I}\\
16& $\rm H^- + H^+         \rightarrow 2H            $ & \citet{1984inch.book.....D}\\
17& $\rm H^- + H^+         \rightarrow H_{2}^{+} + e^- $ & \citet{1978JPhB...11L.671P}\\
18& $\rm H_{2}^{+} + e^-   \rightarrow 2H              $ & \citet{1994ApJ...424..983S}\\
19& $\rm H_{2}^{+} + H^-   \rightarrow H + H_2         $ & \citet{1987IAUS..120..109D}\\
\hline
20& $\rm H + \gamma        \rightarrow H^+ + e^-       $ & \citet{1989agna.book.....O}\\
21& $\rm He^+ + \gamma     \rightarrow He^{++} + e^-   $ & \citet{1989agna.book.....O}\\
22& $\rm He + \gamma       \rightarrow He^{+ } + e^-   $ & \citet{1989agna.book.....O}\\
23& $\rm H^- + \gamma      \rightarrow H + e^-         $ & \citet{1972AA....20..263D,1987ApJ...318...32S}\\
24& $\rm H_{2}^{+} + \gamma \rightarrow H + H^+        $ & \citet{1968PhRv..172....1D}\\
25& $\rm H_2 + \gamma      \rightarrow H_{2}^{+} + e^- $ & \citet{1978JChPh..69.2126O}\\
26& $\rm H_{2}^{+} + \gamma \rightarrow 2H^{+} + e^-   $ & \citet{1968JPhB....1..543B}\\
\hline
27& $\rm H_2 + \gamma      \rightarrow 2H             $ & Section \ref{sub:ss}; \citet{1996ApJ...468..269D}\\
\hline

\end{tabular}
\caption{Reactions and the corresponding references.
  }
\label{table:rates}
\end{table*}

\section{Code tests}
We now extend the description of our code test problems in Section
\ref{sec:codetest} and show the results.

(A) The self-similar, spherical, cosmological infall problem
\citep{1985ApJS...58...39B}: A point mass, if placed in an
unperturbed Einstein-de Sitter universe, will make all particles
around it to be gravitationally bound, leading to a successive
turnaround and collapse of spherically shells. Infalling matter will
be shocked and form a virialized structure, whose profiles are well
described by a self-similar solution. We restrict ourselves to purely
baryonic fluid with the ratio of specific heats $\gamma = 5/3$.

\begin{figure*}
 \includegraphics[%
   width=168mm]{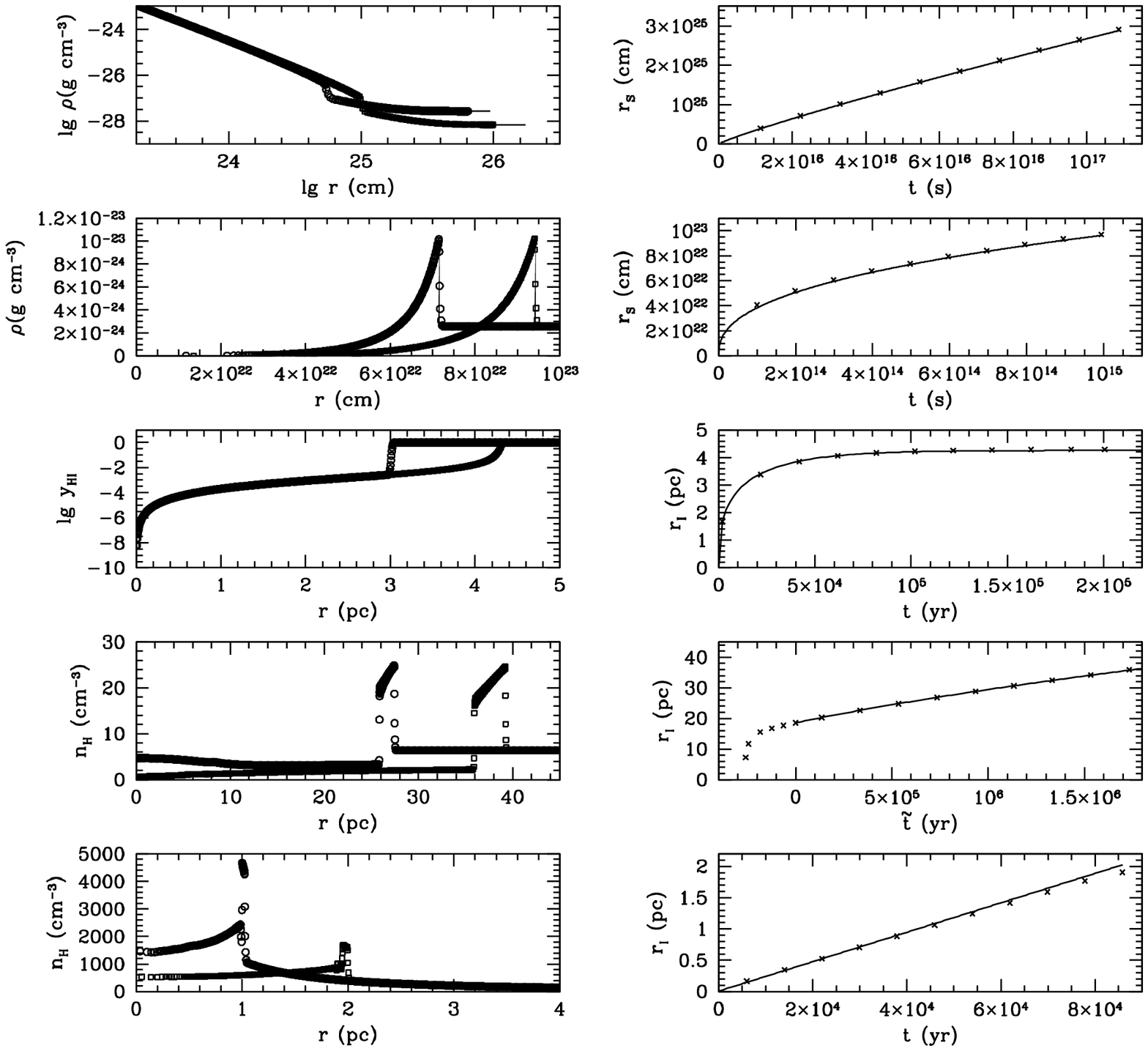}

 \caption{Code test results. From top to bottom, simulation results of
   (A) the self-similar, spherical, cosmological infall problem
   \citep{1985ApJS...58...39B}, 
   (B) the self-similar blast wave from a strong, adiabatic point
   explosion in a uniform gas \citep{1959sdmm.book.....S},
   (C) the propagation of an I-front from a steady point-source in a
   uniform, static medium,
   (D) the gas-dynamical expansion of an H II region from a point source
   in a uniform gas \citep{1966ApJ...143..700L}, and
   (E) the gas-dynamical expansion-phase of the H II region from a
   point-source in a nonuniform gas whose density varies with distance
   $r$ from the source as $r^{-w}$, $w=3/2$
   \citep*{1990ApJ...349..126F} are displayed, respectively, as
   described in text.
   In each
   row, some early (circle) and late (square) snapshots of density
   profiles (neutral fraction profile in case 
   C) are shown in the left panel, while the evolution (cross) of shock radius
   ($r_s$) or I-front ($r_I$) are  shown in the right panel. Data
   points (circle, square, cross) are compared with the analytical
   prediction (solid line), in case analytical solutions
   exist. 
   Note that for test (D), $r_{\rm I}$ is plotted against
   $\tilde{t}=t-t_c$, where $t_c$ is the time when $dr_{\rm 
  I}/dt = c_{\rm I}$. In this case, the analytical solution for
   $r_{\rm I}$ is valid only for $\tilde{t}\ge 0$.
   On left panels, snapshots are shown at $t=5.5\times 10^{16}
   \rm s$ and $1.1\times 
   10^{17} \rm s$ for (A), $t= 5\times 10^{14} \rm s$ and $ 10^{15} \rm
   s$ for (B), 
   $t=1.4\times 10^{4} \rm yr$ and $6.1\times 10^{5} \rm yr$ for (C), $t=
   9\times 10^{5} \rm yr$ and $2\times 10^{6} \rm yr$ for (D), and $t=4.3\times
   10^{4} \rm yr$ and $8.6\times 10^{4} \rm yr$ for (E).
 \label{fig-test}}
 \end{figure*}

The turnaround radius $r_{\rm ta}$, at which the Lagrangian proper velocity
of a shell is zero, evolves as
\begin{equation}
r_{\rm ta}(t)=\left( \frac{3\pi}{4}\right)^{-8/9} \left({\delta_{i}
  R_{i}^{3}}\right)^{1/3} (t/t_{i})^{8/9},
\label{eq:bert-rta}
\end{equation}
where $\delta_i R_{i}^{3}$ defines the seed mass $\delta m$ added to
the Einstein-de Sitter universe,
\begin{equation}
\delta m = \frac{4}{3}\pi \rho_{{\rm H},i} \delta_i R_{i}^{3},
\label{eq:bert-seed}
\end{equation}
where the initial cosmic mean density $\rho_{{\rm H},i}=1/(6\pi G
t_{i}^{2})$ at $t=t_{i}$.
The shock radius $r_s$ is a constant fraction of $r_{\rm ta}$: $r_s (t)=
0.338976 \,r_{\rm ta} (t)$ for $\gamma = 5/3$. The dimensionless radius
$\lambda \equiv r/r_{\rm ta}$ and the dimensionless density
$D\equiv \rho / \rho_{\rm H}$, where the cosmic mean density
$\rho_{\rm H}=1/(6\pi G t^2)$, satisfy the unique Bertschinger solution.
In Fig. \ref{fig-test}, we show
the density profiles and $r_{\rm s}(t)$, obtained from the simulation with
$\delta_i R_{i}^{3} = 1.84\times 10^{71} \,\rm cm^3$, $t_{i}=5.572\times
10^{14} \,\rm s$.

(B) The self-similar blast wave from a strong, adiabatic point
explosion in a uniform gas \citep{1959sdmm.book.....S}:
A point explosion drives a
self-similar blast wave through the initially static, uniform
medium. A strong shock is generated, and $r_{\rm s} (t) = \xi_0 \left(
  \frac{E}{\rho_0} \right)^{1/5} t^{2/5}$, where $E$ is the thermal
energy of explosion, $\rho_0$ is the initial density, and $\xi_0$ is a
dimensionless constant determined by $\gamma$. For $\gamma=5/3$, $\xi_0
= 1.152$. We 
use $E=1.053\times 10^{61} \,\rm erg$, $\gamma=5/3$, and $\rho_0 =
2.5626\times 10^{-24} \,\rm cm^{-3}$ for simulation results displayed in
Fig. \ref{fig-test}.

(C) The propagation of an I-front from a steady point-source in a
uniform, static medium: This is the
case where the classical description of the Str\"{o}mgren radius is
plausible, since gas is forced to remain static, and photoionization
and recombination are the only physical processes determining the
ionized fraction. The I-front from a point source with $N_*$ number of
ionizing photons evolves as
\begin{equation}
r_{\rm I} (t) = R_{\rm S} \left(1-\exp(-t/t_{\rm rec}) \right)^{1/3},
\label{eq:pure-ift}
\end{equation}
where $R_{\rm S}\equiv \left[ 3N_{*}/(4\pi n_{\rm H}^2 \alpha) \right]^{1/3}$
is the Str\"{o}mgren radius, $t_{\rm rec}\equiv 1/(n_{\rm H} \alpha)$
is the recombination time, and $\alpha$ is the recombination rate
coefficient. 
We adopt $N_* = 10^{47} \,\rm s^{-1}$, $n_{\rm H}=10 \,\rm cm^{-3}$,
and $\alpha = 1.05\times 10^{-13} \,\rm cm^{3} \,s^{-1}$. For this test, we
use a monochromatic light whose frequency is slightly above the
hydrogen ionization threshold.

(D) the gas-dynamical expansion of an H II region from a point source
in a uniform gas \citep{1966ApJ...143..700L}: The I-front, initially
propagating as a 
weak R-type front into a uniform medium, slows down and travels as a
D-type front, developing a shock front ahead of it. The I-front
evolves as
\begin{equation}
r_{\rm I} (\tilde{t}) = R_{\rm S,I} \left(1+\frac{7}{4}
  \frac{\tilde{t}}{t_{\rm sc}}\right)^{4/7}, 
\label{eq:lasker}
\end{equation}
where $R_{\rm S,I}$ is the initial Str\"{o}mgren radius,
  $\tilde{t}\equiv t-t_c$
  is the time measured from the moment $t_c$ when $dr_{\rm 
  I}/dt = c_{\rm I}$, and $c_{\rm I}\equiv (p/\rho)^{1/2}$ is the
  isothermal sound speed of the ionized gas \citep{1978ppim.book.....S}. 
We adopt $N_* = 2.45\times 10^{48} \,\rm s^{-1}$ and $n_{\rm H}=6.4\,\rm
cm^{-3}$. 
Following
  \citet{1966ApJ...143..700L}, we force temperature of the ionized gas 
to be $10^4 \,\rm K$, which gives $c_{\rm I}=12.86 \,\rm km/s$. 

(E) The gas-dynamical expansion-phase of the H II region from a
point-source in a nonuniform gas whose density varies with distance
$r$ from the source as $r^{-w}$, $w=3/2$
\citep*{1990ApJ...349..126F}: This case is similar to the
case (D), except that the density follows a power law, $n_{\rm H} \propto
r^{-w}$. Inside the core radius $r_c$, the density is constant at
$n_{{\rm H},c}$. The I-front evolves as 
\begin{equation}
r_{\rm I} (t) = R_{w}
\left[1+\frac{7-2w}{4}\left(\frac{12}{9-4w}\right)^{1/2}\frac{c_{\rm I}
    t}{R_w} \right],
\label{eq:franco}
\end{equation}
where $R_w$ is the size of the initial H II region obtained by
equating the ionization rate and the recombination rate. For
instance, when $w=3/2$,
\begin{equation}
R_{3/2} = r_c \exp
 \left\{\frac{1}{3} \left[ 
       \left( \frac{R_{\rm S}}{r_c} \right)^3 -1
     \right] \right\},
\end{equation}
where $R_{\rm S}\equiv \left[ 3N_{*}/(4\pi n_{{\rm H},c}^2 \alpha)
\right]^{1/3}$.

If $w\le 3/2$, the shock front always travels ahead of the I-front. If
$w> 3/2$, however, the shock front is overtaken by the I-front, which
soon runs to infinity in this ``champagne'' phase. We restrict
ourselves to this critical exponent $w=3/2$. From equation
(\ref{eq:franco}), we obtain $r_{\rm I} (t) = R_{3/2}\left( 1+2c_{\rm I}/
  R_{3/2}\right)$. In our simulation, we use $n_{{\rm
  H},c}=2\times 10^6 \,\rm cm^{-3}$, $r_c = 2.1\times 10^{16} \,\rm
cm$, $N_* = 5\times 10^{49}\,\rm s^{-1}$, and $\alpha = 2.6\times
10^{-13}\,\rm cm^3 s^{-1}$. Temperature of the ionized gas is set at
$T=8000\,\rm K$, such that $c_{\rm I} = 11.5\,\rm km/s$.

%

\end{document}